\documentclass[aps,pra,groupedaddress,twocolumn,notitlepage,superscriptaddress,10pt]{revtex4-1}

\usepackage{float}
\usepackage{graphicx}  
\usepackage{xcolor}
\usepackage{dcolumn}   
\usepackage{bm}        
\usepackage{braket}
\usepackage{amssymb}   
\usepackage{epstopdf}
\usepackage{xfrac}
\usepackage{cancel}
\usepackage{soul}
\usepackage{amsmath}
\usepackage{amsthm}
\usepackage{amsfonts}
\usepackage{bbm}
\definecolor{darkblue}{rgb}{0.1,0.2,0.6}
\definecolor{darkred}{rgb}{0.8,0.1,0.2}
\definecolor{darkgreen}{rgb}{0.31,0.62,0.24}
\usepackage[colorlinks,citecolor=darkblue,linkcolor=darkblue,urlcolor=darkblue]{hyperref} 
\usepackage{tikz}
\usepackage{enumerate}
\usepackage{setspace}
\usepackage{url}  
\usepackage{mathrsfs}

\newcommand{\rT}{{\rho^{T_2}} }
\newcommand{\rTa}{{\rho^{T_2}_A} }
\newcommand{\rTc}{{\rho^{T_2}} }
\newcommand{\Hi}{\mathcal{H}}
\newcommand{\Ni}{\mathcal{N}}

\newcommand{\Tr}{\text{Tr}}
\newcommand{\tr}{\text{Tr}}

\newcommand{\norm}[1]{\left\lVert#1\right\rVert}

\newcommand{\bigket}[1]{ {\big|{#1}\big\rangle} }

\newcommand{\bigbra}[1]{ {\big\langle{#1}\big|} }

\numberwithin{equation}{section}
\renewcommand\theequation{\arabic{section}.\arabic{equation}}

\usepackage{tikz}
\usetikzlibrary{positioning}
\usetikzlibrary{patterns}

\usepackage{tikzit}

\tikzstyle{basic}=[fill=white, draw=black, shape=circle]
\tikzstyle{basic rect}=[fill=white, draw=black, shape=rectangle]
\tikzstyle{medium box}=[fill=white, draw=black, shape=rectangle, minimum width=0.75cm, minimum height=0.75cm]
\tikzstyle{large box}=[fill=white, draw=black, shape=rectangle, minimum height=1.5cm, minimum width=1.5cm]

\tikzstyle{dash}=[-, dashed]
\tikzstyle{dotted}=[-, densely dotted, tikzit draw=magenta, thick]

\usepackage[normalem]{ulem}

\begin{document}

\title{Entanglement negativity spectrum of random mixed states: A diagrammatic approach}

\author{Hassan Shapourian}
 \affiliation{Microsoft Station Q, Santa Barbara, CA 93106, USA}
\affiliation{Department of Physics, Harvard University, Cambridge, MA~02138, USA}
 \affiliation{Department of Physics, Massachusetts Institute of Technology,
 Cambridge, MA~02139, USA}
 \author{Shang Liu}
 \affiliation{Department of Physics, Harvard University, Cambridge, MA~02138, USA}
\author{Jonah Kudler-Flam}
\affiliation{Kadanoff Center for Theoretical Physics, University of Chicago, Chicago, IL~60637, USA}
\author{Ashvin Vishwanath}
 \affiliation{Department of Physics, Harvard University, Cambridge, MA~02138, USA}

\date{\today}

\begin{abstract}

The entanglement properties of random pure states are relevant to a variety of problems ranging from chaotic quantum dynamics to black hole physics. The averaged bipartite entanglement entropy of such states admits a volume law and upon increasing the subregion size follows the Page curve.
In this paper, we generalize this setup to random mixed states by coupling the system to a bath
and use the partial transpose to study their entanglement properties. 
We develop a diagrammatic method to incorporate partial transpose within random matrix theory and formulate a  perturbation theory  in $1/L$, the inverse of the Hilbert space dimension. We compute several quantities including the spectral density of partial transpose (or entanglement negativity spectrum), two-point correlator of eigenvalues, and the logarithmic negativity.
As long as the bath is smaller than the system, we find that upon sweeping the subregion size, the logarithmic negativity shows an initial increase and a final decrease similar to the Page curve, while it admits a plateau in the intermediate regime where the logarithmic negativity only depends on the size of the system and of the bath but not on how the system is partitioned. 
This intermediate phase has no analog in random pure states, and is separated from the two other regimes by a critical point. We further show that when the bath is larger than the system by at least two extra qubits the logarithmic negativity is identically zero which implies that there is no distillable entanglement. Using the diagrammatic approach, we provide a simple derivation of the semi-circle law of the entanglement negativity spectrum in the latter two regimes. We show that despite the appearance of a semicircle distribution, reminiscent of Gaussian unitary ensemble (GUE), the higher order corrections to the negativity spectrum and two-point correlator deviate from those of GUE.
\end{abstract}

\maketitle


\section{Introduction}

Dynamics of strongly interacting quantum systems
is an interdisciplinary research frontier across various fields of physics including quantum computation, condensed matter and high energy physics.
Recent developments suggest that 
 scrambling of the quantum information and the emergence of thermalization as a result of dynamics in a closed quantum system are intimately related, which is in turn a consequence of what is generally known as quantum chaos~\cite{ETH_rev}.
It turned out that several universal properties of such interacting quantum chaotic systems can be reproduced via random matrix theory~\cite{BGS1984,GUHR1998,Berry_spectral_rigidity,2018PhRvX...8b1062K}.
As a result, random matrix theory can be used as an effective description where one can easily carry out calculations by utilizing standard techniques available in random matrix theory and gain further insights into universal features of dynamics of strongly correlated systems. 

Recently, it was realized that signatures of quantum chaos and thermalization can be understood from the reduced density matrix of a single many-body wave function~\cite{ChenLudwig}. In particular, the spectral properties of the reduced density matrix can be captured by the Wishart
random matrix theory.
A Wishart matrix can be obtained from a random pure state (Page or Haar state) $\ket{\Psi}$ in a bipartite Hilbert space $\Hi= \Hi_{A}\otimes \Hi_{B}$ via partial tracing $\rho_A=\Tr_B\ket{\Psi}\bra{\Psi}$. Here, the word random means uniformly distributed on unit norm states.
As is well-know from the pioneering work of Page~\cite{Page1993}, a typical random pure state obeys a volume law bipartite entanglement entropy.
Moreover, the entanglement spectrum is given by the Mar{\v c}enko-Pastur law~\cite{Forrester}.
A more recent study~\cite{Shenker2019} uses this setup 
as a toy model of black hole evaporation to shed light on the information paradox \footnote{Technically, the model in~\cite{Shenker2019} only parallels Wishart random matrix theory when the black hole is in the microcanonical ensemble.}.

\begin{figure}
\centering
\includegraphics[scale=.7]{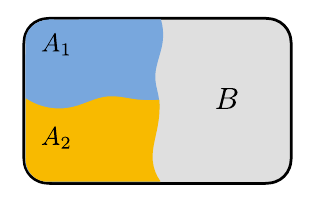}
\caption{\label{fig:geometry} Schematic representation of the system under study.
The goal here is to investigate 
 the typical amount of entanglement between $A_1$ and $A_2$ which are in a random mixed state due to
 coupling to subsystem $B$. The global state of the system $A\cup B$ is obtained from an ensemble of random pure (Page) states. In this setup, there is no notion of locality.}
\end{figure}

\begin{figure*}
\includegraphics[scale=0.58]{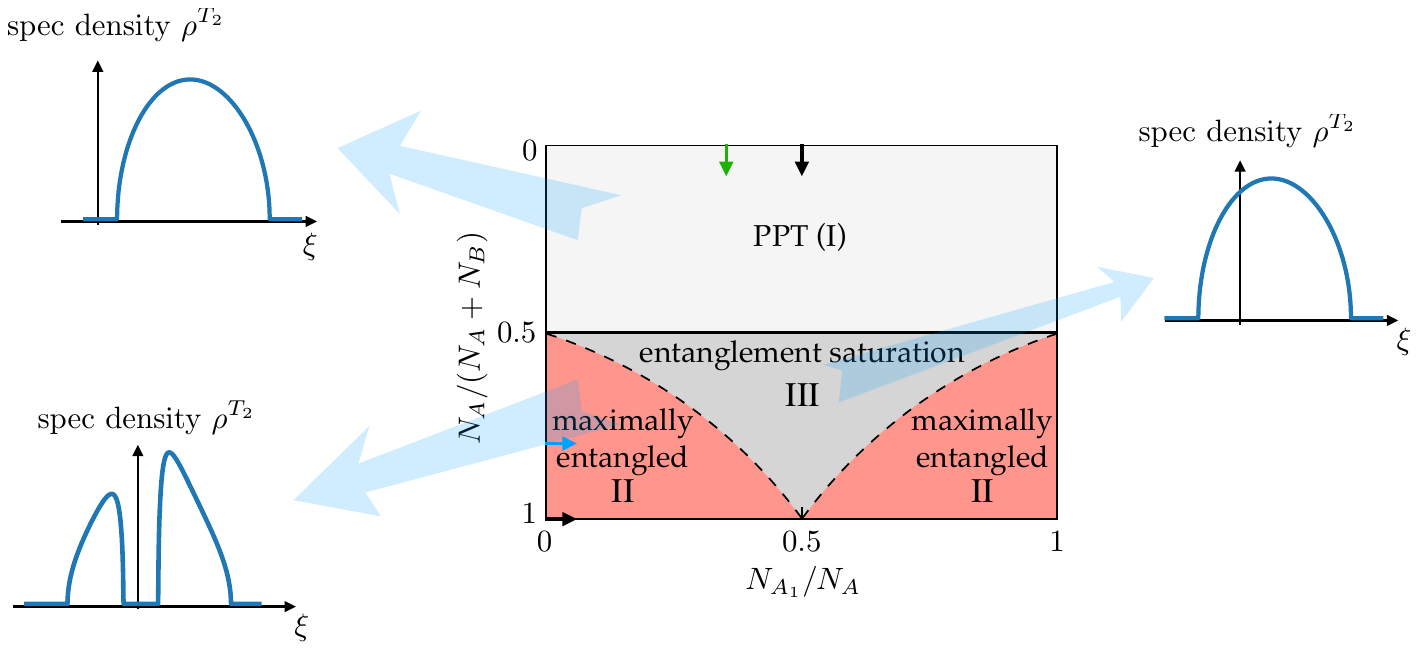}
\caption{\label{fig:phasediag} Phase diagram of reduced density matrix $\rho_A$ obtained from random pure (Page) states. The size of the environment $B$ grows along the vertical axis, while the horizontal axis characterizes the relative size of $A$ partitions. The horizontal line at the bottom $N_B=0$ corresponds to pure states where the Page transition occurs at $N_{A_1}=N_{A_2}=0.5 N_A$. A general form of the spectral density of $\rT$ is shown for each phase.
Arrows (color coded) indicate the paths along which LN is plotted in Fig.~\ref{fig:Neg_vs_L}.}
\end{figure*}

In this paper, we would like to build on the above observation and  further uncover entanglement structures in random mixed states. For this purpose, we consider a tripartite geometry where subsystem $A$ is further decomposed into two subsystems $A_1$ and $A_2$ (see Fig.~\ref{fig:geometry}) and investigate the entanglement (encoded in $\rho_A$) between $A_1$ and $A_2$.
Alternatively, the subsystem $B$ can be viewed as external degrees of freedom (or an environment) for subsystem $A$.
From this viewpoint, this construction provides a tuning parameter $q=L_B/L_A$ for generating random mixed states where $L_s= \dim {\cal H}_s$ for $s=A,B$. We shall call such density matrices random \emph{induced} mixed states. Intuitively, when $q\ll 1$, the bath is small and not capable of thermalizing the system. Hence, we expect the density matrix to behave similarly to Page states where there is a volume-law entanglement. In the other extreme limit, $q\gg 1$, the system is almost fully entangled with the bath (which has many degrees of freedom), and we expect $\rho_A$ to be maximally mixed (or thermal) where there is little quantum correlation between $A_1$ and $A_2$. Comparing the two extreme limits, we then anticipate an entanglement phase transition 
from a volume law mixed state to an unentangled (separable) state.
In this work, we develop a large-$L$ perturbation theory to quantitatively establish this intuitive picture.
In fact, our calculations reveal a richer picture than this intuition suggests. The density matrix in the volume law limit can further be divided into two types: saturated and maximally entangled states (see the lower region of the phase diagram in Fig.~\ref{fig:phasediag}). We explain this fact in greater detail near the end of this section after we introduce our methodology.

To quantify the entanglement in mixed states, we use the partial transpose (PT)~\cite{Peres1996,Horodecki1996,Simon2000,PhysRevLett.86.3658,PhysRevLett.87.167904,Zyczkowski1,Zyczkowski2}, 
and the associated entanglement measure, the logarithmic negativity (LN).
Our choice of PT is motivated by the following reasons:
First, PT exclusively diagnoses 
quantum correlations, as opposed to the usual pure state entanglement measures such as von Neumann and R\'enyi entropies which capture both quantum and classical correlations. Instead, one may use the mutual information $I_{A_1:A_2}=S_{A_1}+S_{A_2}-S_A$ to quantify the mixed state entanglement. However, the mutual information is not an entanglement measure either and overestimates the entanglement of specific type of classically correlated states called separable to be defined below.
Second, the partially transposed density matrix is a single operator which characterizes the state and we can study not only the LN but also its spectral properties, a.k.a.\ the negativity spectrum~\cite{Ruggiero_2,Shap_spectrum}, which can be thought of as the analog of pure state entanglement spectrum for mixed states.
Lastly, LN can be computed numerically efficiently and, as we show in this paper, can also be calculated analytically for the case of random states in the large Hilbert space limit.

The PT of random induced mixed states was previously studied~\cite{Marko2007,Aubrun2010,Aubrun2014,Bhosale,Aubrun2012,Fukuda2013,Collins2012,Szyma_ski_2017,Collins_rev,Gray2018}. In particular, it was numerically found that in the large Hilbert space limit and when the two subsystems are of identical size, $L_{A_1}=L_{A_2}$, the negativity spectrum obeys a
semi-circle law and there exists a transition from positive partial transpose (PPT) states to negative partial transpose (NPT) states at $L_B=4L_A$~\cite{Marko2007,Aubrun2014,Bhosale}. Remarkably, Refs.~\cite{Aubrun2010, Banica_resolvent, Fukuda2013} used the combinatorics of non-crossing partitions in free probability theory to put this result on firm grounds by showing that $\rTa$ (which denotes partially transposed $\rho_A$) converges in moments to a semicircular distribution~\footnote{ 
A technical remark is that PPT states are not necessarily separable. Interestingly, Refs.~\cite{Aubrun2012} showed that in order for $\rho_A$ to be separable, we need to have an environment $B$ larger than $A$ by at least $L_B\geq L_A^{3/2}$, or in terms of qubits $N_B> 3N_A/2$ or $N_A< 2N/5$. This implies that a typical $\rho_A$ in the intermediate regime $0.4 < N_A/N < 0.5$ (right above the horizontal line in the phase diagram) is a PPT state but inseparable (i.e., bound entangled).}. In this work, we complement the earlier analyses by developing a graphical representation for partial transpose and present a systematic way to derive the moments and the resolvent function which is used to calculate the negativity spectrum. Our results match those of Refs.~\cite{Aubrun2010, Banica_resolvent, Fukuda2013} where they overlap. 
We further evaluate the $1/L$ corrections to the resolvent function as well as the two-point correlation function of eigenvalues of $\rTa$ 
to establish that $\rTa$ does not belong to the Gaussian unitary ensemble (GUE) despite the fact that its spectral density approaches a semi-circle law.
Compared to previous methods~\cite{Aubrun2010,Banica_resolvent}, our diagrammatic approach has the advantage of not requiring familiarity in specialized areas such as free probability theory and being more accessible to physicists and potentially generalizable to other quantum systems of great interest in the physics community such as the black hole information problem~\cite{Shenker2019} and equilibrated pure states satisfying the eigenstate thermalization hypothesis~\cite{Vardhan2020}. 
Last but not least, we use our quantitative results from random matrix theory to highlight some physical observations regarding tripartite entanglement in random pure states as we briefly explain below.

Our main results in this paper are summarized in the phase diagram of Fig.~\ref{fig:phasediag} which is obtained analytically in the thermodynamic limit where total number of qubits $N_A+N_B$ are infinitely large while each subsystem contains a finite portion of the full system~\footnote{Note that because of the Hilbert space dimension ($L_{s}$) exponential dependence on number of qubits $L_s=2^{N_s}$, for instance, the limits $N_{A}>N_B$ or $N_{A_1}>N_{A_2}$ translate into $L_A\gg L_B$ or $L_{A_1}\gg L_{A_2}$ regimes, respectively, in the thermodynamic limit.}. 
 This $2d$ phase diagram can be thought of as an extension of the $1d$ phase diagram of Page states, which corresponds to the horizontal line in the bottom with $\rho_A$ pure,
by adding a vertical axis to parameterize how mixed $\rho_A$ is.
In other words, the regions labeled by II in the phase diagram are remnants of the two phases in the pure state limit and the two phase boundaries denoted by dashed lines are remnants of the Page transition with a diverging spectral density at zero.

\begin{figure}
\centering
\includegraphics[scale=0.6]{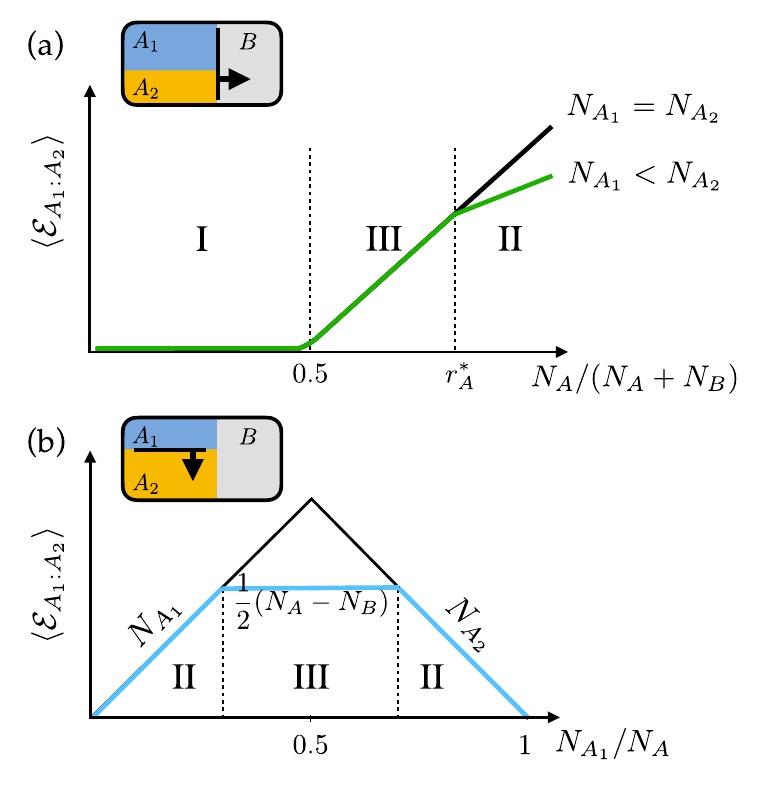}
\caption{\label{fig:Neg_vs_L} Logarithmic negativity between subsystems $A_1$ and $A_2$ (a) as we increase the size of $A$ while keeping the ratio ${N_{A_1}}/{N_{A_2}}$ fixed and (b) as we increase the size of $A_1$ while keeping $N_A/N_B$ fixed. The corresponding paths (vertical for (a) and horizontal for (b)) are shown by arrows in Fig.~\ref{fig:phasediag}.
For reference in panel (b), we also show the Page curve in black.}
\end{figure}

As we move vertically from top to bottom in the phase diagram, as long as $L_B>4 L_A$, i.e., the environment is larger than subsystem $A$ at least by two more qubits ($N_B> N_A+2$), $\rho_A$ is typically a PPT state with a vanishing LN (phase I in Fig.~\ref{fig:phasediag})~\footnote{Our diagrammatic approach correctly captures a factor of four at the critical point between PPT and NPT states $L_B=4L_A$ in agreement with what was previously observed numerically~\cite{Aubrun2010,Bhosale}.
However, the two extra qubits in the description in terms of qubits in an infinite system may seem to be unimportant since the ratio of number of qubits at the critical point is what matters and approaches $N_A/N_B \to 1 + O(1/N)$.}. Moreover, the negativity spectrum is given by a semi-circle over positive domain. 
On the other hand, as we go further down and reach the regime $L_B< 4 L_A$, the density matrix is typically an NPT state (phase II and III) and LN is non-zero. 
This trend is explicitly shown in Fig.~\ref{fig:Neg_vs_L}(a).
The NPT states can further be divided into two sub-categories depending on the scaling behavior of the LN. To see this, it is easier to follow the trend of LN (plotted in Fig.~\ref{fig:Neg_vs_L}(b)) as we move horizontally from left to right in the NPT regime of the phase diagram. When subsystem $A_1$ is much smaller than $A_2$, the amount of entanglement between $A_1$ and $A_2$ is bounded by the size of $A_1$, i.e., ${\cal E}_{A_1:A_2}\sim N_{A_1}$. Similarly, in the other extreme regime when $A_2$ is very small, we have a similar phenomenon for $A_2$ and ${\cal E}_{A_1:A_2}\sim N_{A_2}$. Therefore, LN initially increases linearly in $N_{A_1}$ and decreases to zero eventually as shown in Fig.~\ref{fig:Neg_vs_L}(b). 
We call this regime maximally entangled since the smaller subsystem is fully entangled to its complement within subsystem $A$. This limit can also be thought of as bipartite entangled since there is no entanglement between the smaller subsystem of $A$ and $B$.
The negativity spectrum in this regime consists of two disjoint Mar{\v c}enko-Pastur law distributions, one with negative support and one with positive.
For the intermediate regime where both ${A_1}$
 and ${A_2}$ are comparable in size and occupy less than half of the total system, LN is given by ${\cal E}_{A_1:A_2}\sim \frac{1}{2}(N_{A}-N_B)$ which is independent of the ratio $\frac{N_{A_1}}{N_{A_2}}$. The plateau in Fig.~\ref{fig:Neg_vs_L}(b) corresponds to this regime and because of that, we call it entanglement saturation. We should note that the state in this regime is tripartite entangled since all three $A_1$, $A_2$, and $B$ parties are mutually entangled.
 The negativity spectrum in this case is given by a semi-circle law which partly covers the negative domain.

The rest of our paper is organized as follows:
In Sec.~\ref{sec:Preliminary},
we review some background materials about the partial transpose and its applications.
In Sec.~\ref{sec:numerics}, 
we present the setup by which we generate random mixed states.
Section~\ref{sec:diagrammatic} contains the central result of this paper, where we propose how PT can be incorporated in the graphical approach to the random matrix theory. We further run some sanity checks for our proposal including computing the R\'enyi negativity in this section.
In Sec.~\ref{sec:PTWishartSpectralDensity}, we use this diagrammatic approach to calculate the resolvent function and the negativity spectrum where we map out the phase diagram in Fig.~\ref{fig:phasediag}. 
We further derive the higher order corrections to the semi-circle law.
Next, in Sec.~\ref{sec:2pt function}, we discuss the connected diagrams associated with the two-point function of eigenvalues and show that it is different from that of GUE.
Finally, we finish our paper by several closing remarks and future directions in Sec.~\ref{sec:discussion}. A summary of notation used throughout the paper is given in Table~\ref{tab:notation}.

\begin{table}[]
    \centering
{\footnotesize 
\renewcommand{\arraystretch}{1.2}
\begin{tabular}{cc}
    \hline
    Variable & Description \\
    \hline
    \hline
    $L$ &  Total Hilbert space dim.
     \\
    $L_s$ & Hilbert space dim. of subsys. $s$
     \\
    $N_s$ & $\log_2 L_s$
     \\
    \hline
    $q$ & ${L_B}/{L_A}$
     \\
    $\eta$ & $L_{A_1}/L_{A_2}$
     \\
    $\alpha $ & $L_B/L_{A_1}$
     \\
    $\beta$ & $L_B L_{A_2}/L_{A_1}$
     \\
    \hline
\end{tabular}
}
\renewcommand{\arraystretch}{1}
    \caption{Summary of notations in the paper.}
    \label{tab:notation}
\end{table}

\section{Review of the partial transpose}
\label{sec:Preliminary}

In this section, we briefly review some basics about the PT and LN. 
Expert readers may skip this part.
 Historically, LN has been shown to be useful in studying various quantum many-body systems including 
harmonic oscillator chains~\cite{PhysRevA.66.042327,PhysRevLett.100.080502,PhysRevA.78.012335,Anders2008,PhysRevA.77.062102,PhysRevA.80.012325,Eisler_Neq,Sherman2016,dct-16}, quantum spin models~\cite{PhysRevA.80.010304,PhysRevLett.105.187204,PhysRevB.81.064429,PhysRevLett.109.066403,Ruggiero_1,PhysRevA.81.032311,PhysRevA.84.062307,Mbeng,Gray2019,Grover2018,java-2018}, (1+1)d conformal and integrable field theories~\cite{Calabrese2012,Calabrese2013,Calabrese_Ft2015,Ruggiero_2,Alba_spectrum,kpp-14,fournier-2015,bc-16,Wald_2020,PhysRevB.101.064207,lu2019structure,angelramelli2020logarithmic,Juh_sz_2018,Schreiber,rosz2020entanglement,Shapourian_FS,wu2019entanglement,10.21468/SciPostPhys.8.4.063},  topologically phases of matter~\cite{Wen2016_1,Wen2016_2,PhysRevA.88.042319,PhysRevA.88.042318,hc-18,YeJePark,PhysRevB.101.085136},
and in out-of-equilibrium dynamics \cite{ctc-14,ez-14,hb-15,ac-18b,wen-2015,gh-18,alba2020spreading,Gruber_2020,PhysRevB.101.245130,Kudler-Flam2019,2020JHEP...04..074K,2020arXiv200811266K}, 
 as well as holographic theories~\cite{Rangamani2014,2014JHEP...09..010K,PhysRevD.99.106014,PhysRevLett.123.131603} and variational states~\cite{Clabrese_network2013,Alba2013,PhysRevB.90.064401,Nobili2015}.
More recently, PT was used to construct many-body order parameters for symmetry protected topological phases protected by anti-unitary symmetries~\cite{Pollmann_Turner2012,Shiozaki_Ryu2017,Shap_unoriented,Shiozaki_antiunitary,Kobayashi} and there are experimental proposals to measure it with ion traps and cold atoms~\cite{Elben2019,gbbs-17,csg-18,Elben2020,PhysRevLett.125.200502}.
 
The PT of a density matrix written in a local orthonormal basis  $\{\bigket{e_1^{(k)}}, \bigket{e_2^{(j)}}  \}$,
\begin{align}
\rho_A=\sum_{ijkl} \rho_{ijkl} \bigket{e_1^{(i)}, e_2^{(j)}}   
 \bigbra{e_1^{(k)}, e_2^{(l)}},
\end{align}
is defined by exchanging the indices of subsystem $A_1$ (or $A_2$) as in
\begin{align} \label{eq:rTb}
\rho_A^{T_2}=\sum_{ijkl} \rho_{ijkl} \Ket{e_1^{(i)}, e_2^{(l)}}  \Bra{e_1^{(k)}, e_2^{(j)}}.
\end{align}
We can understand the effect of PT by comparing it with the full transpose $\rho^T_A$. Recall that full transpose is a Hermitian/trace preserving map which leads to a new operator $\rho^T_A=\rho^\ast_A$ with identical eigenvalues to those of $\rho_A$ (hence, transpose is a completely positive map). Similarly, PT is a Hermitian/trace preserving map, which implies that the eigenvalues of $\rTa$ are all real. However, unlike the full transpose, PT is not a completely positive map, i.e., $\rTa$ may have negative eigenvalues. The existence of negative eigenvalues for $\rTa$ turns out to be an indicator of quantum correlations in $\rho_A$.
Based on this fact, the PT test is designed to distinguish quantum correlated mixed states from classically correlated ones. This test is followed by checking whether or not $\rTa$ contains any negative eigenvalues.
The negative eigenvalues of $\rTa$ can then be summed over to construct a measure of entanglement~\cite{PlenioEisert1999,Vidal2002,Plenio2005}
 \begin{align} 
 \label{eq:neg_def}
{\cal N}_{A_1:A_2} &= \frac{\norm{\rTa }_1 -1}{2}, 
\end{align}
in which $\norm{O}_1= \Tr \sqrt{OO^\dag}$ is the trace norm. Since $\rTa$ is Hermitian, the trace norm is simply the sum of the absolute value of its eigenvalues.
In other words, the above quantity directly measures how negative the eigenvalues are. For this reason, it is called the negativity.
Another useful quantity is called the logarithmic negativity (LN) and can be constructed as
 \begin{align} 
 \label{eq:logneg_def}
{\cal E}_{A_1:A_2} &= \log_2 \norm{\rTa }_1.
\end{align}
LN and the negativity are related via ${\cal E}=\log (2\Ni+1)$.
To further characterize $\rho_A$, one can study the distribution of the eigenvalues of $\rTa$ instead of ${\cal E}_{A_1:A_2}$ (or ${\cal N}_{A_1:A_2}$) which are single numbers.
It can be defined formally by
\begin{align} \label{eq:PTdist}
P_\Gamma (\xi)=\sum_{i=1}^{L_A} \delta(\xi-\xi_i),
\end{align}
where the eigenvalues of $\rT$ are denoted by $\{ \xi_i \}$. 
This quantity is usually referred to as the entanglement negativity spectrum (or in short the negativity spectrum) in literature~\cite{Ruggiero_2,Shap_spectrum}.
From the definition of LN (\ref{eq:neg_def}), it is easy to see that
\begin{align}
\label{eq:neg_dist}
\braket{ {\cal N}_{A_1:A_2} } =  -\int_{\xi<0} d\xi\ \xi\, P_\Gamma(\xi).
\end{align}

An important application of the PT is to identify separable states. A separable state is a completely classical state which can be written in the following form 
\begin{align}
\label{eq:separable}
\rho_{\text{sep}} = \sum_{i,j} p_{ij} \rho_1^{(i)}\otimes \rho_2^{(j)},
\qquad p_{ij}>0.
\end{align}
By definition, $\rho_{\text{sep}}$ remains positive semi-definite even after PT since the transposed operators $(\rho_2^{(j)})^T$ (or $(\rho_1^{(i)})^T$ in the case of $\rho_{\text{sep}}^{T_1}$) are positive semi-definite.
This observation indicates that separable states are a subset of PPT states~\footnote{A technical point is that the converse is not true, i.e., the PPT condition does not imply separability, since there exists a family of inseparable states with positive PT (the so-called  bound entangled states)~\cite{Horodecki1997}. Nonetheless, the entanglement in these states cannot be distilled to carry out quantum computing processes~\cite{Horodecki1998}}, or equivalently, an NPT state cannot be separable.
It is known that PPT implies a lack of distillable entanglement. The converse, i.e. that NPT necessarily implies a finite amount of distillable entanglement, remains to be shown, although no counterexamples are known at present~\cite{Zyczkowski2020}. In this paper, we only distinguish PPT and NPT states without reference to distillability.

\section{Random induced mixed states}
\label{sec:numerics}

In this section, we introduce our setup which we use to carry out various calculations in the coming sections. We would like to study the entanglement in random mixed states. As mentioned in the previous section, we choose to use random induced mixed states for this purpose.
An ensemble of random induced mixed states $\{ \rho_A \}$ corresponding to the Hilbert space $\Hi_A=\Hi_{A_1}\otimes \Hi_{A_2}$ is generated by reduced density matrices which are obtained by partial tracing random pure states (or Page states~\cite{Page1993}) 
in a composite Hilbert space $\Hi_{A}\otimes \Hi_{B}$.
It is more convenient to represent such a random pure state in
a tensor product basis as in
\begin{align}
\label{eq:page_tri}
\ket{\Psi}=\sum_{i=1}^{L_A} \sum_{\alpha=1}^{L_B} X_{i\alpha} \ |\Psi^{(i)}_{A} \rangle\otimes|\Psi^{(\alpha)}_B\rangle,
\end{align}
in terms of a $L_A \times L_B$ rectangular random matrix $X$
whose elements ($X_{i \alpha}$) are independent Gaussian random complex variables where the joint probability density is defined by
\begin{align}
    \label{eq:Xprob}
    P\left( \{X_{i\alpha} \} \right) = {\cal Z}^{-1} \exp \left\{- L_A L_B \Tr(XX^\dag) \right\}.
\end{align}
Here, $L_{A}=L_{A_1}\times L_{A_2}$ and $L_B$ denote the size of $\Hi_A$ and $\Hi_B$, respectively, and $ {\cal Z}$ is the overall normalization constant.
Throughout this paper, we consider $A$ and $B$ systems to be comprised of qubits. In other words,  $L_s=2^{N_s}$ where $s=A_1,A_2,B$ and $N_s$ is the number of qubits. This choice is not a necessary ingredient for our calculations and is mainly meant as a physical description of the system.

The random reduced density matrix of system $A$ is then given by
\begin{align}
\label{eq:red_den}
\rho_{A}=\frac{XX^{\dag}}{\tr (X X^\dag)}.
\end{align}
We note that $\rho_A$ is a $L_A \times L_A$ square matrix, and  the denominator (which is also a random variable) is there to enforce the normalization condition $\Tr \rho_A=1$.

{\bf Entanglement spectrum:} The eigenvalues of $\rho_A$ contains information about the entanglement between $A$ and its bath $B$. In the limit $L_A, L_B \to \infty$ while the ratio $L_A/L_B$ is finite (which we will refer to as the large $L$ limit), 
the joint probability density function of these eigenvalues can be derived~\cite{Lloyd1988, Zyczkowski2001}. In doing so, the crucial point is that the normalization factor $\Tr (XX^\dag)= 1+ \delta$ in Eq.\,(\ref{eq:red_den}) is a random variable whose fluctuations about its mean value $1$ is negligible to the leading order in $\frac{1}{L_A L_B}$. Hence, to the leading order, the denominator can be replaced by its mean value, and we may write
\begin{align}
 \rho_A \approx XX^\dag,
 \label{eq:Wishart}
\end{align}
  which is the celebrated  Wishart-Laguerre ensemble~\cite{Forrester} and is extensively studied in the random matrix theory literature. From this observation, one can infer several properties of $\rho_A$ in the large $L$ limit.
  First, the ensemble-averaged von Neumann entanglement entropy is given by
\begin{align}
\label{eq:EE_page}
\langle S_A \rangle = -\braket{ \Tr (\rho_A \ln \rho_A) } =\ln L_{A}  -\frac{L_{A} }{2L_{B}}.
\end{align}
which in terms of number or qubits grows as $N_{A}$ to the leading order, i.e., a volume law.
  Second, the spectral density of the eigenvalues $\{\lambda_i \}$ is given by an appropriately scaled Mar{\v c}enko-Pastur (MP) law \cite{Forrester},
\begin{align}
P(\lambda)&  =
\sum_{i=1}^{L_A} \delta(\lambda-\lambda_i)
= \frac{q L_A^2}{2\pi } \frac{\sqrt{(\lambda_{+} -\lambda )(\lambda- \lambda_{-} )}}{\lambda},\\
\lambda_{\pm} &= \frac{1}{L_A} (1\pm 1/\sqrt{q})^2, 
\label{eq:MPdist}
\end{align}
where $\lambda\in [\lambda_{-}, \lambda_{+} ]$, $q=L_B/ L_A \geq 1$. 
To name a few properties of the above expression, let us mention that for $q=1$, $P(\lambda)$ diverges at the origin as ${\lambda}^{-1/2}$ and for $q\neq 1$ the eigenvalues are bounded away 
from zero. When $q < 1$, $\rho_A$ is rank deficient and the entanglement spectrum includes a delta-function at the origin in addition to the MP distribution.

In the following section, we develop a diagrammatic method to graphically represent the density matrix (\ref{eq:Wishart}) and its PT and introduce a systematic way to compute their moments. As we will see, one benefit of this approach is that it is easy to find the general trend of dominant diagrams in various regimes.


\section{Partial transpose in diagrammatic approach}
\label{sec:diagrammatic}

In this section, we introduce our
diagrammatic approach to treat 
random induced mixed states.
 This method is based on the `t~Hooft $1/L$ (double line) perturbation theory which was also used recently in the context of black hole information problems~\cite{Shenker2019}.
 We propose a diagrammatic implementation for the PT of random matrices. The latter was inspired by a recent development in a seemingly disjoint topic: anyonic PT~\cite{Shap_anyons} that was proposed to diagnose 
entanglement in unitary modular tensor categories, which are an effective description of topologically ordered phases of matter~\cite{Kitaev2006}.

We begin by reviewing the diagrammatic approach to random induced states~\cite{Jurkiewicz,Brezin:1993qg,Zee1995}.
A matrix element of the pure state density matrix associated with the Page state (\ref{eq:page_tri}) is denoted as
\begin{align}
    \left[\ket{\Psi}\bra{\Psi}\right]_{i\alpha,j\beta}= X_{i\alpha}^\ast X_{j\beta}
    =
    \,
    \tikz[baseline=-0.5ex]{
    \draw[dashed] (0,0.2) node[align=center, above] {\footnotesize $\alpha$} -- (0,-0.2);
    \draw[dashed] (1,0.2) node[align=center, above] {\footnotesize $\beta$} -- (1,-0.2);
    \draw (-0.2,0.2) node[align=center, above] {\footnotesize $i$} -- (-0.2,-0.15);
    \draw (1.2,0.2) node[align=center, above] {\footnotesize $j$} -- (1.2,-0.15);
    }\ ,
\end{align}
where the left (right) pair of lines represent a bra (ket) state, and solid (dashed) lines correspond to subsystem $A$ ($B$). Note that each line carries an index. The lower end of the diagrams are reserved for matrix manipulations such as tracing and multiplication, while the upper ends of the lines are used for ensemble averaging.

Therefore, a matrix element of the reduced density matrix is represented by
\begin{align}
    \label{eq:rho_diag}
    [\rho_A]_{i,j}= 
    {\sum_{\alpha=1}^{L_B} X_{i\alpha}^\ast X_{j\alpha}}
= 
\,
    \tikz[baseline=-0.5ex]{
    \draw[dashed] (0,0.2) node[align=center, above] {\footnotesize $\alpha$} -- (0,0);
    \draw[dashed] (0,0)  -- (1,0);
    \draw[dashed]  (1,0.2) node[align=center, above] {\footnotesize $\alpha$} -- (1,0);
    \draw (-0.2,0.2) node[align=center, above] {\footnotesize $i$} -- (-0.2,-0.15);
    \draw (1.2,0.2) node[align=center, above] {\footnotesize $j$} -- (1.2,-0.15);
    }\ .
\end{align}
For brevity, from now on we drop the subscript $A$ and set  $\rho = \rho_A$ 
unless stated otherwise.
Similarly, tracing over the subsystem $A$ degrees of freedom leads to the following diagram, 
\begin{align}
    \Tr\rho=
    \sum_{i=1}^{L_A} \sum_{\alpha=1}^{L_B} |X_{i\alpha}|^2
    =
    \,
    \tikz[baseline=-0.5ex]{
    \draw[dashed] (0,0.2) node[align=center, above] {\footnotesize $\alpha$} -- (0,0);
    \draw[dashed] (0,0) -- (1,0);
    \draw[dashed]  (1,0.2) node[align=center, above] {\footnotesize $\alpha$} -- (1,0);
    \draw (-0.2,0.2) node[align=center, above] {\footnotesize $i$} -- (-0.2,-0.15)--(1.2,-0.15)-- (1.2,0.2) node[align=center, above] {\footnotesize $i$};
    }\ .
\end{align}
The ensemble averaging over the probability distribution (\ref{eq:Xprob}) is accounted by connecting the diagrams (from the top) with a double line as in
\begin{align}
\tikz[baseline=-0.1ex,scale=0.9]{
    \draw[dashed] (1.0,0) arc (0:180:0.5);
    \draw (1.2,0.) arc (0:180:0.7);
    }
    \ \,
    := \braket{X_{i\alpha}^\ast X_{j \beta}}=  \frac{1}{L_A L_B}\, \delta_{ij} \delta_{\alpha\beta},
    \label{eq:doubleline1}
\end{align}
where the braket $\braket{\ \cdot\ }$ denotes the ensemble average over the probability distribution (\ref{eq:Xprob}).
For example, we may write
\begin{align}
    \braket{\Tr\rho}=
    \,
    \tikz[baseline=0ex]{
    \draw[dashed] (1.0,0) arc (0:180:0.5);
    \draw[dashed] (0,0) -- (1,0);
    \draw (1.2,0.) arc (0:180:0.7);
    \draw (-0.2,0.0)-- (-0.2,-0.15)--(1.2,-0.15)-- (1.2,0.0);
    }\ = 1,
\end{align}
where for every solid (dashed) loop we multiply by a factor of $L_A$ ($L_B$).
This ensures the correct normalization on average. 
R\'enyi entropies can be computed similarly. From now on, we omit the matrix indices in diagrams for simplicity. 

{\bf Review of R\'enyi Entropy Calculations:} We now review how second and third R\'enyi entropies are calculated and briefly touch upon the leading order diagrams in the two regimes $L_A\gg L_B$ and $L_A\ll L_B$. These would constitute a useful reference for our later comparison with R\'enyi negativities.
The purity (or second R\'enyi entropy) is evaluated as follows,
\begin{align}
\label{eq:tr_r2}
    \Tr\rho^2= 
    \,
    \tikz[baseline=0ex]{
    \draw[dashed] (0,0.2)-- (0,0);
    \draw[dashed] (0,0) -- (1,0);
    \draw[dashed]  (1,0.2) -- (1,0);
    \draw (-0.2,0.2)-- (-0.2,-0.25);
    \draw (1.2,-0.1)-- (1.2,0.2);
    \draw[dashed] (2,0.2)-- (2,0);
    \draw[dashed] (2,0) -- (3,0);
    \draw[dashed]  (3,0.2) -- (3,0);
    \draw (1.8,0.2)-- (1.8,-0.1);
    \draw (3.2,-0.25)-- (3.2,0.2);
    \draw (1.2,-0.1)-- (1.8,-0.1);
    \draw (-0.2,-0.25) -- (3.2,-0.25);
    }\ ,
\end{align}
and its ensemble average is found to be
\begin{align}
    \braket{\Tr\rho^2} &=
    \,
    \tikz[scale=0.8,baseline=0.5ex]{
    \draw[dashed] (0,0) -- (1,0);
    \draw (-0.2,0.)-- (-0.2,-0.25);
    \draw (1.2,-0.1)-- (1.2,0.);
    \draw[dashed] (2,0) -- (3,0);
    \draw (1.8,0.)-- (1.8,-0.1);
    \draw (3.2,-0.25)-- (3.2,0.);
    \draw (1.2,-0.1)-- (1.8,-0.1);
    \draw (-0.2,-0.25) -- (3.2,-0.25);
    \draw[dashed] (1.0,0) arc (0:180:0.5);
    \draw[dashed] (3.0,0) arc (0:180:0.5);
    \draw (1.2,0) arc (0:180:0.7);
    \draw (3.2,0) arc (0:180:0.7);
    }
\ \,
+
\ \,
    \tikz[scale=0.7,baseline=0.5ex]{
    \draw[dashed] (0,0) -- (1,0);
    \draw (-0.2,0.)-- (-0.2,-0.25);
    \draw (1.2,-0.1)-- (1.2,0.);
    \draw[dashed] (2,0) -- (3,0);
    \draw (1.8,0.)-- (1.8,-0.1);
    \draw (3.2,-0.25)-- (3.2,0.);
    \draw (1.2,-0.1)-- (1.8,-0.1);
    \draw (-0.2,-0.25) -- (3.2,-0.25);
    \draw[dashed] (2.0,0) arc (0:180:0.5);
    \draw[dashed] (3.0,0) arc (0:180:1.5);
    \draw (1.8,0) arc (0:180:0.3);
    \draw (3.2,0) arc (0:180:1.7);
    }
\ \,
\nonumber
\\
&=\frac{1}{L_A} + \frac{1}{L_B},
\end{align}
which matches the exact result~\cite{Lubkin},
$\frac{L_A+L_B}{L_A L_B+1}$,
in the large $L_AL_B$ limit.

The third moment of $\rho$ is given by,
\begin{align}
\label{eq:tr_r3}
    \braket{\Tr\rho^3} =&
    \
    \tikz[scale=0.45,baseline=0.5ex]{
    \draw[dashed] (0,0) -- (1,0);
    \draw (-0.2,0.)-- (-0.2,-0.35);
    \draw (1.2,-0.15)-- (1.2,0.);
    \draw[dashed] (2,0) -- (3,0);
    \draw (1.8,0.)-- (1.8,-0.15);
    \draw (3.2,-0.15)-- (3.2,0.);
    \draw[dashed] (4,0) -- (5,0);
    \draw (3.8,0.)-- (3.8,-0.15);
    \draw (5.2,-0.35)-- (5.2,0.);
    \draw (1.2,-0.15)--(1.8,-0.15);
    \draw (3.2,-0.15)--(3.8,-0.15);
    \draw (-0.2,-0.35)-- (5.2,-0.35);
    \draw[dashed] (4.0,0) arc (0:180:0.5);
    \draw[dashed] (2.0,0) arc (0:180:0.5);
    \draw[dashed] (5.0,0) arc (0:180:2.5);
    \draw (3.8,0) arc (0:180:0.3);
    \draw (1.8,0) arc (0:180:0.3);
    \draw (5.2,0) arc (0:180:2.7);
    }  \,
    +
    3\times\,
\tikz[scale=0.45,baseline=0.5ex]{
    \draw[dashed] (0,0) -- (1,0);
    \draw (-0.2,0.)-- (-0.2,-0.35);
    \draw (1.2,-0.15)-- (1.2,0.);
    \draw[dashed] (2,0) -- (3,0);
    \draw (1.8,0.)-- (1.8,-0.15);
    \draw (3.2,-0.15)-- (3.2,0.);
    \draw[dashed] (4,0) -- (5,0);
    \draw (3.8,0.)-- (3.8,-0.15);
    \draw (5.2,-0.35)-- (5.2,0.);
    \draw (1.2,-0.15)--(1.8,-0.15);
    \draw (3.2,-0.15)--(3.8,-0.15);
    \draw (-0.2,-0.35)-- (5.2,-0.35);
    \draw[dashed] (2.0,0) arc (0:180:0.5);
    \draw[dashed] (3.0,0) arc (0:180:1.5);
    \draw[dashed] (5.0,0) arc (0:180:0.5);
    \draw (1.8,0) arc (0:180:0.3);
    \draw (3.2,0) arc (0:180:1.7);
    \draw (5.2,0) arc (0:180:0.7);
    }  
    \nonumber \\ \nonumber \\
   & +
    \,
    \tikz[scale=0.5,baseline=-0.5ex]{
    \draw[dashed] (0,0) -- (1,0);
    \draw (-0.2,0.)-- (-0.2,-0.35);
    \draw (1.2,-0.15)-- (1.2,0.);
    \draw[dashed] (2,0) -- (3,0);
    \draw (1.8,0.)-- (1.8,-0.15);
    \draw (3.2,-0.15)-- (3.2,0.);
    \draw[dashed] (4,0) -- (5,0);
    \draw (3.8,0.)-- (3.8,-0.15);
    \draw (5.2,-0.35)-- (5.2,0.);
    \draw (1.2,-0.15)--(1.8,-0.15);
    \draw (3.2,-0.15)--(3.8,-0.15);
    \draw (-0.2,-0.35)-- (5.2,-0.35);
    \draw[dashed] (1.0,0) arc (0:180:0.5);
    \draw[dashed] (3.0,0) arc (0:180:0.5);
    \draw[dashed] (5.0,0) arc (0:180:0.5);
    \draw (1.2,0) arc (0:180:0.7);
    \draw (3.2,0) arc (0:180:0.7);
    \draw (5.2,0) arc (0:180:0.7);
    } \,
     \nonumber \\ 
    =& 
    \frac{1}{ L_B^2} 
    + \frac{3}{L_{A} L_B}
    + \frac{1}{L_{A}^2},
\end{align}
which also matches with the exact results~\cite{Collins_rev} in the large $L_A L_B$ limit.

Besides a systematic way of evaluating R\'enyi entropies, another advantage of the graphical representation is that it is easy to find the pattern of dominant diagrams in a certain regime of parameters.
For instance, when subsystem $A$ is much larger than subsystem $B$, i.e., $L_A \gg L_B$, we get
\begin{align}
    \braket{\Tr\rho^n} \approx
    \tikz[scale=0.45,baseline=0.5ex]{
    \draw[dashed] (0,0) -- (1,0);
    \draw (-0.2,0.)-- (-0.2,-0.35);
    \draw (1.2,-0.15)-- (1.2,0.);
    \draw[dashed] (2,0) -- (3,0);
    \draw (1.8,0.)-- (1.8,-0.15);
    \draw (3.2,-0.15)-- (3.2,0.);
    \draw[dashed] (4,0) -- (4.35,0);
    \draw (3.8,0.)-- (3.8,-0.15);
    \draw[dashed] (7,0) -- (8,0);
    \draw[dashed] (5.7,0) -- (6,0);
    \draw (6.8,0.)-- (6.8,-0.15);
    \draw (6.2,0.)-- (6.2,-0.15);
    \draw (8.2,-0.35)-- (8.2,0.);
    \draw (1.2,-0.15)--(1.8,-0.15);
    \draw (3.2,-0.15)--(3.8,-0.15);
    \draw (6.2,-0.15)--(6.8,-0.15);
    \draw (-0.2,-0.35)-- (8.2,-0.35);
    \draw[dashed] (4.0,0) arc (0:180:0.5);
    \draw[dashed] (2.0,0) arc (0:180:0.5);
    \draw[dashed] (7,0) arc (0:180:0.5);
    \draw[dashed] (8,0) arc (0:180:4);
    \node[] at (5,0.2) {$\cdots$};
    \draw (3.8,0) arc (0:180:0.3);
    \draw (1.8,0) arc (0:180:0.3);
    \draw (6.8,0) arc (0:180:0.3);
    \draw (8.2,0) arc (0:180:4.2);
    }  
    = L_B^{1-n},
    \label{eq:LAgg_Renyi}
\end{align}
while in the opposite regime  $L_A \ll L_B$, we obtain
\begin{align}
    \braket{\Tr\rho^n} \approx
    \tikz[scale=0.5,baseline=-0.5ex]{
    \draw[dashed] (0,0) -- (1,0);
    \draw (-0.2,0.)-- (-0.2,-0.35);
    \draw (1.2,-0.15)-- (1.2,0.);
    \draw[dashed] (2,0) -- (3,0);
    \draw (1.8,0.)-- (1.8,-0.15);
    \draw (3.2,-0.15)-- (3.2,0.);
    \draw[dashed] (5,0) -- (6,0);
    \draw (4.8,0.)-- (4.8,-0.15);
    \draw (6.2,-0.35)-- (6.2,0.);
    \draw (1.2,-0.15)--(1.8,-0.15);
    \draw (3.2,-0.15)--(3.45,-0.15);
    \draw (-0.2,-0.35)-- (6.2,-0.35);
    \draw (4.5,-0.15)--(4.8,-0.15);
    \node[] at (4.,0.2) {$\cdots$};
    \draw[dashed] (1.0,0) arc (0:180:0.5);
    \draw[dashed] (3.0,0) arc (0:180:0.5);
    \draw[dashed] (6,0) arc (0:180:0.5);
    \draw (1.2,0) arc (0:180:0.7);
    \draw (3.2,0) arc (0:180:0.7);
    \draw (6.2,0) arc (0:180:0.7);
    } 
    = L_A^{1-n}.
 \label{eq:LAll_Renyi}
\end{align}
Another useful way to view the leading terms is by assigning a genus number to the diagrams. As explained in Appendix~\ref{app:genus} both terms above have zero genus.
On a side note, these diagrams can be viewed as the presence and absence of replica wormholes, respectively, in the gravitation interpretation of the random matrix theory where each term correspond to a saddle point solution~\cite{Shenker2019}. It further provides a simple way to derive the leading term in Page's formula (\ref{eq:EE_page}) for the entanglement entropy.


{\bf Implementing Partial Transpose:} We now incorporate the PT as an operation on the density matrix (\ref{eq:rho_diag}).
Let us recall that subsystem $A$ is further partitioned into $A_1$ and $A_2$. So, we define a tripartite vector $X_{(i_1 i_2)\alpha}$ and trace out susbystem $B$ to obtain the following reduced density matrix
\begin{align}
    [\rho]_{i_1i_2,j_1j_2}= \sum_\alpha X^\ast_{(i_1i_2),\alpha}X_{(j_1j_2),\alpha}
    =
    \,
    \tikz[scale=0.8,baseline=-0.5ex]{
    \draw[dashed] (0,0.2)-- (0,0);
    \draw[dashed] (0,0) -- (1,0);
    \draw[dashed]  (1,0.2) -- (1,0);
    \draw (-0.2,0.2)-- (-0.2,-0.15);
    \draw (1.2,-0.15)-- (1.2,0.2);
    \draw[densely dotted,thick] (-0.3,0.2)-- (-0.3,-0.15);
    \draw[densely dotted,thick] (1.3,-0.15)-- (1.3,0.2);
    }\ ,
\end{align}
where the dotted and solid lines correspond to subsystems $A_1$ and $A_2$, respectively. 
We define PT diagrammatically by
\begin{align}
	\label{eq:rT_diag}
    [\rho^{T_2}]_{i_1i_2,j_1j_2}= \sum_\alpha X^\ast_{(i_1\textcolor{red}{j_2}),\alpha}X_{(j_1\textcolor{red}{i_2}),\alpha}
    =\,
    \tikz[scale=0.8,baseline=-1.5ex]{
    \draw[dashed] (0,0.2)-- (0,0);
    \draw[dashed] (0,0) -- (1,0);
    \draw[dashed]  (1,0.2) -- (1,0);
    \draw (-0.2,0.2)-- (-0.2,-0.15);
    \draw (-0.2,-0.15)-- (1.,-0.7);
    \draw (1.2,-0.15)-- (1.2,0.2);
    \draw (1.2,-0.15)-- (0,-0.7);
    \draw[densely dotted,thick] (-0.3,0.2)-- (-0.3,-0.4);
    \draw[densely dotted,thick] (1.3,-0.4)-- (1.3,0.2);
    }\ .
\end{align}
The underlying operation in the above diagram is to swap the indices for one subsystem as emphasized by indices highlighted in red. Such implementation of the PT is indeed not limited to random density matrices and can be applied to deterministic density matrices of spin chains~\footnote{In the latter context, we need to impose certain rules on the diagrammatic representation of the PT to make it consistent with time-reversal symmetry. For instance, Penrose binor calculus~\cite{1996PhRvD..54.2664D,Rovelli:2004tv} 
provides such a consistent representation which can be used  to derive the $\mathbb{Z}_2$ topological invariant for topological phases of spin chains protected by time-reversal symmetry~\cite{Shap_anyons}.}.

We now run two sanity checks on our proposed diagrammatic representation of the PT.
Both conditions must be met even without ensemble averaging.
First, the PT is trace preserving,
\begin{align}
\label{eq:tr_r}
    \Tr\rho^{T_2} &=
    \,
    \tikz[scale=0.8,baseline=-1.5ex]{
    \draw[dashed] (0,0.2)-- (0,0);
    \draw[dashed] (0,0) -- (1,0);
    \draw[dashed]  (1,0.2) -- (1,0);
    \draw (-0.2,0.2)-- (-0.2,-0.15);
    \draw (-0.2,-0.15)-- (1.,-0.7);
    \draw (1.2,-0.15)-- (1.2,0.2);
    \draw (1.2,-0.15)-- (0,-0.7);
    \draw (1.,-0.7)-- (0,-0.7);
    \draw[densely dotted,thick] (-0.3,0.2)-- (-0.3,-0.9);
    \draw[densely dotted,thick] (-0.3,-0.9)-- (1.3,-0.9);
    \draw[densely dotted,thick] (1.3,-0.9)-- (1.3,0.2);
    }
    = \Tr \rho
    \nonumber
\end{align}
which clearly holds since the crossings are only meant to rearrange the way matrix indices are contracted and can be removed or lifted by moving the lines around. 
Second, PT has to obey the identity $\Tr[(\rho^{T_2})^2]=\Tr \rho^2$~\footnote{This is because
\begin{align*}
    \Tr[(\rho^{T_2})^2] &= 
    \sum_{i_1 i_2 j_1 j_2} [\rho^{T_2}]_{i_1i_2,j_1j_2}
    [\rho^{T_2}]_{j_1j_2,i_1i_2} \\
    &=    \sum_{i_1 i_2 j_1 j_2} [\rho]_{i_1j_2,j_1i_2}
    [\rho]_{j_1i_2,i_1j_2} =\Tr \rho^2.
\end{align*}}, 
\begin{align}
    \Tr[(\rho^{T_2})^2]&=
    \,
    \tikz[scale=0.8,baseline=-1.5ex]{
    \draw[dashed] (0,0.2)-- (0,0);
    \draw[dashed] (0,0) -- (1,0);
    \draw[dashed]  (1,0.2) -- (1,0);
    \draw (-0.2,0.2)-- (-0.2,-0.15);
    \draw (-0.2,-0.15)-- (1.,-0.7);
    \draw (1.2,-0.15)-- (1.2,0.2);
    \draw (1.2,-0.15)-- (0,-0.8);
    \draw[densely dotted,thick] (-0.3,0.2)-- (-0.3,-0.95);
    \draw[densely dotted,thick] (1.3,-0.1)-- (1.3,0.2);
    \draw[dashed] (2,0.2)-- (2,0);
    \draw[dashed] (2,0) -- (3,0);
    \draw[dashed]  (3,0.2) -- (3,0);
    \draw (1.8,0.2)-- (1.8,-0.15);
    \draw (1.8,-0.15)-- (3.,-0.8);
    \draw (3.2,-0.15)-- (3.2,0.2);
    \draw (3.2,-0.15)-- (2,-0.7);
    \draw[densely dotted,thick] (1.7,0.2)-- (1.7,-0.1);
    \draw[densely dotted,thick] (3.3,-0.95)-- (3.3,0.2);
    \draw[densely dotted,thick] (1.3,-0.1)--(1.7,-0.1);
    \draw[densely dotted,thick](-0.3,-0.95)--(3.3,-0.95);
    \draw (0,-0.8)--(3.,-0.8);
    \draw (1.0,-0.7)-- (2,-0.7);
    } 
    = \Tr \rho^2
\end{align}
which is manifestly the case upon comparing with Eq.\,(\ref{eq:tr_r2}) and the fact that lines only represent how indices are contracted and can be moved freely. 

{\bf Calculating R\'enyi Negativity:} The lowest non-trivial moment of the PT which differs from the R\'enyi entropies is the third power, which is given by
\begin{align}
    \Tr[(\rho^{T_2})^3]&=
    \tikz[scale=0.8,baseline=-1.5ex]{
    \draw[dashed] (0,0.2)-- (0,0);
    \draw[dashed] (0,0) -- (1,0);
    \draw[dashed]  (1,0.2) -- (1,0);
    \draw (-0.2,0.2)-- (-0.2,-0.15);
    \draw (-0.2,-0.15)-- (1.,-0.7);
    \draw (1.2,-0.15)-- (1.2,0.2);
    \draw (1.2,-0.15)-- (0,-0.8);
    \draw[densely dotted,thick] (-0.3,0.2)-- (-0.3,-0.95);
    \draw[densely dotted,thick] (1.3,-0.1)-- (1.3,0.2);
    \draw[dashed] (2,0.2)-- (2,0);
    \draw[dashed] (2,0) -- (3,0);
    \draw[dashed]  (3,0.2) -- (3,0);
    \draw (1.8,0.2)-- (1.8,-0.15);
    \draw (1.8,-0.15)-- (3.,-0.7);
    \draw (3.2,-0.15)-- (3.2,0.2);
    \draw (3.2,-0.15)-- (2,-0.7);
    \draw[densely dotted,thick] (1.7,0.2)-- (1.7,-0.1);
    \draw[densely dotted,thick] (3.3,-0.1)-- (3.3,0.2);
    \draw[dashed] (4,0.2)-- (4,0);
    \draw[dashed] (4,0) -- (5,0);
    \draw[dashed]  (5,0.2) -- (5,0);
    \draw (3.8,0.2)-- (3.8,-0.15);
    \draw (3.8,-0.15)-- (5.,-0.8);
    \draw (5.2,-0.15)-- (5.2,0.2);
    \draw (5.2,-0.15)-- (4,-0.7);
    \draw[densely dotted,thick] (3.7,0.2)-- (3.7,-0.1);
    \draw[densely dotted,thick] (5.3,-0.95)-- (5.3,0.2);
    \draw[densely dotted,thick] (1.3,-0.1)--(1.7,-0.1);
    \draw[densely dotted,thick](3.3,-0.1)--(3.7,-0.1);
    \draw[densely dotted,thick](-0.3,-0.95)--(5.3,-0.95);
    \draw (3,-0.7)--(4.,-0.7);
    \draw (0,-0.8)--(5.,-0.8);
    \draw (1.0,-0.7)-- (2,-0.7);
    } \, .
\end{align}
The ensemble average of this quantity is found to be
\begin{align}
\label{eq:tr_rT3}
    \braket{\Tr \left(\rho^{T_2}\right)^3 }=&
    \
    \tikz[scale=0.45,baseline=0.5ex]{
    \draw[dashed] (0,0) -- (1,0);
    \draw (-0.2,0.)-- (-0.2,-0.15);
    \draw (-0.2,-0.15)-- (1.,-0.7);
    \draw (1.2,-0.15)-- (1.2,0.);
    \draw (1.2,-0.15)-- (0,-0.8);
    \draw[densely dotted,thick] (-0.3,0)-- (-0.3,-0.95);
    \draw[densely dotted,thick] (1.3,-0.1)-- (1.3,0.);
    \draw[dashed] (2,0) -- (3,0);
    \draw (1.8,0.)-- (1.8,-0.15);
    \draw (1.8,-0.15)-- (3.,-0.7);
    \draw (3.2,-0.15)-- (3.2,0.);
    \draw (3.2,-0.15)-- (2,-0.7);
    \draw[densely dotted,thick] (1.7,0.)-- (1.7,-0.1);
    \draw[densely dotted,thick] (3.3,-0.1)-- (3.3,0.);
    \draw[dashed] (4,0) -- (5,0);
    \draw (3.8,0.)-- (3.8,-0.15);
    \draw (3.8,-0.15)-- (5.,-0.8);
    \draw (5.2,-0.15)-- (5.2,0.);
    \draw (5.2,-0.15)-- (4,-0.7);
    \draw[densely dotted,thick] (3.7,0.)-- (3.7,-0.1);
    \draw[densely dotted,thick] (5.3,-0.95)-- (5.3,0.);
    \draw[densely dotted,thick] (1.3,-0.1)--(1.7,-0.1);
    \draw[densely dotted,thick](3.3,-0.1)--(3.7,-0.1);
    \draw[densely dotted,thick](-0.3,-0.95)--(5.3,-0.95);
    \draw (3,-0.7)--(4.,-0.7);
    \draw (0,-0.8)--(5.,-0.8);
    \draw (1.0,-0.7)-- (2,-0.7);
    \draw[densely dotted,thick] (5.3,0) arc (0:180:2.8);
    \draw[densely dotted,thick] (1.7,0) arc (0:180:0.2);
    \draw[densely dotted,thick] (3.7,0) arc (0:180:0.2);
    \draw[dashed] (4.0,0) arc (0:180:0.5);
    \draw[dashed] (2.0,0) arc (0:180:0.5);
    \draw[dashed] (5.0,0) arc (0:180:2.5);
    \draw (3.8,0) arc (0:180:0.3);
    \draw (1.8,0) arc (0:180:0.3);
    \draw (5.2,0) arc (0:180:2.7);
    } \
    +
    3\times\,
\tikz[scale=0.45,baseline=0.5ex]{
    \draw[dashed] (0,0) -- (1,0);
    \draw (-0.2,0.)-- (-0.2,-0.15);
    \draw (-0.2,-0.15)-- (1.,-0.7);
    \draw (1.2,-0.15)-- (1.2,0.);
    \draw (1.2,-0.15)-- (0,-0.8);
    \draw[densely dotted,thick] (-0.3,0)-- (-0.3,-0.95);
    \draw[densely dotted,thick] (1.3,-0.1)-- (1.3,0.);
    \draw[dashed] (2,0) -- (3,0);
    \draw (1.8,0.)-- (1.8,-0.15);
    \draw (1.8,-0.15)-- (3.,-0.7);
    \draw (3.2,-0.15)-- (3.2,0.);
    \draw (3.2,-0.15)-- (2,-0.7);
    \draw[densely dotted,thick] (1.7,0.)-- (1.7,-0.1);
    \draw[densely dotted,thick] (3.3,-0.1)-- (3.3,0.);
    \draw[dashed] (4,0) -- (5,0);
    \draw (3.8,0.)-- (3.8,-0.15);
    \draw (3.8,-0.15)-- (5.,-0.8);
    \draw (5.2,-0.15)-- (5.2,0.);
    \draw (5.2,-0.15)-- (4,-0.7);
    \draw[densely dotted,thick] (3.7,0.)-- (3.7,-0.1);
    \draw[densely dotted,thick] (5.3,-0.95)-- (5.3,0.);
    \draw[densely dotted,thick] (1.3,-0.1)--(1.7,-0.1);
    \draw[densely dotted,thick](3.3,-0.1)--(3.7,-0.1);
    \draw[densely dotted,thick](-0.3,-0.95)--(5.3,-0.95);
    \draw (3,-0.7)--(4.,-0.7);
    \draw (0,-0.8)--(5.,-0.8);
    \draw (1.0,-0.7)-- (2,-0.7);
    \draw[densely dotted,thick] (3.3,0) arc (0:180:1.8);
    \draw[densely dotted,thick] (5.3,0) arc (0:180:0.8);
    \draw[densely dotted,thick] (1.7,0) arc (0:180:0.2);
    \draw[dashed] (2.0,0) arc (0:180:0.5);
    \draw[dashed] (3.0,0) arc (0:180:1.5);
    \draw[dashed] (5.0,0) arc (0:180:0.5);
    \draw (1.8,0) arc (0:180:0.3);
    \draw (3.2,0) arc (0:180:1.7);
    \draw (5.2,0) arc (0:180:0.7);
    }  
    \nonumber \\
    &+
   \
    \tikz[scale=0.5,baseline=0.5ex]{
    \draw[dashed] (0,0) -- (1,0);
    \draw (-0.2,0.)-- (-0.2,-0.15);
    \draw (-0.2,-0.15)-- (1.,-0.7);
    \draw (1.2,-0.15)-- (1.2,0.);
    \draw (1.2,-0.15)-- (0,-0.8);
    \draw[densely dotted,thick] (-0.3,0)-- (-0.3,-0.95);
    \draw[densely dotted,thick] (1.3,-0.1)-- (1.3,0.);
    \draw[dashed] (2,0) -- (3,0);
    \draw (1.8,0.)-- (1.8,-0.15);
    \draw (1.8,-0.15)-- (3.,-0.7);
    \draw (3.2,-0.15)-- (3.2,0.);
    \draw (3.2,-0.15)-- (2,-0.7);
    \draw[densely dotted,thick] (1.7,0.)-- (1.7,-0.1);
    \draw[densely dotted,thick] (3.3,-0.1)-- (3.3,0.);
    \draw[dashed] (4,0) -- (5,0);
    \draw (3.8,0.)-- (3.8,-0.15);
    \draw (3.8,-0.15)-- (5.,-0.8);
    \draw (5.2,-0.15)-- (5.2,0.);
    \draw (5.2,-0.15)-- (4,-0.7);
    \draw[densely dotted,thick] (3.7,0.)-- (3.7,-0.1);
    \draw[densely dotted,thick] (5.3,-0.95)-- (5.3,0.);
    \draw[densely dotted,thick] (1.3,-0.1)--(1.7,-0.1);
    \draw[densely dotted,thick](3.3,-0.1)--(3.7,-0.1);
    \draw[densely dotted,thick](-0.3,-0.95)--(5.3,-0.95);
    \draw (3,-0.7)--(4.,-0.7);
    \draw (0,-0.8)--(5.,-0.8);
    \draw (1.0,-0.7)-- (2,-0.7);
    \draw[densely dotted,thick] (1.3,0) arc (0:180:0.8);
    \draw[densely dotted,thick] (3.3,0) arc (0:180:0.8);
    \draw[densely dotted,thick] (5.3,0) arc (0:180:0.8);
    \draw[dashed] (1.0,0) arc (0:180:0.5);
    \draw[dashed] (3.0,0) arc (0:180:0.5);
    \draw[dashed] (5.0,0) arc (0:180:0.5);
    \draw (1.2,0) arc (0:180:0.7);
    \draw (3.2,0) arc (0:180:0.7);
    \draw (5.2,0) arc (0:180:0.7);
    }  \
    +
    \ 
    \tikz[scale=0.45,baseline=0.5ex]{
    \draw[dashed] (0,0) -- (1,0);
    \draw (-0.2,0.)-- (-0.2,-0.15);
    \draw (-0.2,-0.15)-- (1.,-0.7);
    \draw (1.2,-0.15)-- (1.2,0.);
    \draw (1.2,-0.15)-- (0,-0.8);
    \draw[densely dotted,thick] (-0.3,0)-- (-0.3,-0.95);
    \draw[densely dotted,thick] (1.3,-0.1)-- (1.3,0.);
    \draw[dashed] (2,0) -- (3,0);
    \draw (1.8,0.)-- (1.8,-0.15);
    \draw (1.8,-0.15)-- (3.,-0.7);
    \draw (3.2,-0.15)-- (3.2,0.);
    \draw (3.2,-0.15)-- (2,-0.7);
    \draw[densely dotted,thick] (1.7,0.)-- (1.7,-0.1);
    \draw[densely dotted,thick] (3.3,-0.1)-- (3.3,0.);
    \draw[dashed] (4,0) -- (5,0);
    \draw (3.8,0.)-- (3.8,-0.15);
    \draw (3.8,-0.15)-- (5.,-0.8);
    \draw (5.2,-0.15)-- (5.2,0.);
    \draw (5.2,-0.15)-- (4,-0.7);
    \draw[densely dotted,thick] (3.7,0.)-- (3.7,-0.1);
    \draw[densely dotted,thick] (5.3,-0.95)-- (5.3,0.);
    \draw[densely dotted,thick] (1.3,-0.1)--(1.7,-0.1);
    \draw[densely dotted,thick](3.3,-0.1)--(3.7,-0.1);
    \draw[densely dotted,thick](-0.3,-0.95)--(5.3,-0.95);
    \draw (3,-0.7)--(4.,-0.7);
    \draw (0,-0.8)--(5.,-0.8);
    \draw (1.0,-0.7)-- (2,-0.7);
    \draw[densely dotted,thick] (3.3,0) arc (0:180:1.8);
    \draw[densely dotted,thick] (5.3,0) arc (0:180:1.8);
    \draw[densely dotted,thick] (3.7,2) arc (0:180:1.2);
    \draw[densely dotted,thick]	(3.7,0) -- (3.7,2);
    \draw[densely dotted,thick]	(1.3,0) -- (1.3,2);
    \draw[dashed] (5,0) arc (0:180:1.5);
    \draw[dashed] (4,2) arc (0:180:1.5);
    \draw[dashed] (4,0) -- (4,2);
    \draw[dashed] (1,0) -- (1,2);
    \draw[dashed] (3,0) arc (0:180:1.5);
    \draw (3.8,2) arc (0:180:1.3);
    \draw (3.8,0) -- (3.8,2);
    \draw (1.2,0) -- (1.2,2);
    \draw (5.2,0) arc (0:180:1.7);
    \draw (3.2,0) arc (0:180:1.7);
    } 
    \nonumber \\ \nonumber \\
    = &
    \frac{L_{A_1}^2 + 3 L_A L_B + L_B^2 + L_ {A_2}^2}{(L_A L_B)^2},
\end{align}
which again matches the exact results~\cite{Bhosale,Fukuda2013},
\begin{align}
\label{eq:tr3cmplx}
\braket{ \Tr \big(\rho^{T_2}\big)^3} =\dfrac{L_{A_1}^2+L_{A_2}^2+L_{B}^2+3L_A L_B}{
(L_A L_B+1)(L_A L_B+2)},
\end{align}
 in the large $L_A L_B$ limit. 
The difference between the exact expression and our large-$L$ formula is  at least of order $\frac{1}{(L_A L_B)^2}$ which is due to normalization factor in the denominator of Eq.~(\ref{eq:red_den}).
 We will address a systematic way to evaluate these corrections later in Sec.~\ref{sec:1/L corrections}.
  It is worth noting that unlike the R\'enyi entropies which are computed for the untransposed density matrix and only contain planar diagrams, the partially transposed one may have contributions in terms of crossed diagrams in the upper half, since we already have crossings in the lower half of the diagram. For instance, in the above calculation we see that there are four terms and the last term involves crossings in the upper half. This, however, does not occur in the untransposed case as shown in Eq.\,(\ref{eq:tr_r3}).
 
Comparing with the third R\'enyi entropy (\ref{eq:tr_r3}), we notice that
\begin{align}
\braket{ \Tr \big(\rho^3 \big) \,-\, \Tr \big(\rho ^{T_2}\big)^3 }  =
L_B^{-2} (1- L_{A_1}^{-2}- L_{A_2}^{-2}),
\end{align}
which means on average the third moment after PT is smaller than that before. 
Furthermore, it is interesting to note that the third R\'enyi negativity (\ref{eq:tr_rT3}) has a permutation symmetry among the three parties upon exchanging $A_1$, $A_2$, and $B$, which also holds for the exact expression (\ref{eq:tr3cmplx}).

Using the diagrammatic rules, We can write down the exact expression  
for the $n$-th R\'enyi negativity of Wishart matrices: 
\begin{align}
    \braket{ \Tr (\rho^{T_2})^n } &=\frac{1}{(L_A L_B)^n} \sum_{P\in S_n}L_B^{c(P)} L_{A_1}^{c(P_+\circ P)} L_{A_2}^{c(P_-\circ P)}. 
    \label{ExactExpression}
\end{align}
Here, $P\in S_n$ is a permutation of $n$ indices and $c(P)$ is the number of cycles in $P$, including trivial ones; for example, $Q=(12)(3)\in S_3$ has $2$ cycles and $c(Q)=2$. $P_\pm$ are two special permutations defined as $P_\pm(i)=(i\pm 1) \mod n$, i.e.,  cyclic (anti-cyclic) permutations. $P_1 \circ P_2$ is a composite permutation which performs permutation $P_1$ after permuting by $P_2$. 
This expression was also produced by free probability theory techniques in Ref.~\cite{Banica_resolvent}.

In the remainder of this section, we use an approximate form to the above expression by finding the leading order term in two possible regimes: $L_A\ll L_B$ and $L_A\gg L_B$, based on which, we compute the logarithmic negativity by analytical continuation as in
\begin{align}
    \braket{{\cal E}_{A_1:A_2}}= \lim_{n\to \frac{1}{2}} \log_2 \braket{\Tr(\rT)^{2n}}.
    \label{eq:replicalimit}
\end{align}
First, in the regime when the subsystem $B$ is larger than $A$, we get 
\begin{align}
    \braket{\Tr \left(\rho^{T_2}\right)^{n} } \approx&
    \, \
    \tikz[scale=0.5,baseline=-1ex]{
    \draw[dashed] (0,0) -- (1,0);
    \draw (-0.2,0.)-- (-0.2,-0.15);
    \draw (-0.2,-0.15)-- (1.,-0.7);
    \draw (1.2,-0.15)-- (1.2,0.);
    \draw (1.2,-0.15)-- (0,-0.8);
    \draw[densely dotted,thick] (-0.3,0)-- (-0.3,-0.95);
    \draw[densely dotted,thick] (1.3,-0.1)-- (1.3,0.);
    \draw[dashed] (2,0) -- (3,0);
    \draw (1.8,0.)-- (1.8,-0.15);
    \draw (1.8,-0.15)-- (3.,-0.7);
    \draw (3.2,-0.15)-- (3.2,0.);
    \draw (3.2,-0.15)-- (2,-0.7);
    \draw[densely dotted,thick] (1.7,0.)-- (1.7,-0.1);
    \draw[densely dotted,thick] (3.3,-0.1)-- (3.3,0.);
    \draw[dashed] (6,0) -- (7,0);
    \draw (5.8,0.)-- (5.8,-0.15);
    \draw (5.8,-0.15)-- (7.,-0.8);
    \draw (7.2,-0.15)-- (7.2,0.);
    \draw (7.2,-0.15)-- (6,-0.7);
    \draw (5.4,-0.7)-- (6,-0.7);
    \draw[densely dotted,thick] (5.7,0.)-- (5.7,-0.1);
    \draw[densely dotted,thick] (5.4,-0.1)-- (5.7,-0.1);
    \draw[densely dotted,thick] (7.3,-0.95)-- (7.3,0.);
    \draw[densely dotted,thick] (1.3,-0.1)--(1.7,-0.1);
    \draw[densely dotted,thick](3.3,-0.1)--(3.7,-0.1);
    \draw[densely dotted,thick](-0.3,-0.95)--(7.3,-0.95);
    \draw (3,-0.7)--(3.7,-0.7);
    \draw (0,-0.8)--(7.,-0.8);
    \draw (1.0,-0.7)-- (2,-0.7);
    \draw[densely dotted,thick] (1.3,0) arc (0:180:0.8);
    \draw[densely dotted,thick] (3.3,0) arc (0:180:0.8);
    \draw[densely dotted,thick] (7.3,0) arc (0:180:0.8);
    \draw[dashed] (1.0,0) arc (0:180:0.5);
    \draw[dashed] (3.0,0) arc (0:180:0.5);
    \draw[dashed] (7.0,0) arc (0:180:0.5);
    \draw (1.2,0) arc (0:180:0.7);
    \draw (3.2,0) arc (0:180:0.7);
    \draw (7.2,0) arc (0:180:0.7);
    \node[] at (4.6,0.3) {$\cdots$};
    } 
    = L_A^{1-n}\, ,
    \label{eq:LAll}
\end{align}
regardless of $n$ being odd or even, where the corrections are smaller at least by a factor of $\tfrac{L_A}{L_B}$. Taking the replica limit of this formula, we obtain $\braket{{\cal E}_{A_1:A_2}}=0$.
Second, we consider a regime where $L_{A} \gg L_{B}$. This regime can in turn be divided into two subregimes depending on whether or not $A_1$ and $A_2$ are smaller than half the total system.
When each party is smaller than half the total system, i.e., $L_{A_s} \ll L_B L_{A_{\bar s}}$ where $s=1,2$ and $A_{\bar s}$ denotes complement of $A_s$ in subsystem $A$, we obtain from Eq.\,(\ref{ExactExpression}) that
\begin{align}
    \label{eq:RN_catalan}
    \braket{\Tr \left(\rho^{T_2}\right)^{n} } \approx \ 
    \left\{
    \begin{array}{cl}
     \frac{C_k L_A}{(L_A L_B)^k}  \ \ \ \ & n=2k, \\
     \\
     \frac{(2k+1) C_k}{(L_A L_B)^k} \ \ \ \ & n=2k+1,
     \end{array}
     \right.
\end{align}
which come from diagrams of the following types
\begin{align}
    \tikz[scale=0.4,baseline=1.5ex]{
    \draw[dashed] (0,0) -- (1,0);
    \draw (-0.2,0.)-- (-0.2,-0.15);
    \draw (-0.2,-0.15)-- (1.,-0.7);
    \draw (1.2,-0.15)-- (1.2,0.0);
    \draw (1.2,-0.15)-- (0,-0.95);
    \draw[densely dotted,thick] (-0.3,0.)-- (-0.3,-1.3);
    \draw[densely dotted,thick] (1.3,-0.1)-- (1.3,0.);
    \draw[dashed] (2,0) -- (3,0);
    \draw (1.8,0.)-- (1.8,-0.15);
    \draw (1.8,-0.15)-- (3.,-0.7);
    \draw (3.2,-0.15)-- (3.2,0.);
    \draw (3.2,-0.15)-- (2,-0.7);
    \draw[densely dotted,thick] (1.7,0.)-- (1.7,-0.1);
    \draw[densely dotted,thick] (3.3,-0.1)-- (3.3,0.);
    \draw[dashed] (4,0) -- (5,0);
    \draw (3.8,0.)-- (3.8,-0.15);
    \draw (3.8,-0.15)-- (5.,-0.7);
    \draw (5.2,-0.15)-- (5.2,0.);
    \draw (5.2,-0.15)-- (4,-0.7);
    \draw[densely dotted,thick] (3.7,0.)-- (3.7,-0.1);
    \draw[densely dotted,thick] (5.3,-0.1)-- (5.3,0.);
    \draw[dashed] (6,0) -- (7,0);
    \draw (5.8,0.)-- (5.8,-0.15);
    \draw (5.8,-0.15)-- (7.,-0.8);
    \draw (7.3,-0.8)-- (7.,-0.8);
    \draw (7.2,-0.15)-- (7.2,0.);
    \draw (7.2,-0.15)-- (6,-0.7);
    \draw[densely dotted,thick] (5.7,0.)-- (5.7,-0.1);
    \draw[densely dotted,thick] (7.3,0)-- (7.3,-0.1);
    \draw[densely dotted,thick] (7.3,-0.1)-- (7.7,-0.1);
    \draw[densely dotted,thick] (1.3,-0.1)--(1.7,-0.1);
    \draw[densely dotted,thick] (3.3,-0.1)--(3.7,-0.1);
    \draw[densely dotted,thick] (5.3,-0.1)--(5.7,-0.1);
    \draw (1.0,-0.7)-- (2,-0.7);
    \draw (3.0,-0.7)-- (4,-0.7);
    \draw (5.0,-0.7)-- (6,-0.7);
    \draw[densely dotted,thick] (3.3,0) arc (0:180:1.8);
    \draw[densely dotted,thick] (1.7,0) arc (0:180:0.2);
    \draw[dashed] (2.0,0) arc (0:180:0.5);
    \draw[dashed] (3.0,0) arc (0:180:1.5);
    \draw (1.8,0) arc (0:180:0.3);
    \draw (3.2,0) arc (0:180:1.7);
    \draw[densely dotted,thick] (7.3,0) arc (0:180:1.8);
    \draw[densely dotted,thick] (5.7,0) arc (0:180:0.2);
    \draw[dashed] (6.0,0) arc (0:180:0.5);
    \draw[dashed] (7.0,0) arc (0:180:1.5);
    \draw (5.8,0) arc (0:180:0.3);
    \draw (7.2,0) arc (0:180:1.7);
    \draw[dashed] (10,0) -- (11,0);
    \draw (9.8,0.)-- (9.8,-0.15);
    \draw (9.8,-0.15)-- (11,-0.7);
    \draw (11.2,-0.15)-- (11.2,0.0);
    \draw (11.2,-0.15)-- (10,-0.8);
    \draw (9.7,-0.8)-- (10,-0.8);
    \draw[densely dotted,thick] (9.7,0.)-- (9.7,-0.1);
    \draw[densely dotted,thick] (9.3,-0.1)-- (9.7,-0.1);
    \draw[densely dotted,thick] (11.3,-0.1)-- (11.3,0.);
    \draw[dashed] (12,0) -- (13,0);
    \draw (11.8,0.)-- (11.8,-0.15);
    \draw (11.8,-0.15)-- (13.,-0.95);
    \draw (13.2,-0.15)-- (13.2,0.);
    \draw (13.2,-0.15)-- (12,-0.7);
    \draw[densely dotted,thick] (11.7,0.)-- (11.7,-0.1);
    \draw[densely dotted,thick] (13.3,-1.3)-- (13.3,0.);
    \draw[densely dotted,thick] (11.3,-0.1)--(11.7,-0.1);
    \draw (11.0,-0.7)-- (12,-0.7);
    \draw[densely dotted,thick] (13.3,0) arc (0:180:1.8);
    \draw[densely dotted,thick] (11.7,0) arc (0:180:0.2);
    \draw[dashed] (12.0,0) arc (0:180:0.5);
    \draw[dashed] (13.0,0) arc (0:180:1.5);
    \draw (11.8,0) arc (0:180:0.3);
    \draw (13.2,0) arc (0:180:1.7);
    \node[] at (8.6,0.5) {$\cdots$};
    \draw[densely dotted,thick] (-0.3,-1.3)-- (13.3,-1.3);
    \draw (0,-0.95) -- (13,-0.95);
 }\, ,
     \label{eq:LAgg}
\end{align}
and
\begin{align}
    \tikz[scale=0.4,baseline=1.5ex]{
    \draw[dashed] (0,0) -- (1,0);
    \draw (-0.2,0.)-- (-0.2,-0.15);
    \draw (-0.2,-0.15)-- (1.,-0.7);
    \draw (1.2,-0.15)-- (1.2,0.0);
    \draw (1.2,-0.15)-- (0,-0.95);
    \draw[densely dotted,thick] (-0.3,0.)-- (-0.3,-1.3);
    \draw[densely dotted,thick] (1.3,-0.1)-- (1.3,0.);
    \draw[dashed] (2,0) -- (3,0);
    \draw (1.8,0.)-- (1.8,-0.15);
    \draw (1.8,-0.15)-- (3.,-0.7);
    \draw (3.2,-0.15)-- (3.2,0.);
    \draw (3.2,-0.15)-- (2,-0.7);
    \draw[densely dotted,thick] (1.7,0.)-- (1.7,-0.1);
    \draw[densely dotted,thick] (3.3,-0.1)-- (3.3,0.);
    \draw[dashed] (4,0) -- (5,0);
    \draw (3.8,0.)-- (3.8,-0.15);
    \draw (3.8,-0.15)-- (5.,-0.7);
    \draw (5.2,-0.15)-- (5.2,0.);
    \draw (5.2,-0.15)-- (4,-0.7);
    \draw[densely dotted,thick] (3.7,0.)-- (3.7,-0.1);
    \draw[densely dotted,thick] (5.3,-0.1)-- (5.3,0.);
    \draw[dashed] (6,0) -- (7,0);
    \draw (5.8,0.)-- (5.8,-0.15);
    \draw (5.8,-0.15)-- (7.,-0.8);
    \draw (7.3,-0.8)-- (7.,-0.8);
    \draw (7.2,-0.15)-- (7.2,0.);
    \draw (7.2,-0.15)-- (6,-0.7);
    \draw[densely dotted,thick] (5.7,0.)-- (5.7,-0.1);
    \draw[densely dotted,thick] (7.3,0)-- (7.3,-0.1);
    \draw[densely dotted,thick] (7.3,-0.1)-- (7.7,-0.1);
    \draw[densely dotted,thick] (1.3,-0.1)--(1.7,-0.1);
    \draw[densely dotted,thick] (3.3,-0.1)--(3.7,-0.1);
    \draw[densely dotted,thick] (5.3,-0.1)--(5.7,-0.1);
    \draw (1.0,-0.7)-- (2,-0.7);
    \draw (3.0,-0.7)-- (4,-0.7);
    \draw (5.0,-0.7)-- (6,-0.7);
    \draw[densely dotted,thick] (3.3,0) arc (0:180:1.8);
    \draw[densely dotted,thick] (1.7,0) arc (0:180:0.2);
    \draw[dashed] (2.0,0) arc (0:180:0.5);
    \draw[dashed] (3.0,0) arc (0:180:1.5);
    \draw (1.8,0) arc (0:180:0.3);
    \draw (3.2,0) arc (0:180:1.7);
    \draw[densely dotted,thick] (7.3,0) arc (0:180:1.8);
    \draw[densely dotted,thick] (5.7,0) arc (0:180:0.2);
    \draw[dashed] (6.0,0) arc (0:180:0.5);
    \draw[dashed] (7.0,0) arc (0:180:1.5);
    \draw (5.8,0) arc (0:180:0.3);
    \draw (7.2,0) arc (0:180:1.7);
    \draw[dashed] (10,0) -- (11,0);
    \draw (9.8,0.)-- (9.8,-0.15);
    \draw (9.8,-0.15)-- (11,-0.7);
    \draw (11.2,-0.15)-- (11.2,0.0);
    \draw (11.2,-0.15)-- (10,-0.8);
    \draw (9.7,-0.8)-- (10,-0.8);
    \draw[densely dotted,thick] (9.7,0.)-- (9.7,-0.1);
    \draw[densely dotted,thick] (9.3,-0.1)-- (9.7,-0.1);
    \draw[densely dotted,thick] (11.3,-0.1)-- (11.3,0.);
    \draw[dashed] (12,0) -- (13,0);
    \draw (11.8,0.)-- (11.8,-0.15);
    \draw (11.8,-0.15)-- (13.,-0.8);
    \draw (13.2,-0.15)-- (13.2,0.);
    \draw (13.2,-0.15)-- (12,-0.7);
    \draw[densely dotted,thick] (11.7,0.)-- (11.7,-0.1);
    \draw[densely dotted,thick] (11.3,-0.1)--(11.7,-0.1);
    \draw (11.0,-0.7)-- (12,-0.7);
    \draw (13,-0.8) -- (14.,-0.8);
    \draw[densely dotted,thick] (13.3,0) arc (0:180:1.8);
    \draw[densely dotted,thick] (11.7,0) arc (0:180:0.2);
    \draw[dashed] (12.0,0) arc (0:180:0.5);
    \draw[dashed] (13.0,0) arc (0:180:1.5);
    \draw (11.8,0) arc (0:180:0.3);
    \draw (13.2,0) arc (0:180:1.7);
    \draw[dashed] (14,0) -- (15,0);
    \draw (13.8,0.)-- (13.8,-0.15);
    \draw (13.8,-0.15)-- (15,-0.95);
    \draw (15.2,-0.15)-- (15.2,0.);
    \draw (15.2,-0.15)-- (14,-0.8);
    \draw[densely dotted,thick] (15.3,-1.3)-- (15.3,0.);
    \draw[densely dotted,thick](13.7,-0.1)--(13.3,-0.1);
    \draw[densely dotted,thick] (15.3,0) arc (0:180:0.8);
    \draw[dashed] (15.0,0) arc (0:180:0.5);
    \draw (15.2,0) arc (0:180:0.7);
    \node[] at (8.6,0.5) {$\cdots$};
    \draw[densely dotted,thick] (-0.3,-1.3)-- (15.3,-1.3);
    \draw (0,-0.95) -- (15.,-0.95);
 }\, .
      \label{eq:LAgg-odd}
\end{align}
for even and odd $n$, respectively.
Here, the corrections are smaller at least by a factor of $\tfrac{L_{A_s}}{L_B L_{A_{\bar s}}}$ or $\tfrac{L_{B}}{L_A}$,
and $C_n$ is the $n$th Catalan number 
\begin{align}
    \label{eq:Catalan-def}
    C_n = \frac{\binom{2n}{n} }{n+1},
\end{align}
where $\binom{n}{k}$ denotes $n$ choose $k$ which comes from counting the number of leading order diagrams by using Eq.~(\ref{ExactExpression}) (See Appendix~\ref{app:catalan} for a derivation). 
Using free probability theory techniques, a similar formula was derived in Refs.~\cite{Aubrun2010,Collins2012}.
The appearance of the Catalan numbers is the first hint that the spectral density is related to the semi-circle law~\cite{Forrester}. This is because
\begin{align}
    \braket{ \Tr(\rTc)^n } = \int d\xi\, \xi^n P_{\Gamma}(\xi),
\end{align}
and we know that moments of a semi-circle probability distribution is given by the Catalan numbers,
\begin{align}
    \frac{1}{2\pi} \int_{-2}^{2} dx\, x^{2n} \sqrt{4-x^2} = C_n.
\end{align}
Using (\ref{eq:replicalimit}), we take the replica limit of Eq.~(\ref{eq:RN_catalan}) to obtain
\begin{align}
\braket{ {\cal E}_{A_1:A_2} } \approx \frac{1}{2} \left( \log_2 L_A  -\log_2 L_B \right) + c_1,
\label{eq:LN_smallq}
\end{align}
where $c_1= \log (8/3\pi)$ comes from taking the replica limit of the Catalan number (\ref{eq:Catalan-def}).

Lastly, we consider the subregime of $L_A\gg L_B$ where subsystem $A_1$ is larger than half the total system, i.e., $L_{A_1}\gg L_B L_{A_2}$. The dominant terms in (\ref{ExactExpression}) are given by
\begin{align}
    \label{eq:RN_largeLA1}
    \braket{\Tr \left(\rho^{T_2}\right)^{n} } \approx \ 
    \left\{
    \begin{matrix}
    L_{B}^{1-n} L_{A_2}^{2-n} & \qquad  n=2k,
    \\
    \\
    (L_{B}L_{A_2})^{1-n} & \qquad  n=2k+1.
    \end{matrix}
    \right.
\end{align}
For instance, the corresponding diagrams for $n=4$ and $n=5$ look as follows
\begin{align}
\label{eq:RT4-diag-LA1gg}
\braket{\Tr \left(\rho^{T_2}\right)^{4} } \approx\ 
\tikz[scale=0.4,baseline=1.5ex]{
    \draw[dashed] (0,0) -- (1,0);
    \draw (-0.2,0.)-- (-0.2,-0.15);
    \draw (-0.2,-0.15)-- (1.,-0.7);
    \draw (1.2,-0.15)-- (1.2,0.0);
    \draw (1.2,-0.15)-- (0,-0.8);
    \draw[densely dotted,thick] (-0.3,0.)-- (-0.3,-0.9);
    \draw[densely dotted,thick] (1.3,-0.1)-- (1.3,0.);
    \draw[dashed] (2,0) -- (3,0);
    \draw (1.8,0.)-- (1.8,-0.15);
    \draw (1.8,-0.15)-- (3.,-0.7);
    \draw (3.2,-0.15)-- (3.2,0.);
    \draw (3.2,-0.15)-- (2,-0.7);
    \draw[densely dotted,thick] (1.7,0.)-- (1.7,-0.1);
    \draw[densely dotted,thick] (3.3,-0.1)-- (3.3,0.);
    \draw[dashed] (4,0) -- (5,0);
    \draw (3.8,0.)-- (3.8,-0.15);
    \draw (3.8,-0.15)-- (5.,-0.7);
    \draw (5.2,-0.15)-- (5.2,0.);
    \draw (5.2,-0.15)-- (4,-0.7);
    \draw[densely dotted,thick] (3.7,0.)-- (3.7,-0.1);
    \draw[densely dotted,thick] (5.3,-0.1)-- (5.3,0.);
    \draw[dashed] (6,0) -- (7,0);
    \draw (5.8,0.)-- (5.8,-0.15);
    \draw (5.8,-0.15)-- (7.,-0.8);
    \draw (7.2,-0.15)-- (7.2,0.);
    \draw (7.2,-0.15)-- (6,-0.7);
    \draw[densely dotted,thick] (5.7,0.)-- (5.7,-0.1);
    \draw[densely dotted,thick] (7.3,-0.9)-- (7.3,0.);
    \draw[densely dotted,thick] (1.3,-0.1)--(1.7,-0.1);
    \draw[densely dotted,thick] (3.3,-0.1)--(3.7,-0.1);
    \draw[densely dotted,thick] (5.3,-0.1)--(5.7,-0.1);
    \draw[densely dotted,thick] (-0.3,-0.9)--(7.3,-0.9);
    \draw (1.0,-0.7)-- (2,-0.7);
    \draw (3.0,-0.7)-- (4,-0.7);
    \draw (5.0,-0.7)-- (6,-0.7);
    \draw (0.0,-0.8)-- (7,-0.8);
    \draw[densely dotted,thick] (7.3,0) arc (0:180:3.8);
    \draw[densely dotted,thick] (1.7,0) arc (0:180:0.2);
    \draw[densely dotted,thick] (3.7,0) arc (0:180:0.2);
    \draw[densely dotted,thick] (5.7,0) arc (0:180:0.2);
    \draw[dashed] (6.0,0) arc (0:180:0.5);
    \draw[dashed] (4.0,0) arc (0:180:0.5);
    \draw[dashed] (2.0,0) arc (0:180:0.5);
    \draw[dashed] (7.0,0) arc (0:180:3.5);
    \draw (5.8,0) arc (0:180:0.3);
    \draw (3.8,0) arc (0:180:0.3);
    \draw (1.8,0) arc (0:180:0.3);
    \draw (7.2,0) arc (0:180:3.7);
 }\, ,
\end{align}
\begin{align}
\label{eq:RT5-diag-LA1gg}
\braket{\Tr \left(\rho^{T_2}\right)^{5} } \approx\ \tikz[scale=0.4,baseline=1.5ex]{
    \draw[dashed] (0,0) -- (1,0);
    \draw (-0.2,0.)-- (-0.2,-0.15);
    \draw (-0.2,-0.15)-- (1.,-0.7);
    \draw (1.2,-0.15)-- (1.2,0.0);
    \draw (1.2,-0.15)-- (0,-0.8);
    \draw[densely dotted,thick] (-0.3,0.)-- (-0.3,-0.9);
    \draw[densely dotted,thick] (1.3,-0.1)-- (1.3,0.);
    \draw[dashed] (2,0) -- (3,0);
    \draw (1.8,0.)-- (1.8,-0.15);
    \draw (1.8,-0.15)-- (3.,-0.7);
    \draw (3.2,-0.15)-- (3.2,0.);
    \draw (3.2,-0.15)-- (2,-0.7);
    \draw[densely dotted,thick] (1.7,0.)-- (1.7,-0.1);
    \draw[densely dotted,thick] (3.3,-0.1)-- (3.3,0.);
    \draw[dashed] (4,0) -- (5,0);
    \draw (3.8,0.)-- (3.8,-0.15);
    \draw (3.8,-0.15)-- (5.,-0.7);
    \draw (5.2,-0.15)-- (5.2,0.);
    \draw (5.2,-0.15)-- (4,-0.7);
    \draw[densely dotted,thick] (3.7,0.)-- (3.7,-0.1);
    \draw[densely dotted,thick] (5.3,-0.1)-- (5.3,0.);
    \draw[dashed] (6,0) -- (7,0);
    \draw (5.8,0.)-- (5.8,-0.15);
    \draw (5.8,-0.15)-- (7.,-0.7);
    \draw (7.2,-0.15)-- (7.2,0.);
    \draw (7.2,-0.15)-- (6,-0.7);
    \draw[densely dotted,thick] (5.7,0.)-- (5.7,-0.1);
    \draw[densely dotted,thick] (7.3,-0.1)-- (7.3,0.);
    \draw[dashed] (8,0) -- (9,0);
    \draw (7.8,0.)-- (7.8,-0.15);
    \draw (7.8,-0.15)-- (9.,-0.8);
    \draw (9.2,-0.15)-- (9.2,0.);
    \draw (9.2,-0.15)-- (8,-0.7);
    \draw[densely dotted,thick] (7.7,0.)-- (7.7,-0.1);
    \draw[densely dotted,thick] (9.3,-0.9)-- (9.3,0.);
    \draw[densely dotted,thick] (1.3,-0.1)--(1.7,-0.1);
    \draw[densely dotted,thick] (3.3,-0.1)--(3.7,-0.1);
    \draw[densely dotted,thick] (5.3,-0.1)--(5.7,-0.1);
    \draw[densely dotted,thick] (7.3,-0.1)--(7.7,-0.1);
    \draw[densely dotted,thick] (-0.3,-0.9)--(9.3,-0.9);
    \draw (1.0,-0.7)-- (2,-0.7);
    \draw (3.0,-0.7)-- (4,-0.7);
    \draw (5.0,-0.7)-- (6,-0.7);
    \draw (7.0,-0.7)-- (8,-0.7);
    \draw (.0,-0.8)-- (9,-0.8);
    \draw[densely dotted,thick] (9.3,0) arc (0:180:4.8);
    \draw[densely dotted,thick] (1.7,0) arc (0:180:0.2);
    \draw[densely dotted,thick] (3.7,0) arc (0:180:0.2);
    \draw[densely dotted,thick] (5.7,0) arc (0:180:0.2);
    \draw[densely dotted,thick] (7.7,0) arc (0:180:0.2);
    \draw[dashed] (6.0,0) arc (0:180:0.5);
    \draw[dashed] (4.0,0) arc (0:180:0.5);
    \draw[dashed] (2.0,0) arc (0:180:0.5);
    \draw[dashed] (8.0,0) arc (0:180:0.5);
    \draw[dashed] (9.0,0) arc (0:180:4.5);
    \draw (5.8,0) arc (0:180:0.3);
    \draw (3.8,0) arc (0:180:0.3);
    \draw (1.8,0) arc (0:180:0.3);
    \draw (7.8,0) arc (0:180:0.3);
    \draw (9.2,0) arc (0:180:4.7);
 }\, .
\end{align}
One should notice that there are always $n$ loops of $A_1$, one loop of $B$ regardless of the parity of $n$. In contrast, the number of loops of $A_2$ depends on $n$ being odd or even: There is only one loop of $A_2$ in the odd moments while there are two loops in the even moments. The other limit, $L_{A_2}\gg L_B L_{A_1}$, can be calculated similarly.
Analytically continuing the even moments implies that
\begin{align}
    \label{eq:LN_ME}
    \braket{{\cal E}_{A_1:A_2}} \approx
     \min (\log_2 L_{A_1},\log_2 L_{A_2}).
\end{align}
We can further justify our expansion by another way of distinguishing leading terms from subleading ones. As usual in random matrix theory, we can assign a genus number to a diagram here, too. As we explained in Appendix~\ref{app:genus}, the diagrams for the moments of $\rT$ can be characterized by two genus numbers (as opposed to bipartite geometry where there is only one genus) associated with the subgraphs composed of $A_1B$ and $A_2B$. The PPT and semi-circle law regimes correspond to zero genus numbers, i.e., planar diagrams with respect to both subgraphs, whereas the dominant diagram in the maximally entangled phase is characterized by $g_1 = 0$ and $g_2 = [\frac{k-1}{2}]$ (see details in Appendix~\ref{app:genus}).

\begin{figure}
    \centering
    \includegraphics[scale=0.7]{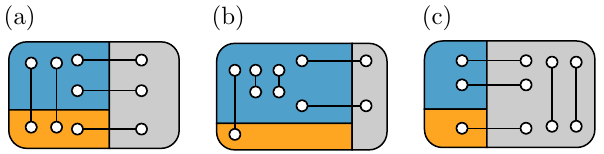}
    \caption{A schematic representation of tripartite random pure state in terms of Bell pairs (see the main text for more explanation). Each circle represents a qubit, and two qubits connected by a line represent a Bell pair. The three phases in Fig.~\ref{fig:phasediag} are illustrated for $10$ qubits: (a) Entanglement saturation regime, (b) maximally entangled regime, and (c) PPT regime.}
    \label{fig:Bell_pair}
\end{figure}

To sum up, let us rephrase all the results in terms of number of qubits within each subsystem. We find that
\begin{align}
    \braket{{\cal E}_{A_1:A_2}}  \approx  0, & \qquad  N_{A} < N_{B},
    \label{eq:LN_PPT}
\end{align}
while if $N_A > N_{B}$,
\begin{align}
    \label{eq:LN_regimes}
     \braket{{\cal E}_{A_1:A_2}}  \approx \ 
    \left\{
    \begin{matrix}
    \frac{1}{2} (N_A-N_B) + c_1 , &   N_{A_s} < \frac{N}{2},
    \\ \\
    \min(N_{A_1},N_{A_2}), &   \text{otherwise}.
    \end{matrix}
    \right.
\end{align}
The above expression can be  understood heuristically in terms of a collection of Bell pairs. We denote the number of Bell pairs shared between $s$ and $s'$ by $n_{s:s'}$. Figure~\ref{fig:Bell_pair} shows examples for each regime when $N_A+N_B=10$.
Using Page's formula for the bipartite entanglement entropy (\ref{eq:EE_page}), we may write 
\begin{align}
    n_{A_1:A_2} + n_{A_1:B} &= N_{A_1}, \\
    n_{A_1:A_2} + n_{A_2:B} &= N_{A_2}, \\
    n_{A_1:B} + n_{A_2:B} &= N_{B}, 
\end{align}
in the saturated entangled regime where $A_1$ and $A_2$ are comparable in size with each other (and crucially larger than $\frac{1}{2}(N_A-N_B)$).
This in turn gives
\begin{align}
    {\cal E}_{A_1:A_2} = n_{A_1:A_2} = \frac{1}{2}(N_A-N_B).
\end{align}
A schematic example in this regime is shown in Fig.~\ref{fig:Bell_pair}(a). In this representation, moving the entanglement cut between $A_1$ and $A_2$ is described by moving one of the Bell pairs shared between $A$ and $B$ from $A_1$ to $A_2$ or vice versa. It is evident that such process does not change ${\cal E}_{A_1:A_2}$.

In contrast, in the maximally entangled regime where $A_2$ is larger than half the system (Fig.~\ref{fig:Bell_pair}(b)), from Page's formula we have
\begin{align}
    n_{A_1:A_2} + n_{A_1:B} &= N_{A_1}, \\
    n_{A_1:A_2} + n_{A_2:B} &= N_{A_1}+N_{B}, \\
    n_{A_1:B} + n_{A_2:B} &= N_{B}, 
\end{align}
which implies
\begin{align}
    {\cal E}_{A_1:A_2} = n_{A_1:A_2} = N_{A_1}.
\end{align}
A similar derivation leads to ${\cal E}_{A_1:A_2} = N_{A_2}$ when $N_{A_1}>N_{A_2}$.

The PPT regime is realized when subsystem $B$ is larger than subsystem $A$ (Fig.~\ref{fig:Bell_pair}(c)). Here, we may write
\begin{align}
    n_{A_1:A_2} + n_{A_1:B} &= N_{A_1}, \\
    n_{A_1:A_2} + n_{A_2:B} &= N_{A_2}, \\
    n_{A_1:B} + n_{A_2:B} &= N_{A}, 
\end{align}
which implies
\begin{align}
    {\cal E}_{A_1:A_2} = n_{A_1:A_2} = 0.
\end{align}

The above observation means that the Bell-pair-type entanglement can reproduce the correct scaling form for the leading order term of the logarithmic negativity, which can be a consequence of the fact that tripartite random states are rarely multipartite entangled. This is an interesting technical point which is beyond the scope of this paper and would be a subject of a future study. A similar phenomenon was already pointed out in stabilizer states~\cite{Smith-Leung,Nezami-Walter}.

Now that we have established the basic properties of our graphical representation of the PT, we take one step further to investigate the spectral properties of $\rT$ in the next two sections.
We should also note that the above derivation of the logarithmic negativity is based on analytical continuation of the leading order terms of the R\'enyi negativity in various regimes. However, this process may have some issues especially how we analytically continue the R\'enyi index.
In the next part, we find the resolvent function and calculate the full spectral density of $\rT$ which provides an unambiguous way to derive the logarithmic negativity and justifies Eqs.~(\ref{eq:LN_PPT}) and (\ref{eq:LN_regimes}).


\section{Negativity spectrum}
\label{sec:PTWishartSpectralDensity}

Our central goal in this section is to find the spectral density of the partially transposed density matrix as defined in Eq.\,(\ref{eq:PTdist}).
Let $G(z)$ be a resolvent function (or one-point Green function) for a random matrix $H$,
\begin{align}
    G(z)=\frac{1}{L_A} \left\langle \Tr\left( \frac{1}{z-H}\right) \right\rangle,
\end{align}
in terms of which we can compute the spectral density
\begin{align}
    \label{eq:dos}
    P(\xi)= - \frac{L_A}{\pi} \text{Im} \lim_{\epsilon\to 0} G(z)\big|
    _{z=\xi + i\epsilon},
\end{align}
since
\begin{align}
    \lim_{\epsilon\to 0} \frac{1}{\lambda + i\epsilon} = \text{PV}\, \frac{1}{\lambda} - i \pi \delta(\lambda).
\end{align}
Diagrammatically, it is more convenient to work with the untraced quantity, i.e., the inverse matrix,
\begin{align}
\label{eq:Ghat-def}
\hat G(z) = \braket{(z-H)^{-1}},
\end{align}
which is related to $G(z)$ via
\begin{align}
\label{eq:Ghat-G}
    [\hat G(z)]_{ij} = G(z) \delta_{ij}, 
\end{align}
thanks to the disorder averaging (\ref{eq:doubleline1}). Because of this property (also known as the Haar symmetry), we can use the matrix $\hat G(z)$ or the scalar $G(z)$ interchangeably.

The diagrammatic approach follows by expanding $G(z)$ in inverse powers of $z$ 
\begin{align}
	\label{eq:Gzdef}
    G(z) &= \frac{1}{L_A} \sum_{n=0} \frac{1}{z^{n+1}} \braket{\Tr(H^n)} \\
     &=  \frac{1}{L_A} \left\langle \Tr\left(\frac{1}{z} + \frac{1}{z} H \frac{1}{z}  + \frac{1}{z} H \frac{1}{z} H \frac{1}{z} + \cdots
     \right) \right\rangle. \nonumber
\end{align}
In our graphical representation of a given term, we insert the diagram (\ref{eq:rT_diag}) for every power of $H$. 

In what follows, we evaluate the above infinite sum by finding a recursive relation for $G(z)$, a.k.a.\  the Schwinger-Dyson (SD) equation. First, we show that the SD equation in the regime where each subsystem of $A$ is smaller than half the total system, i.e., $L_{A_s}\ll L_{A_{\bar s}} L_B$, takes a quadratic form and the spectral density is given by a semi-circle law. In this limit, we neglect certain diagrams. Next, we explain the lowest order  corrections to the semi-circle law distribution.
Finally, we discuss a more general calculation which only assumes $L_{A_1}\gg L_{A_2}$ (with no constraint on $L_B$) and includes the diagrams which we neglected in the semi-circle regime. This calculation results in a cubic SD equation which is used to map out the phase diagram in Fig.~\ref{fig:phasediag}.

\subsection{Semi-circle law}

Here, we find the resolvent function in the regime where each subsystem $A_1$, $A_2$ is smaller than half of the full system, $N_{A_i}<\frac{N}{2}$. Mathematically, this regime corresponds to the limit where $\frac{L_{A_s}}{L_B L_{A_{\bar s}}}\ll 1$ where $A_{\bar s}$ denotes the complement of $A_s$ within subsystem $A$. To carry out calculations systematically such that our perturbative expansion makes sense, we use a different normalization than that of Eq.\,(\ref{eq:doubleline1}). The actual normalization will be included via rescaling after the spectral density is evaluated. To the leading order, we neglect the denominator in Eq.\,(\ref{eq:red_den}) and consider the Wishart matrix
\begin{align}
    H = L_B (X X^\dag)^{T_2},
    \label{eq:Wishart_normal}
\end{align}
in the $G(z)$ expansion (\ref{eq:Gzdef}). 
The ensemble average is then represented by triple lines with amplitude $\tfrac{1}{L_A}$ (given the above normalization),
\begin{align}
\tikz[baseline=-0.1ex,scale=0.9]{
    \draw[dashed] (1.0,0) arc (0:180:0.5);
    \draw (1.1,0.) arc (0:180:0.6);
    \draw[densely dotted,thick] (1.2,0.) arc (0:180:0.7);
    }
    \ \,
    =\frac{1}{L_A}\, \delta_{i_1j_1} \delta_{i_2j_2} \delta_{\alpha\beta}.
\end{align}
As before, close loop of subsystem $s=A_1,A_2$, or $B$ gives a factor of $L_s$. Hence, we obtain
\begin{widetext}
\begin{align}
\tikz[baseline=-0.5ex]{
    \draw[densely dotted,thick] (0.35,0.05)--(0.65,0.05);
    \draw[densely dotted,thick] (-0.35,0.05)--(-0.65,0.05);
    \draw (0.35,-0.05)--(0.65,-0.05);
    \draw (-0.35,-0.05)--(-0.65,-0.05);
    \draw[thick] (0,0) circle (10pt) node[anchor=center] {$G$};
    }
= \ \,   \tikz[baseline=-0.5ex]{
    \draw[densely dotted,thick] (-0.5,0.05)--(0.5,0.05);
    \draw (-0.5,-0.05)--(0.5,-0.05);
    }
 &+ \ \,
\tikz[baseline=-1.5ex]{
    \draw[dashed] (0,0) -- (1,0);
    \draw (-0.2,0.0)-- (-0.2,-0.15);
    \draw (-0.2,-0.15)-- (1.,-0.7);
    \draw (1.2,-0.15)-- (1.2,0.0);
    \draw (1.2,-0.15)-- (0,-0.7);
    \draw[densely dotted,thick] (-0.3,0.0)-- (-0.3,-0.4);
    \draw[densely dotted,thick] (1.3,-0.4)-- (1.3,0.0);
    \draw[dashed] (1,0) arc (0:180:0.5);
    \draw (1.2,0) arc (0:180:0.7);
    \draw[densely dotted,thick] (1.3,0) arc (0:180:0.8);
    }
\ \,
+
\ \,
\tikz[scale=0.8,baseline=-0.5ex]{
    \draw[dashed] (0,0) -- (1,0);
    \draw (-0.2,0.)-- (-0.2,-0.15);
    \draw (-0.2,-0.15)-- (1.,-0.7);
    \draw (1.2,-0.15)-- (1.2,0.0);
    \draw (1.2,-0.15)-- (0,-0.8);
    \draw[densely dotted,thick] (-0.3,0.)-- (-0.3,-0.4);
    \draw[densely dotted,thick] (1.3,-0.1)-- (1.3,0.);
    \draw[dashed] (2,0) -- (3,0);
    \draw (1.8,0.)-- (1.8,-0.15);
    \draw (1.8,-0.15)-- (3.,-0.8);
    \draw (3.2,-0.15)-- (3.2,0.);
    \draw (3.2,-0.15)-- (2,-0.7);
    \draw[densely dotted,thick] (1.7,0.)-- (1.7,-0.1);
    \draw[densely dotted,thick] (3.3,-0.4)-- (3.3,0.);
    \draw[densely dotted,thick] (1.3,-0.1)--(1.7,-0.1);
    \draw (1.0,-0.7)-- (2,-0.7);
    \draw[dashed] (1,0) arc (0:180:0.5);
    \draw (1.2,0) arc (0:180:0.7);
    \draw[densely dotted,thick] (1.3,0) arc (0:180:0.8);
    \draw[dashed] (3,0) arc (0:180:0.5);
    \draw (3.2,0) arc (0:180:0.7);
    \draw[densely dotted,thick] (3.3,0) arc (0:180:0.8);
 }
 \ \,
+
\cdots
\nonumber \\
&+ 
\ \,
\tikz[scale=0.7,baseline=-0.5ex]{
    \draw[dashed] (0,0) -- (1,0);
    \draw (-0.2,0.)-- (-0.2,-0.15);
    \draw (-0.2,-0.15)-- (1.,-0.7);
    \draw (1.2,-0.15)-- (1.2,0.0);
    \draw (1.2,-0.15)-- (0,-0.8);
    \draw[densely dotted,thick] (-0.3,0.)-- (-0.3,-0.4);
    \draw[densely dotted,thick] (1.3,-0.1)-- (1.3,0.);
    \draw[dashed] (2,0) -- (3,0);
    \draw (1.8,0.)-- (1.8,-0.15);
    \draw (1.8,-0.15)-- (3.,-0.8);
    \draw (3.2,-0.15)-- (3.2,0.);
    \draw (3.2,-0.15)-- (2,-0.7);
    \draw[densely dotted,thick] (1.7,0.)-- (1.7,-0.1);
    \draw[densely dotted,thick] (3.3,-0.4)-- (3.3,0.);
    \draw[densely dotted,thick] (1.3,-0.1)--(1.7,-0.1);
    \draw (1.0,-0.7)-- (2,-0.7);
    \draw[densely dotted,thick] (3.3,0) arc (0:180:1.8);
    \draw[densely dotted,thick] (1.7,0) arc (0:180:0.2);
    \draw[dashed] (2.0,0) arc (0:180:0.5);
    \draw[dashed] (3.0,0) arc (0:180:1.5);
    \draw (1.8,0) arc (0:180:0.3);
    \draw (3.2,0) arc (0:180:1.7);
 }
\ \,
+ 
 \ \,
 \tikz[scale=0.7,baseline=1.5ex]{
    \draw[dashed] (0,0) -- (1,0);
    \draw (-0.2,0.)-- (-0.2,-0.15);
    \draw (-0.2,-0.15)-- (1.,-0.7);
    \draw (1.2,-0.15)-- (1.2,0.0);
    \draw (1.2,-0.15)-- (0,-0.8);
    \draw[densely dotted,thick] (-0.3,0.)-- (-0.3,-0.4);
    \draw[densely dotted,thick] (1.3,-0.1)-- (1.3,0.);
    \draw[dashed] (2,0) -- (3,0);
    \draw (1.8,0.)-- (1.8,-0.15);
    \draw (1.8,-0.15)-- (3.,-0.7);
    \draw (3.2,-0.15)-- (3.2,0.);
    \draw (3.2,-0.15)-- (2,-0.7);
    \draw[densely dotted,thick] (1.7,0.)-- (1.7,-0.1);
    \draw[densely dotted,thick] (3.3,-0.1)-- (3.3,0.);
    \draw[dashed] (4,0) -- (5,0);
    \draw (3.8,0.)-- (3.8,-0.15);
    \draw (3.8,-0.15)-- (5.,-0.7);
    \draw (5.2,-0.15)-- (5.2,0.);
    \draw (5.2,-0.15)-- (4,-0.7);
    \draw[densely dotted,thick] (3.7,0.)-- (3.7,-0.1);
    \draw[densely dotted,thick] (5.3,-0.1)-- (5.3,0.);
    \draw[dashed] (6,0) -- (7,0);
    \draw (5.8,0.)-- (5.8,-0.15);
    \draw (5.8,-0.15)-- (7.,-0.8);
    \draw (7.2,-0.15)-- (7.2,0.);
    \draw (7.2,-0.15)-- (6,-0.7);
    \draw[densely dotted,thick] (5.7,0.)-- (5.7,-0.1);
    \draw[densely dotted,thick] (7.3,-0.4)-- (7.3,0.);
    \draw[densely dotted,thick] (1.3,-0.1)--(1.7,-0.1);
    \draw[densely dotted,thick] (3.3,-0.1)--(3.7,-0.1);
    \draw[densely dotted,thick] (5.3,-0.1)--(5.7,-0.1);
    \draw (1.0,-0.7)-- (2,-0.7);
    \draw (3.0,-0.7)-- (4,-0.7);
    \draw (5.0,-0.7)-- (6,-0.7);
    \draw[densely dotted,thick] (3.3,0) arc (0:180:1.8);
    \draw[densely dotted,thick] (1.7,0) arc (0:180:0.2);
    \draw[dashed] (2.0,0) arc (0:180:0.5);
    \draw[dashed] (3.0,0) arc (0:180:1.5);
    \draw (1.8,0) arc (0:180:0.3);
    \draw (3.2,0) arc (0:180:1.7);
    \draw[densely dotted,thick] (7.3,0) arc (0:180:1.8);
    \draw[densely dotted,thick] (5.7,0) arc (0:180:0.2);
    \draw[dashed] (6.0,0) arc (0:180:0.5);
    \draw[dashed] (7.0,0) arc (0:180:1.5);
    \draw (5.8,0) arc (0:180:0.3);
    \draw (7.2,0) arc (0:180:1.7);
 }
\ \,
+ \cdots
\nonumber \\
&+ 
 \ \,
\tikz[scale=0.5,baseline=2ex]{
    \draw[dashed] (0,0) -- (1,0);
    \draw (-0.2,0.)-- (-0.2,-0.15);
    \draw (-0.2,-0.15)-- (1.,-0.7);
    \draw (1.2,-0.15)-- (1.2,0.0);
    \draw (1.2,-0.15)-- (0,-0.8);
    \draw[densely dotted,thick] (-0.3,0.)-- (-0.3,-0.4);
    \draw[densely dotted,thick] (1.3,-0.1)-- (1.3,0.);
    \draw[dashed] (2,0) -- (3,0);
    \draw (1.8,0.)-- (1.8,-0.15);
    \draw (1.8,-0.15)-- (3.,-0.7);
    \draw (3.2,-0.15)-- (3.2,0.);
    \draw (3.2,-0.15)-- (2,-0.7);
    \draw[densely dotted,thick] (1.7,0.)-- (1.7,-0.1);
    \draw[densely dotted,thick] (3.3,-0.1)-- (3.3,0.);
    \draw[dashed] (4,0) -- (5,0);
    \draw (3.8,0.)-- (3.8,-0.15);
    \draw (3.8,-0.15)-- (5.,-0.7);
    \draw (5.2,-0.15)-- (5.2,0.);
    \draw (5.2,-0.15)-- (4,-0.7);
    \draw[densely dotted,thick] (3.7,0.)-- (3.7,-0.1);
    \draw[densely dotted,thick] (5.3,-0.1)-- (5.3,0.);
    \draw[dashed] (6,0) -- (7,0);
    \draw (5.8,0.)-- (5.8,-0.15);
    \draw (5.8,-0.15)-- (7.,-0.8);
    \draw (7.2,-0.15)-- (7.2,0.);
    \draw (7.2,-0.15)-- (6,-0.7);
    \draw[densely dotted,thick] (5.7,0.)-- (5.7,-0.1);
    \draw[densely dotted,thick] (7.3,-0.4)-- (7.3,0.);
    \draw[densely dotted,thick] (1.3,-0.1)--(1.7,-0.1);
    \draw[densely dotted,thick] (3.3,-0.1)--(3.7,-0.1);
    \draw[densely dotted,thick] (5.3,-0.1)--(5.7,-0.1);
    \draw (1.0,-0.7)-- (2,-0.7);
    \draw (3.0,-0.7)-- (4,-0.7);
    \draw (5.0,-0.7)-- (6,-0.7);
    \draw[densely dotted,thick] (3.7,0) arc (0:180:0.2);
    \draw[densely dotted,thick] (7.3,0) arc (0:180:3.8);
    \draw[densely dotted,thick] (5.7,0) arc (0:180:2.2);
    \draw[densely dotted,thick] (5.3,0) arc (0:180:1.8);
    \draw[dashed] (6.0,0) arc (0:180:2.5);
    \draw[dashed] (5.,0) arc (0:180:1.5);
    \draw[dashed] (4.,0) arc (0:180:0.5);
    \draw[dashed] (7.0,0) arc (0:180:3.5);
    \draw (5.8,0) arc (0:180:2.3);
    \draw (5.2,0) arc (0:180:1.7);
    \draw (3.8,0) arc (0:180:0.3);
    \draw (7.2,0) arc (0:180:3.7);
 } \ \,
 + \cdots
\label{eq:Gz_expansion}
\end{align}
\end{widetext}

We should note that in the right hand side of the above equation terms are ordered such that they contribute as $q$, $q^2$, $q^3$, $\cdots$ (except the first term in the first row which is $1/z$),
where $q=L_B/L_A$ as defined earlier.
Also, the diagrams of the following sort (which appear in the diagrammatic expansion of the propagator for  $\rho$~\cite{Jurkiewicz}) are subleading here, since
\begin{align}
\tikz[scale=0.6,baseline=1.5ex]{
    \draw[dashed] (0,0) -- (1,0);
    \draw (-0.2,0.)-- (-0.2,-0.15);
    \draw (-0.2,-0.15)-- (1.,-0.7);
    \draw (1.2,-0.15)-- (1.2,0.0);
    \draw (1.2,-0.15)-- (0,-0.8);
    \draw[densely dotted,thick] (-0.3,0.)-- (-0.3,-0.4);
    \draw[densely dotted,thick] (1.3,-0.1)-- (1.3,0.);
    \draw[dashed] (2,0) -- (3,0);
    \draw (1.8,0.)-- (1.8,-0.15);
    \draw (1.8,-0.15)-- (3.,-0.7);
    \draw (3.2,-0.15)-- (3.2,0.);
    \draw (3.2,-0.15)-- (2,-0.7);
    \draw[densely dotted,thick] (1.7,0.)-- (1.7,-0.1);
    \draw[densely dotted,thick] (3.3,-0.1)-- (3.3,0.);
    \draw[dashed] (4,0) -- (5,0);
    \draw (3.8,0.)-- (3.8,-0.15);
    \draw (3.8,-0.15)-- (5.,-0.8);
    \draw (5.2,-0.15)-- (5.2,0.);
    \draw (5.2,-0.15)-- (4,-0.7);
    \draw[densely dotted,thick] (3.7,0.)-- (3.7,-0.1);
    \draw[densely dotted,thick] (5.3,-0.4)-- (5.3,0.);
    \draw[densely dotted,thick] (1.3,-0.1)--(1.7,-0.1);
    \draw[densely dotted,thick] (3.3,-0.1)--(3.7,-0.1);
    \draw (1.0,-0.7)-- (2,-0.7);
    \draw (3.0,-0.7)-- (4,-0.7);
    \draw[densely dotted,thick] (5.3,0) arc (0:180:2.8);
    \draw[densely dotted,thick] (1.7,0) arc (0:180:0.2);
    \draw[densely dotted,thick] (3.7,0) arc (0:180:0.2);
    \draw[dashed] (4.0,0) arc (0:180:0.5);
    \draw[dashed] (2.0,0) arc (0:180:0.5);
    \draw[dashed] (5.0,0) arc (0:180:2.5);
    \draw (3.8,0) arc (0:180:0.3);
    \draw (1.8,0) arc (0:180:0.3);
    \draw (5.2,0) arc (0:180:2.7);
 }
 \ \,
 \sim
 \frac{L_B L_{A_1}^2}{L_A^3}
 = \frac{L_B}{L_A}\frac{1}{L_{A_2}^2},
 \label{eq:semicirc_neglect}
\end{align}
which compared to the dominant diagram of $\Tr(\rT)^3$ (that is of order $q^2$) is smaller by a factor of $\tfrac{L_{A_1}}{L_{A_2} L_B}$ (that is our perturbation parameter). We note that all diagrams here are planar with respect to both subgraphs $A_1 B$ and $A_2 B$, or in other words, $g_1=g_2=0$.

Thus, $G$ can be evaluated as a geometric series in terms of a self energy $\Sigma$ as in
\begin{align}
    \tikz[baseline=-0.5ex]{
    \draw[densely dotted,thick] (0.35,0.05)--(0.65,0.05);
    \draw[densely dotted,thick] (-0.35,0.05)--(-0.65,0.05);
    \draw (0.35,-0.05)--(0.65,-0.05);
    \draw (-0.35,-0.05)--(-0.65,-0.05);
    \draw[thick] (0,0) circle (10pt) node[anchor=center] {$G$};
    }
    &=
   \tikz[baseline=-0.5ex]{
    \draw[densely dotted,thick] (-0.4,0.05)--(0.35,0.05);
    \draw (-0.4,-0.05)--(0.35,-0.05);
    }
    +
    \tikz[baseline=-0.5ex]{
    \draw[densely dotted,thick] (0.35,0.05)--(0.65,0.05);
    \draw[densely dotted,thick] (-0.35,0.05)--(-0.65,0.05);
    \draw (0.35,-0.05)--(0.65,-0.05);
    \draw (-0.35,-0.05)--(-0.65,-0.05);
    \draw[thick] (0,0) circle (10pt) node[anchor=center] {$\Sigma$};
    }
    +
    \tikz[baseline=-0.5ex]{
    \draw[densely dotted,thick] (0.35,0.05)--(0.65,0.05);
    \draw[densely dotted,thick] (-0.35,0.05)--(-0.65,0.05);
    \draw (0.35,-0.05)--(0.65,-0.05);
    \draw (-0.35,-0.05)--(-0.65,-0.05);
    \draw[thick] (0,0) circle (10pt) node[anchor=center] {$\Sigma$};
    \draw[thick] (1.05,0) circle (10pt) node[anchor=center] {$\Sigma$};
    \draw[densely dotted,thick] (1.4,0.05)--(1.7,0.05);
    \draw (1.4,-0.05)--(1.7,-0.05);
    }
    + \cdots \nonumber \\
      &= \frac{1}{z-\Sigma(z)},
      \label{eq:geometric}
\end{align}
where the
self-energy $\Sigma$ consists of the sum of all one-particle irreducible diagrams and reads as 
\begin{align}
\tikz[baseline=-0.5ex]{
    \draw[thick] (0,0) circle (10pt) node[anchor=center] {$\Sigma$};
    }
=& \ \,
\tikz[baseline=-1.5ex]{
    \draw[dashed] (0,0) -- (1,0);
    \draw (-0.2,0.0)-- (-0.2,-0.15);
    \draw (-0.2,-0.15)-- (1.,-0.7);
    \draw (1.2,-0.15)-- (1.2,0.0);
    \draw (1.2,-0.15)-- (0,-0.7);
    \draw[densely dotted,thick] (-0.3,0.0)-- (-0.3,-0.4);
    \draw[densely dotted,thick] (1.3,-0.4)-- (1.3,0.0);
    \draw[dashed] (1,0) arc (0:180:0.5);
    \draw (1.2,0) arc (0:180:0.7);
    \draw[densely dotted,thick] (1.3,0) arc (0:180:0.8);
    }
 \ \,
+
\ \,
\tikz[scale=0.7,baseline=-0.5ex]{
    \draw[dashed] (0,0) -- (1,0);
    \draw (-0.2,0.)-- (-0.2,-0.15);
    \draw (-0.2,-0.15)-- (1.,-0.7);
    \draw (1.2,-0.15)-- (1.2,0.0);
    \draw (1.2,-0.15)-- (0,-0.8);
    \draw[densely dotted,thick] (-0.3,0.)-- (-0.3,-0.4);
    \draw[densely dotted,thick] (1.3,-0.1)-- (1.3,0.);
    \draw[dashed] (2,0) -- (3,0);
    \draw (1.8,0.)-- (1.8,-0.15);
    \draw (1.8,-0.15)-- (3.,-0.8);
    \draw (3.2,-0.15)-- (3.2,0.);
    \draw (3.2,-0.15)-- (2,-0.7);
    \draw[densely dotted,thick] (1.7,0.)-- (1.7,-0.1);
    \draw[densely dotted,thick] (3.3,-0.4)-- (3.3,0.);
    \draw[densely dotted,thick] (1.3,-0.1)--(1.7,-0.1);
    \draw (1.0,-0.7)-- (2,-0.7);
    \draw[densely dotted,thick] (3.3,0) arc (0:180:1.8);
    \draw[densely dotted,thick] (1.7,0) arc (0:180:0.2);
    \draw[dashed] (2.0,0) arc (0:180:0.5);
    \draw[dashed] (3.0,0) arc (0:180:1.5);
    \draw (1.8,0) arc (0:180:0.3);
    \draw (3.2,0) arc (0:180:1.7);
 }
\nonumber \\
&+
 \ \,
\tikz[scale=0.5,baseline=2ex]{
    \draw[dashed] (0,0) -- (1,0);
    \draw (-0.2,0.)-- (-0.2,-0.15);
    \draw (-0.2,-0.15)-- (1.,-0.7);
    \draw (1.2,-0.15)-- (1.2,0.0);
    \draw (1.2,-0.15)-- (0,-0.8);
    \draw[densely dotted,thick] (-0.3,0.)-- (-0.3,-0.4);
    \draw[densely dotted,thick] (1.3,-0.1)-- (1.3,0.);
    \draw[dashed] (2,0) -- (3,0);
    \draw (1.8,0.)-- (1.8,-0.15);
    \draw (1.8,-0.15)-- (3.,-0.7);
    \draw (3.2,-0.15)-- (3.2,0.);
    \draw (3.2,-0.15)-- (2,-0.7);
    \draw[densely dotted,thick] (1.7,0.)-- (1.7,-0.1);
    \draw[densely dotted,thick] (3.3,-0.1)-- (3.3,0.);
    \draw[dashed] (4,0) -- (5,0);
    \draw (3.8,0.)-- (3.8,-0.15);
    \draw (3.8,-0.15)-- (5.,-0.7);
    \draw (5.2,-0.15)-- (5.2,0.);
    \draw (5.2,-0.15)-- (4,-0.7);
    \draw[densely dotted,thick] (3.7,0.)-- (3.7,-0.1);
    \draw[densely dotted,thick] (5.3,-0.1)-- (5.3,0.);
    \draw[dashed] (6,0) -- (7,0);
    \draw (5.8,0.)-- (5.8,-0.15);
    \draw (5.8,-0.15)-- (7.,-0.8);
    \draw (7.2,-0.15)-- (7.2,0.);
    \draw (7.2,-0.15)-- (6,-0.7);
    \draw[densely dotted,thick] (5.7,0.)-- (5.7,-0.1);
    \draw[densely dotted,thick] (7.3,-0.4)-- (7.3,0.);
    \draw[densely dotted,thick] (1.3,-0.1)--(1.7,-0.1);
    \draw[densely dotted,thick] (3.3,-0.1)--(3.7,-0.1);
    \draw[densely dotted,thick] (5.3,-0.1)--(5.7,-0.1);
    \draw (1.0,-0.7)-- (2,-0.7);
    \draw (3.0,-0.7)-- (4,-0.7);
    \draw (5.0,-0.7)-- (6,-0.7);
    \draw[densely dotted,thick] (3.7,0) arc (0:180:0.2);
    \draw[densely dotted,thick] (7.3,0) arc (0:180:3.8);
    \draw[densely dotted,thick] (5.7,0) arc (0:180:2.2);
    \draw[densely dotted,thick] (5.3,0) arc (0:180:1.8);
    \draw[dashed] (6.0,0) arc (0:180:2.5);
    \draw[dashed] (5.,0) arc (0:180:1.5);
    \draw[dashed] (4.,0) arc (0:180:0.5);
    \draw[dashed] (7.0,0) arc (0:180:3.5);
    \draw (5.8,0) arc (0:180:2.3);
    \draw (5.2,0) arc (0:180:1.7);
    \draw (3.8,0) arc (0:180:0.3);
    \draw (7.2,0) arc (0:180:3.7);
 } \ \, 
 + \cdots
\end{align}
%
As a result, we arrive at the following SD relation 
\begin{align}
\label{eq:SD_diag}
\tikz[baseline=-0.5ex]{
    \draw[thick] (0,0) circle (10pt) node[anchor=center] {$\Sigma$};
    }
 =&
 \ \,
\tikz[baseline=1ex]{
    \draw[dashed] (0,0) -- (1,0);
    \draw (-0.2,0.0)-- (-0.2,-0.2);
    \draw (1.2,-0.2)-- (1.2,0.0);
    \draw[densely dotted,thick] (-0.3,0.0)-- (-0.3,-0.2);
    \draw[densely dotted,thick] (1.3,-0.2)-- (1.3,0.0);
    \draw[dashed] (1,0) arc (0:180:0.5);
    \draw (1.2,0) arc (0:180:0.7);
    \draw[densely dotted,thick] (1.3,0) arc (0:180:0.8);
    }
 \ \,
+
\ \,
\tikz[scale=0.7,baseline=1ex]{
    \draw[dashed] (0,0) -- (0.7,0);
    \draw (-0.2,0.)-- (-0.2,-0.2);
    \draw (0.9,-0.05)-- (0.9,0.0);
    \draw[densely dotted,thick] (-0.3,0.)-- (-0.3,-0.2);
    \draw[dashed] (2.3,0) -- (3,0);
    \draw (3.2,-0.2)-- (3.2,0.);
    \draw (2.1,-0.05)-- (2.1,0.0);
    \draw[densely dotted,thick] (3.3,-0.2)-- (3.3,0.);
    \draw[densely dotted,thick] (1,0.05)--(2,0.05);
    \draw (0.9,-0.05)-- (2.1,-0.05);
    \draw[densely dotted,thick] (3.3,0) arc (0:180:1.8);
    \draw[densely dotted,thick] (2.,0.05) arc (10:170:0.5);
    \draw[dashed] (2.3,0) arc (0:180:0.8);
    \draw[dashed] (3.0,0) arc (0:180:1.5);
    \draw (2.1,0) arc (0:180:0.6);
    \draw (3.2,0) arc (0:180:1.7);
    \draw[thick,fill=white] (1.5,0.) circle (0.25) node[anchor=center] {\footnotesize$G$};
    }\ \, ,
\end{align}
which leads to the algebraic relation
\begin{align}
	\label{eq:SD}
    \Sigma= q (1+ G).
\end{align}
Notice that in the graphical representation (\ref{eq:SD_diag}) we resolved the crossings as they are mainly implemented to keep track of matrix index contractions.   
Finally, we combine Eqs.~(\ref{eq:geometric}) and (\ref{eq:SD}) to arrive at
\begin{align}
    \label{eq:sd_semicirc}
   q G^2 - (z-q) G + 1 =0.
\end{align}
The corresponding solution then reads,
\begin{align}
    G(z) =\frac{1}{2 q} \left( (z-q) - \sqrt{(z-q)^2 - 4 q} \right).
    \label{eq:GreensFunction_SemicircleRegime}
\end{align}
Using Eq.\,(\ref{eq:dos}), we get a semi-circle law spectrum
\begin{align}
    \label{eq:spec_diag}
    n (x)= \frac{L_A}{2\pi q} \sqrt{4q - (x-q)^2},
\end{align}
which upon a proper rescaling leads to
\begin{align}
\label{eq:semicirc}
 P_\Gamma (\xi) = \frac{2 L_A}{\pi a^2}\sqrt{a^2-\bigg(\xi-\frac{1}{L_A}\bigg)^2},
 \qquad \left|\xi-\frac{1}{L_A}\right| < a,
\end{align}
where the radius is given by 
\begin{align}
a = \frac{2}{\sqrt{L_A L_B}}.
\end{align} 
The semi-circle law, which we derived here, was used as an empirical fit to numerical simulations in previous works~\cite{Aubrun2012,Bhosale}.
As we see, to the leading order in our pertubation parameter $\frac{L_{A_s}}{L_B L_{A_{\bar{s}}}}$ the spectral density (and hence the log negativity) only depends on $L_A/L_B$ and not on the relative ratio of the subsystem sizes
$L_{A_1}/L_{A_2}$. This means that the entanglement is independent of how we partition $A$ and may be thought of as topological. For this reason, we call $\rho$ a saturated entangled state.

\begin{figure}
\includegraphics[scale=0.65]{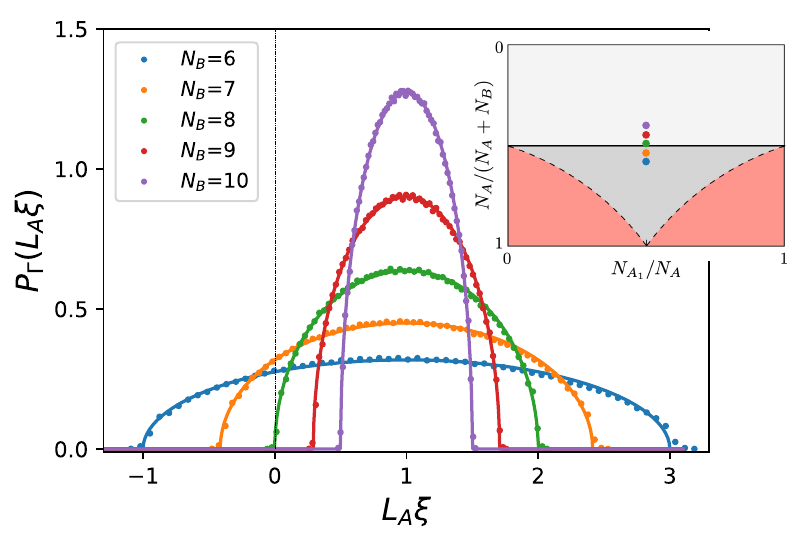}
\caption{\label{fig:semicirc} Negativity spectrum of a random mixed state (\ref{eq:red_den}) in the entanglement saturation regime (cf.~region III in Fig.~\ref{fig:phasediag}). Circles are numerical simulation and solid lines are semi-circle law~(\ref{eq:semicirc}).
Inset shows the location of each curve (color coded) in the phase diagram.
Here, we choose $N_{A_1}=N_{A_2}=3$.
Ensemble averages are performed over $10^4$ samples. Small fluctuations are remnant of the universal oscillations due to finite-size effects.}
\end{figure}

 A comparison of the semi-circle law with the exact numerics is shown in Fig.~\ref{fig:semicirc}, which indicates a good agreement. An immediate result of the semi-circle distribution is that for $a > \frac{1}{L_A}$, there exist some negative eigenvalues, while $ a< \frac{1}{L_A}$ all eigenvalues are positive. This is the anticipated NPT-PPT transition (also known as the entanglement phase  transition~\cite{Aubrun2012} or sudden death of the entanglement in literature~\cite{Calabrese_Ft2015,Sherman2016}) 
 The transition point corresponds to $L_B= 4 L_A$ (or $q=4$) which can be interpreted as when subsystem $B$ has two more qubits than subsystem $A$.

Another interesting observation is that the range of the eigenvalues is the same both before and after PT,
$2 a = \lambda_{+} - \lambda_{-} = \frac{4}{\sqrt{L_A L_B}}$,
where $\lambda_{\pm}$ are the limits of the MP distribution in Eq.\,(\ref{eq:MPdist}). Extreme deviations from this behavior occurs away from the saturated entangled regime.

We plug in the semi-circle law to Eq.\,(\ref{eq:neg_dist}) and derive an approximate form of the average negativity~\cite{Bhosale},
\begin{align}
\braket{ {\cal E}_{A_1:A_2} }_{sm} =
\log_2 \Bigg(\frac{2}{\pi} \sin^{-1}\frac{\sqrt{q}}{2}+
       \frac{ \left(\sqrt{q}+\frac{8}{\sqrt{q}} \right) \sqrt{ 1-\frac{q}{4}} }{3\pi }
       \Bigg), 
\label{eq:logneg}
\end{align}
where $\braket{\cdot}_{sm}$ denotes average using the semi-circle law approximation (\ref{eq:semicirc}). Above the transition point $q\geq 4$, i.e.,  $L_B\geq 4 L_A$, this formula gives zero for the average LN. Furthermore, if we take the limit $q \ll 1$ (deep in the NPT regime) the dominant term comes from the second term inside parenthesis above and yields the same result as Eq.~(\ref{eq:LN_smallq}) including the order-one constant $c_1$.

As mentioned, so far in this regime, neither the spectral density nor the LN depend on how we partition the subsystem $A$. In fact, the dependence on the ratio of subsystems $A_1$ and $A_2$ only shows up at higher orders, which we elaborate below.


\subsubsection{$1/L$ corrections}
\label{sec:1/L corrections}

Here, for simplicity of discussion, we restrict to the concrete large-$L$ limit where $L_A,L_B\rightarrow\infty$ and $q=L_B/L_A$ is held fixed with $0<q<\infty$. We further assume that $\eta=L_{A_1}/L_{A_2}$ is kept fixed with $0<\eta<\infty$ as $L_A\rightarrow \infty$. The one-point Green's function of $H=L_B (XX^\dag)^{T_2}$ therefore has a perturbative expansion in terms of a single small parameter $1/L_A$ with \eqref{eq:GreensFunction_SemicircleRegime} being the leading order result. 
The next-to-leading-order correction to this one-point function is found to be
\begin{align}
    \label{eq:Gz_corr}
	G^{(1)}(z)=- q \left( \frac{ \eta + \eta^{-1}}{L_A} \right) G'(z)G^2(z)(1+G(z)).  
\end{align}
Here, by abusing the notation, $G(z)$ denotes the leading-order one-point function. The key to compute $G^{(1)}$ is to consider the following two diagrams: 
\begin{align}
&\begin{tikzpicture}
	\begin{pgfonlayer}{nodelayer}
		\node [style=none] (0) at (-0.75, 0.25) {};
		\node [style=none] (1) at (-1, -0.5) {};
		\node [style=none] (2) at (1, -0.5) {};
		\node [style=none] (3) at (0.75, 0.25) {};
		\node [style=none] (4) at (-0.5, 0.5) {};
		\node [style=none] (5) at (0.5, 0.5) {};
		\node [style=none] (18) at (-2, -0.5) {};
		\node [style=none] (19) at (-2, 0.25) {};
		\node [style=none] (20) at (-1, 0.25) {};
		\node [style=none] (21) at (-1, 0.5) {};
		\node [style=none] (22) at (-0.75, 0.5) {};
		\node [style=none] (23) at (0.75, 0.5) {};
		\node [style=none] (24) at (1, 0.5) {};
		\node [style=none] (25) at (1, 0.25) {};
		\node [style=none] (26) at (2.25, 0.25) {};
		\node [style=none] (27) at (2, -0.5) {};
		\node [style=none] (28) at (4, -0.5) {};
		\node [style=none] (29) at (3.75, 0.25) {};
		\node [style=none] (30) at (2.5, 0.5) {};
		\node [style=none] (31) at (3.5, 0.5) {};
		\node [style=none] (32) at (2, 0.25) {};
		\node [style=none] (33) at (2, 0.5) {};
		\node [style=none] (34) at (2.25, 0.5) {};
		\node [style=none] (35) at (3.75, 0.5) {};
		\node [style=none] (36) at (4, 0.5) {};
		\node [style=none] (37) at (4, 0.25) {};
		\node [style=none] (38) at (6.25, 0.25) {};
		\node [style=none] (39) at (6, -0.5) {};
		\node [style=none] (40) at (8, -0.5) {};
		\node [style=none] (41) at (7.75, 0.25) {};
		\node [style=none] (42) at (6.5, 0.5) {};
		\node [style=none] (43) at (7.5, 0.5) {};
		\node [style=none] (44) at (6, 0.25) {};
		\node [style=none] (45) at (6, 0.5) {};
		\node [style=none] (46) at (6.25, 0.5) {};
		\node [style=none] (47) at (7.75, 0.5) {};
		\node [style=none] (48) at (8, 0.5) {};
		\node [style=none] (49) at (8, 0.25) {};
		\node [style=none] (50) at (4.5, 0.25) {};
		\node [style=none] (51) at (5.5, 0.25) {};
		\node [style=none] (52) at (9, 0.25) {};
		\node [style=none] (53) at (9, -0.5) {};
		\node [style=none] (54) at (4.5, -0.5) {};
		\node [style=none] (55) at (5.5, -0.5) {};
		\node [style=none] (56) at (4.5, 1) {};
		\node [style=none] (57) at (4.5, 1.25) {};
		\node [style=none] (58) at (4.5, 1.5) {};
		\node [style=none] (59) at (5.5, 1.5) {};
		\node [style=none] (60) at (5.5, 1.25) {};
		\node [style=none] (61) at (5.5, 1) {};
		\node [style=none] (62) at (1, 2.25) {};
		\node [style=none] (63) at (1, 2.5) {};
		\node [style=none] (64) at (1, 2) {};
		\node [style=none] (65) at (6, 2.5) {};
		\node [style=none] (66) at (6, 2.25) {};
		\node [style=none] (67) at (6, 2) {};
		\node [style=none] (68) at (5, 0) {$\cdots$};
		\node [style=none] (69) at (0, -1) {$1$};
		\node [style=none] (70) at (3, -1) {$2$};
		\node [style=none] (71) at (7, -1) {$n$};
	\end{pgfonlayer}
	\begin{pgfonlayer}{edgelayer}
		\draw (0.center) to (2.center);
		\draw (1.center) to (3.center);
		\draw [style=dashed] (4.center) to (5.center);
		\draw (26.center) to (28.center);
		\draw (27.center) to (29.center);
		\draw [style=dashed] (30.center) to (31.center);
		\draw (38.center) to (40.center);
		\draw (39.center) to (41.center);
		\draw [style=dashed] (42.center) to (43.center);
		\draw [style=dotted] (19.center) to (20.center);
		\draw [style=dotted] (25.center) to (32.center);
		\draw [style=dotted] (37.center) to (50.center);
		\draw [style=dotted] (51.center) to (44.center);
		\draw [style=dotted] (49.center) to (52.center);
		\draw (18.center) to (1.center);
		\draw (2.center) to (27.center);
		\draw (40.center) to (53.center);
		\draw (28.center) to (54.center);
		\draw (55.center) to (39.center);
		\draw [style=dotted] (21.center) to (20.center);
		\draw [style=dotted] (24.center) to (25.center);
		\draw [style=dotted] (33.center) to (32.center);
		\draw [style=dotted] (36.center) to (37.center);
		\draw [style=dotted] (45.center) to (44.center);
		\draw [style=dotted] (48.center) to (49.center);
		\draw (22.center) to (0.center);
		\draw (23.center) to (3.center);
		\draw (34.center) to (26.center);
		\draw (35.center) to (29.center);
		\draw (46.center) to (38.center);
		\draw (47.center) to (41.center);
		\draw [style=dotted, bend left=90, looseness=1.75] (24.center) to (33.center);
		\draw [bend left=90, looseness=1.75] (23.center) to (34.center);
		\draw [style=dashed, bend left=90, looseness=1.75] (5.center) to (30.center);
		\draw [style=dashed, bend left=45] (31.center) to (58.center);
		\draw [bend left=45] (35.center) to (57.center);
		\draw [style=dotted, bend left=45] (36.center) to (56.center);
		\draw [style=dotted, bend left=45] (61.center) to (45.center);
		\draw [bend left=45] (60.center) to (46.center);
		\draw [style=dashed, bend left=45] (59.center) to (42.center);
		\draw [bend left=45] (22.center) to (62.center);
		\draw [style=dotted, bend left=45] (21.center) to (63.center);
		\draw [style=dashed, bend left=45] (4.center) to (64.center);
		\draw [bend left=45] (66.center) to (47.center);
		\draw [style=dotted, bend left=45] (65.center) to (48.center);
		\draw [style=dashed, bend left=45] (67.center) to (43.center);
		\draw [style=dotted] (63.center) to (65.center);
		\draw (62.center) to (66.center);
		\draw [style=dashed] (64.center) to (67.center);
	\end{pgfonlayer}
\end{tikzpicture} \nonumber\\
&={q}/{L_{A_2}^{n+{\rm mod}(n,2)-2}}, 
\label{eq:correction1}
\end{align}
\begin{align}
&\begin{tikzpicture}
	\begin{pgfonlayer}{nodelayer}
		\node [style=none] (0) at (-0.75, 0.25) {};
		\node [style=none] (1) at (-1, -0.5) {};
		\node [style=none] (2) at (1, -0.5) {};
		\node [style=none] (3) at (0.75, 0.25) {};
		\node [style=none] (4) at (-0.5, 0.5) {};
		\node [style=none] (5) at (0.5, 0.5) {};
		\node [style=none] (18) at (-2, -0.5) {};
		\node [style=none] (19) at (-2, 0.25) {};
		\node [style=none] (20) at (-1, 0.25) {};
		\node [style=none] (21) at (-1, 0.5) {};
		\node [style=none] (22) at (-0.75, 0.5) {};
		\node [style=none] (23) at (0.75, 0.5) {};
		\node [style=none] (24) at (1, 0.5) {};
		\node [style=none] (25) at (1, 0.25) {};
		\node [style=none] (26) at (2.25, 0.25) {};
		\node [style=none] (27) at (2, -0.5) {};
		\node [style=none] (28) at (7, -0.5) {};
		\node [style=none] (29) at (6.75, 0.25) {};
		\node [style=none] (30) at (2.5, 0.5) {};
		\node [style=none] (31) at (6.5, 0.5) {};
		\node [style=none] (32) at (2, 0.25) {};
		\node [style=none] (33) at (2, 0.5) {};
		\node [style=none] (34) at (2.25, 0.5) {};
		\node [style=none] (35) at (6.75, 0.5) {};
		\node [style=none] (36) at (7, 0.5) {};
		\node [style=none] (37) at (7, 0.25) {};
		\node [style=none] (38) at (9.25, 0.25) {};
		\node [style=none] (39) at (9, -0.5) {};
		\node [style=none] (40) at (11, -0.5) {};
		\node [style=none] (41) at (10.75, 0.25) {};
		\node [style=none] (42) at (9.5, 0.5) {};
		\node [style=none] (43) at (10.5, 0.5) {};
		\node [style=none] (44) at (9, 0.25) {};
		\node [style=none] (45) at (9, 0.5) {};
		\node [style=none] (46) at (9.25, 0.5) {};
		\node [style=none] (47) at (10.75, 0.5) {};
		\node [style=none] (48) at (11, 0.5) {};
		\node [style=none] (49) at (11, 0.25) {};
		\node [style=none] (50) at (7.5, 0.25) {};
		\node [style=none] (51) at (8.5, 0.25) {};
		\node [style=none] (52) at (12, 0.25) {};
		\node [style=none] (53) at (12, -0.5) {};
		\node [style=none] (54) at (7.5, -0.5) {};
		\node [style=none] (55) at (8.5, -0.5) {};
		\node [style=none] (56) at (7.5, 2.5) {};
		\node [style=none] (57) at (7.5, 2.75) {};
		\node [style=none] (58) at (7.5, 3) {};
		\node [style=none] (59) at (8.5, 3) {};
		\node [style=none] (60) at (8.5, 2.75) {};
		\node [style=none] (61) at (8.5, 2.5) {};
		\node [style=none] (62) at (0.75, 3.75) {};
		\node [style=none] (63) at (0.5, 4) {};
		\node [style=none] (64) at (1, 3.5) {};
		\node [style=none] (65) at (9.5, 4) {};
		\node [style=none] (66) at (9.25, 3.75) {};
		\node [style=none] (67) at (9, 3.5) {};
		\node [style=none] (68) at (8, 0) {$\cdots$};
		\node [style=none] (69) at (0, -1) {$1$};
		\node [style=none] (70) at (3, -1) {$2$};
		\node [style=none] (71) at (10, -1) {$n$};
		\node [style=none] (72) at (4, -0.5) {};
		\node [style=none] (73) at (3.75, 0.25) {};
		\node [style=none] (74) at (3.5, 0.5) {};
		\node [style=none] (75) at (3.75, 0.5) {};
		\node [style=none] (76) at (4, 0.5) {};
		\node [style=none] (77) at (4, 0.25) {};
		\node [style=none] (78) at (5.25, 0.25) {};
		\node [style=none] (79) at (5, -0.5) {};
		\node [style=none] (80) at (5.5, 0.5) {};
		\node [style=none] (81) at (5, 0.25) {};
		\node [style=none] (82) at (5, 0.5) {};
		\node [style=none] (83) at (5.25, 0.5) {};
		\node [style=none] (84) at (6, -1) {$3$};
	\end{pgfonlayer}
	\begin{pgfonlayer}{edgelayer}
		\draw (38.center) to (40.center);
		\draw (39.center) to (41.center);
		\draw [style=dashed] (42.center) to (43.center);
		\draw [style=dotted] (19.center) to (20.center);
		\draw [style=dotted] (25.center) to (32.center);
		\draw [style=dotted] (37.center) to (50.center);
		\draw [style=dotted] (51.center) to (44.center);
		\draw [style=dotted] (49.center) to (52.center);
		\draw (18.center) to (1.center);
		\draw (2.center) to (27.center);
		\draw (40.center) to (53.center);
		\draw (28.center) to (54.center);
		\draw (55.center) to (39.center);
		\draw [style=dotted] (21.center) to (20.center);
		\draw [style=dotted] (24.center) to (25.center);
		\draw [style=dotted] (33.center) to (32.center);
		\draw [style=dotted] (36.center) to (37.center);
		\draw [style=dotted] (45.center) to (44.center);
		\draw [style=dotted] (48.center) to (49.center);
		\draw (22.center) to (0.center);
		\draw (23.center) to (3.center);
		\draw (34.center) to (26.center);
		\draw (35.center) to (29.center);
		\draw (46.center) to (38.center);
		\draw (47.center) to (41.center);
		\draw [style=dotted] (77.center) to (81.center);
		\draw (72.center) to (79.center);
		\draw [style=dotted] (76.center) to (77.center);
		\draw [style=dotted] (82.center) to (81.center);
		\draw (75.center) to (73.center);
		\draw (83.center) to (78.center);
		\draw (1.center) to (3.center);
		\draw (26.center) to (72.center);
		\draw (27.center) to (73.center);
		\draw (78.center) to (28.center);
		\draw (79.center) to (29.center);
		\draw [style=dashed] (4.center) to (5.center);
		\draw [style=dashed] (30.center) to (74.center);
		\draw [style=dashed] (80.center) to (31.center);
		\draw [style=dashed, bend left=90, looseness=1.75] (4.center) to (74.center);
		\draw [bend left=90, looseness=1.75] (22.center) to (75.center);
		\draw [style=dotted, bend left=90, looseness=1.75] (21.center) to (76.center);
		\draw (0.center) to (2.center);
		\draw [style=dashed, bend left=90, looseness=1.75] (30.center) to (31.center);
		\draw [bend left=90, looseness=1.75] (34.center) to (35.center);
		\draw [style=dotted, bend left=90, looseness=1.75] (33.center) to (36.center);
		\draw [style=dotted, bend left=45] (82.center) to (58.center);
		\draw [bend left=45] (83.center) to (57.center);
		\draw [style=dashed, bend left=45] (80.center) to (56.center);
		\draw [style=dotted, bend left=45] (59.center) to (48.center);
		\draw [bend left=45] (60.center) to (47.center);
		\draw [style=dashed, bend left=45] (61.center) to (43.center);
		\draw [style=dotted] (64.center) to (24.center);
		\draw (62.center) to (23.center);
		\draw [style=dashed] (63.center) to (5.center);
		\draw [style=dotted] (67.center) to (45.center);
		\draw (66.center) to (46.center);
		\draw [style=dashed] (65.center) to (42.center);
		\draw [style=dashed] (63.center) to (65.center);
		\draw (62.center) to (66.center);
		\draw [style=dotted] (64.center) to (67.center);
	\end{pgfonlayer}
\end{tikzpicture} \nonumber\\
&={q}/{L_{A_1}^{n+{\rm mod}(n,2)-2}}, 
\label{eq:correction2}
\end{align}
both of which contribute to $G^{(1)}(z)$ when $n=3$ or $4$~\footnote{The exact same terms contribute as the leading order in $L_{A_1} > L_B L_{A_2}$ and $L_{A_1} > L_B L_{A_2}$ regimes, respectively, as discussed around Eq.\,(\ref{eq:RN_largeLA1})} and we have neglected them in (\ref{eq:Gz_expansion}) as discussed around Eq.\,(\ref{eq:semicirc_neglect}). In fact, the most general diagrams for $G^{(1)}(z)$ can be obtained by decorating the above diagrams at $n=3,4$ with order-$1$ terms, which consists of two steps. First, each internal bare propagator (not the two ends) should be replaced by $G(z)$ that is the full leading-order propagator. Second, we can take any single diagram of $G(z)$, cut on a bare propagator to separate it into two parts (possibly still connected by contraction lines) and then glue the two parts to the two ends of each order-$1/L_A$ diagrams above. This second step gives a factor $\frac{1}{z}\times (-z G'(z))$ where $-zG'(z)$ counts the number of different ways to separate a $G(z)$ diagram into two and $1/z$ is the additional bare propagator resulting from the cutting. Combining all these contributions leads to Eq.\,(\ref{eq:Gz_corr}) for $G^{(1)}(z)$. This expression is derived for the Wishart random matrix~(\ref{eq:Wishart}), which does not have a fluctuating trace in the denominator as in the random mixed state (\ref{eq:red_den}). However, it turned out that the fluctuations of this normalization factor in (\ref{eq:red_den}) is of order $1/L_A^2$ and therefore contribute at higher orders. We have verified that Eq.\,(\ref{eq:Gz_corr}) coincides with the exact expansion \eqref{ExactExpression} near $z=\infty$ at least up to order $1/z^{11}$ in the large-$L$ limit. 

\begin{figure}
	\centering
	\includegraphics[width=1\linewidth]{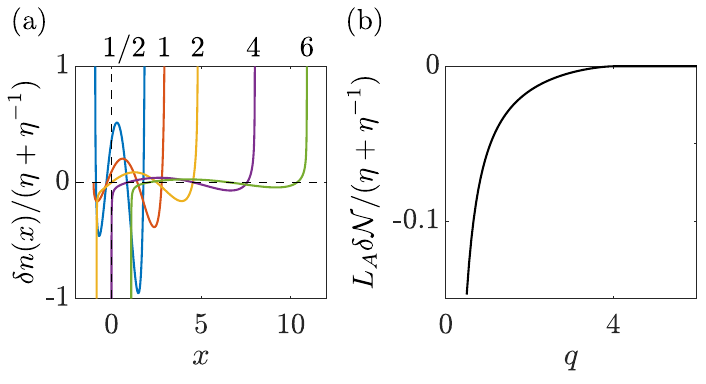}
	\caption{(a) Correction to the entanglement negativity spectrum: $\delta n(x)/(\eta+\eta^{-1})$ for a few different values of $q$ as indicated on the top of the figure. (b) Correction to the negativity: $L_A \delta \mathcal N/(\eta+\eta^{-1})$ as a function of $q=L_B/L_A$. }
	\label{SubleadingCorrections}
\end{figure}

Using (\ref{eq:Gz_corr}), we obtain the relevant correction to the spectral density of $H$ via 
\begin{align}
	\delta n(x)=-\frac{L_A}{\pi}{
		\rm Im}\lim_{\epsilon\rightarrow 0^+}G^{(1)}(x+i\epsilon). 
\end{align}
$\delta n(x)$ is supported on $[a_-,a_+]$ with $a_\pm=q\pm 2\sqrt{q}$, identical to the support of the leading-order spectral density (\ref{eq:spec_diag}). Furthermore, $\delta n$ satisfies the condition $\int dx~x^{0,1,2} \delta n(x)=0$, reflecting the fact that $G^{(1)}(z)$ only contains order $z^{-4}$ or higher terms near $z=\infty$. Consequently, the correction to the spectral density of $\rho^{T_2}$ is given by $dP_\Gamma(\xi)=L_B \delta n(L_B\xi)$, and the correction to the negativity $\mathcal N(\rho)$ in (\ref{eq:neg_def}) is computed via
\begin{align}
	\delta {\cal N} =-\frac{1}{qL_A}\int_{x<0}dx\, x\delta n(x). 
\end{align}
Figure~\ref{SubleadingCorrections} shows $\delta n(x)$ for a few different values of $q$ and $\delta \mathcal {N}$ as a function of $q$. $n(x)$ almost always has power-law divergences with power $-1/2$ near the two ends $a_\pm$ of its support, except for $q=1$ where $n(x)$ diverges at $a_+$ but is finite at $a_-$. We find that the correction to the negativity is always negative when $q<4$ and zero when $q\geq 4$.
The factor $(L_{A_1}^{-2}+L_{A_2}^{-2})= L_A^{-1} (\eta+\eta^{-1})$ then implies that the entanglement between $A_1$ and $A_2$ is maximized when $\eta=1$ provided that $L_A$ and $L_B$ are held fixed. 

We finish this part by some remarks regarding the relation between $\rT$ and the GUE ensemble.
Recall that the leading-order spectral density of $\rho^{T_2}$ is given by a semi-circle law which looks the same as that of a (shifted and properly rescaled) GUE matrix. Such a GUE-type matrix can be defined using a $1$-Hermitian matrix integral 
\begin{align}
    \mathcal{Z}=\int dH~e^{- L\Tr V(H)},
\end{align}
where $H$ is an $L_A\times L_A$ Hermitian matrix and $V(x)$ in general is a real analytic function that directly applies to the eigenvalues of $H$. Our result for $G^{(1)}(z)$ shows that a partially transposed Wishart random matrix is \emph{not} equivalent to a GUE-type matrix for arbitrary $q$ and $\eta$, since there is no correction of order $1/L$ in GUE \cite{SaadJT}. In fact, it is not equivalent to any $1$-Hermitian matrix integral defined above because for such random matrices, the general multi-point Green's functions at all orders are completely determined by the leading order one-point function \cite{SaadJT,Eynard_2004}. 

In the next part, we take on a less restrictive limit $N_{A_1}> N_{A_2}$ and calculate the resolvent function by including all terms of type (\ref{eq:correction1}) in the sum (\ref{eq:Gz_expansion}). The resulting resolvent function captures the right half of the phase diagram in Fig.\,\ref{fig:phasediag} and also reproduces the semi-circle law in the regime studied above.

\subsection{General result}

Here, we derive the resolvent function in the limit where $N_{A_1}>N_{A_2}$. 
To this end, we choose 
the normalization of the random matrix as
$H = L_B L_{A_2} (X X^\dag)^{T_2}$ in Eq.\,(\ref{eq:Gzdef}) and 
carry out a $\frac{1}{L_{A_1}}$ perturbation theory.
Given this normalization, the ensemble average is represented by triple lines with a factor of $\frac{1}{L_{A_1}}$,
\begin{align}
\tikz[baseline=-0.1ex,scale=0.9]{
    \draw[dashed] (1.0,0) arc (0:180:0.5);
    \draw (1.1,0.) arc (0:180:0.6);
    \draw[densely dotted,thick] (1.2,0.) arc (0:180:0.7);
    }
    \ \,
    =\frac{1}{L_{A_1}}\, \delta_{i_1 j_1} \delta_{i_2 j_2} \delta_{\alpha\beta},
\end{align} 
and as usual close loop of subsystem $s=A_1,A_2$, or $B$ gives a factor of $L_s$.

\begin{figure*}
    \centering
    \includegraphics[scale=0.8]{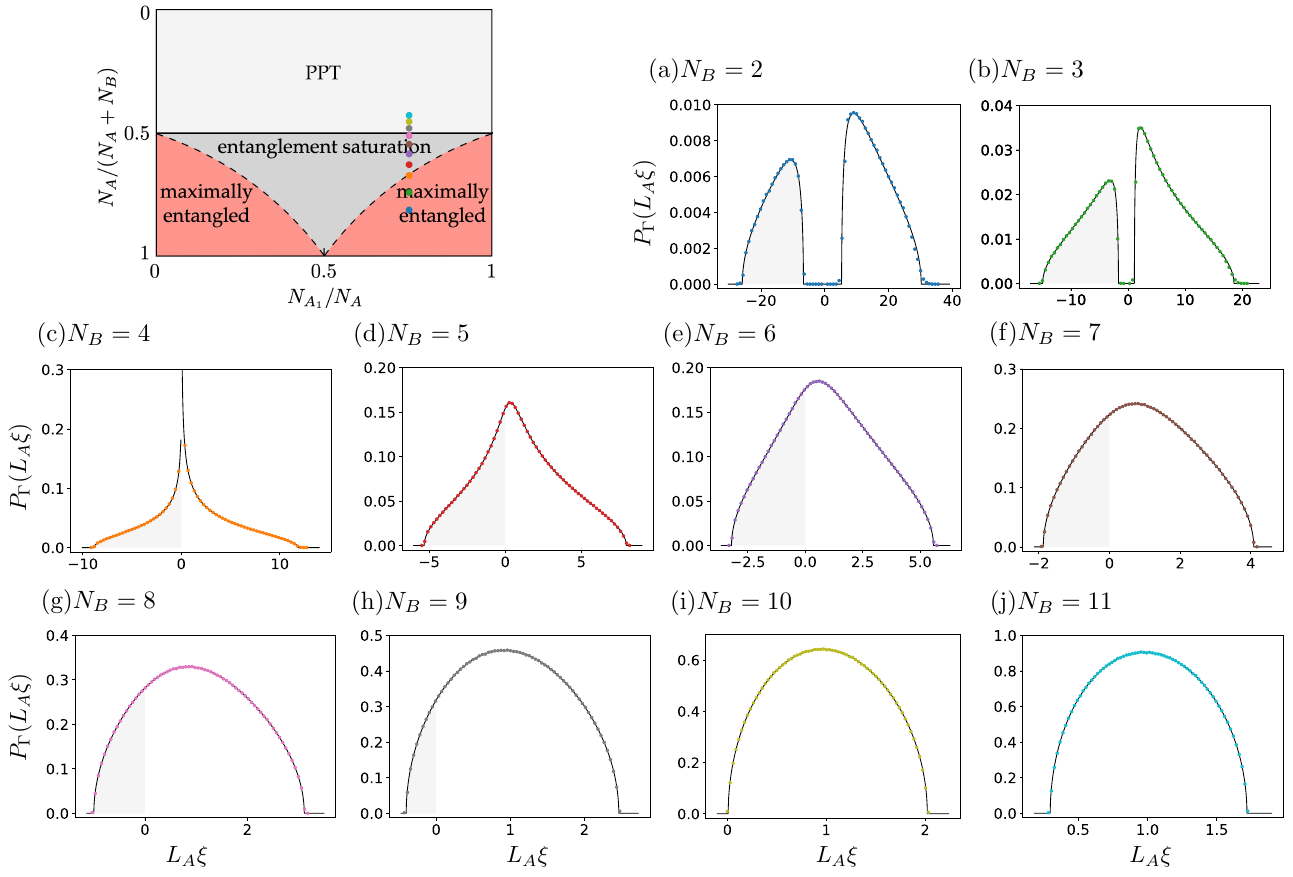}
    \caption{Evolution of the negativity spectrum as the size of subsystem $B$ is increased. This trend corresponds to sweeping a vertical path from bottom to top in the phase diagram as shown in the first panel. The color circles in each panel are numerical simulations (averaged over $10^4$ samples) and solid lines correspond to  Eq.~(\ref{eq:Gz_cubic_sol}).
    Here, we set $N_{A_2}=2$ and $N_{A_1}=6$.
    The shaded regions indicate the domain of negative eigenvalues.
    Two transitions occur as we crank up $N_B$: 
    First, the transition from maximally entangled to a saturated entangled state at $N_B=4$, characterized by diverging spectral density at $\xi=0$, 
    second, the NPT to PPT transition at $N_B=10$ where the semi-circle's support becomes completely non-negative.
    }
    \label{fig:NS_vs_Lb}
\end{figure*}

We note that since there is an even/odd effect for the R\'enyi negativity, we need to consider two self-energy functions in the expansion of the resolvent function
\begin{align}
    \tikz[baseline=-0.5ex]{
    \draw[densely dotted,thick] (0.35,0.05)--(0.65,0.05);
    \draw[densely dotted,thick] (-0.35,0.05)--(-0.65,0.05);
    \draw (0.35,-0.05)--(0.65,-0.05);
    \draw (-0.35,-0.05)--(-0.65,-0.05);
    \draw[thick] (0,0) circle (10pt) node[anchor=center] {$G$};
    }
    =&
   \tikz[baseline=-0.5ex]{
    \draw[densely dotted,thick] (-0.4,0.05)--(0.35,0.05);
    \draw (-0.4,-0.05)--(0.35,-0.05);
    }
    +
    \tikz[baseline=-0.5ex]{
    \draw[densely dotted,thick] (0.35,0.05)--(0.65,0.05);
    \draw[densely dotted,thick] (-0.35,0.05)--(-0.65,0.05);
    \draw (0.35,-0.05)--(0.65,-0.05);
    \draw (-0.35,-0.05)--(-0.65,-0.05);
    \draw[thick] (0,0) circle (10pt) node[anchor=center] {$\Sigma_o$};
    }
    +
    \tikz[baseline=-0.5ex]{
    \draw[densely dotted,thick] (0.35,0.05)--(0.65,0.05);
    \draw[densely dotted,thick] (-0.35,0.05)--(-0.65,0.05);
    \draw (0.35,-0.05)--(0.65,-0.05);
    \draw (-0.35,-0.05)--(-0.65,-0.05);
    \draw[thick] (0,0) circle (10pt) node[anchor=center] {$\Sigma_e$};
    }
    \nonumber \\
    &+
    \tikz[baseline=-0.5ex]{
    \draw[densely dotted,thick] (0.35,0.05)--(0.65,0.05);
    \draw[densely dotted,thick] (-0.35,0.05)--(-0.65,0.05);
    \draw (0.35,-0.05)--(0.65,-0.05);
    \draw (-0.35,-0.05)--(-0.65,-0.05);
    \draw[thick] (0,0) circle (10pt) node[anchor=center] {$\Sigma_o$};
    \draw[thick] (1.05,0) circle (10pt) node[anchor=center] {$\Sigma_e$};
    \draw[densely dotted,thick] (1.4,0.05)--(1.7,0.05);
    \draw (1.4,-0.05)--(1.7,-0.05);
    }
+
   \tikz[baseline=-0.5ex]{
    \draw[densely dotted,thick] (0.35,0.05)--(0.65,0.05);
    \draw[densely dotted,thick] (-0.35,0.05)--(-0.65,0.05);
    \draw (0.35,-0.05)--(0.65,-0.05);
    \draw (-0.35,-0.05)--(-0.65,-0.05);
    \draw[thick] (0,0) circle (10pt) node[anchor=center] {$\Sigma_e$};
    \draw[thick] (1.05,0) circle (10pt) node[anchor=center] {$\Sigma_o$};
    \draw[densely dotted,thick] (1.4,0.05)--(1.7,0.05);
    \draw (1.4,-0.05)--(1.7,-0.05);
    }
    + \cdots \nonumber \\
      &=\frac{1}{z- \Sigma_o(z) - \Sigma_e(z)},
      \label{eq:geometric2}
\end{align}
where 
\begin{align}
\label{eq:So-expansion}
\tikz[baseline=-0.5ex]{
    \draw[thick] (0,0) circle (10pt) node[anchor=center] {$\Sigma_o$};
    }
=& \ \,
\tikz[scale=0.8,baseline=-1.5ex]{
    \draw[dashed] (0,0) -- (1,0);
    \draw (-0.2,0.0)-- (-0.2,-0.15);
    \draw (-0.2,-0.15)-- (1.,-0.7);
    \draw (1.2,-0.15)-- (1.2,0.0);
    \draw (1.2,-0.15)-- (0,-0.7);
    \draw[densely dotted,thick] (-0.3,0.0)-- (-0.3,-0.4);
    \draw[densely dotted,thick] (1.3,-0.4)-- (1.3,0.0);
    \draw[dashed] (1,0) arc (0:180:0.5);
    \draw (1.2,0) arc (0:180:0.7);
    \draw[densely dotted,thick] (1.3,0) arc (0:180:0.8);
    }
 \ \,
+
\tikz[scale=0.5,baseline=1.5ex]{
    \draw[dashed] (0,0) -- (1,0);
    \draw (-0.2,0.)-- (-0.2,-0.15);
    \draw (-0.2,-0.15)-- (1.,-0.7);
    \draw (1.2,-0.15)-- (1.2,0.0);
    \draw (1.2,-0.15)-- (0,-0.8);
    \draw[densely dotted,thick] (-0.3,0.)-- (-0.3,-0.4);
    \draw[densely dotted,thick] (1.3,-0.1)-- (1.3,0.);
    \draw[dashed] (2,0) -- (3,0);
    \draw (1.8,0.)-- (1.8,-0.15);
    \draw (1.8,-0.15)-- (3.,-0.7);
    \draw (3.2,-0.15)-- (3.2,0.);
    \draw (3.2,-0.15)-- (2,-0.7);
    \draw[densely dotted,thick] (1.7,0.)-- (1.7,-0.1);
    \draw[densely dotted,thick] (3.3,-0.1)-- (3.3,0.);
    \draw[dashed] (4,0) -- (5,0);
    \draw (3.8,0.)-- (3.8,-0.15);
    \draw (3.8,-0.15)-- (5.,-0.8);
    \draw (5.2,-0.15)-- (5.2,0.);
    \draw (5.2,-0.15)-- (4,-0.7);
    \draw[densely dotted,thick] (3.7,0.)-- (3.7,-0.1);
    \draw[densely dotted,thick] (5.3,-0.4)-- (5.3,0.);
    \draw[densely dotted,thick] (1.3,-0.1)--(1.7,-0.1);
    \draw[densely dotted,thick] (3.3,-0.1)--(3.7,-0.1);
    \draw (1.0,-0.7)-- (2,-0.7);
    \draw (3.0,-0.7)-- (4,-0.7);
    \draw[densely dotted,thick] (5.3,0) arc (0:180:2.8);
    \draw[densely dotted,thick] (1.7,0) arc (0:180:0.2);
    \draw[densely dotted,thick] (3.7,0) arc (0:180:0.2);
    \draw[dashed] (4.0,0) arc (0:180:0.5);
    \draw[dashed] (2.0,0) arc (0:180:0.5);
    \draw[dashed] (5.0,0) arc (0:180:2.5);
    \draw (3.8,0) arc (0:180:0.3);
    \draw (1.8,0) arc (0:180:0.3);
    \draw (5.2,0) arc (0:180:2.7);
 }
 \ \,
+
\cdots,
\end{align}
\begin{align}
\label{eq:Se-expansion}
\tikz[baseline=-0.5ex]{
    \draw[thick] (0,0) circle (10pt) node[anchor=center] {$\Sigma_e$};
    }
=& \ \,
\tikz[scale=0.45,baseline=-0.5ex]{
    \draw[dashed] (0,0) -- (1,0);
    \draw (-0.2,0.)-- (-0.2,-0.15);
    \draw (-0.2,-0.15)-- (1.,-0.7);
    \draw (1.2,-0.15)-- (1.2,0.0);
    \draw (1.2,-0.15)-- (0,-0.8);
    \draw[densely dotted,thick] (-0.3,0.)-- (-0.3,-0.4);
    \draw[densely dotted,thick] (1.3,-0.1)-- (1.3,0.);
    \draw[dashed] (2,0) -- (3,0);
    \draw (1.8,0.)-- (1.8,-0.15);
    \draw (1.8,-0.15)-- (3.,-0.8);
    \draw (3.2,-0.15)-- (3.2,0.);
    \draw (3.2,-0.15)-- (2,-0.7);
    \draw[densely dotted,thick] (1.7,0.)-- (1.7,-0.1);
    \draw[densely dotted,thick] (3.3,-0.4)-- (3.3,0.);
    \draw[densely dotted,thick] (1.3,-0.1)--(1.7,-0.1);
    \draw (1.0,-0.7)-- (2,-0.7);
    \draw[densely dotted,thick] (3.3,0) arc (0:180:1.8);
    \draw[densely dotted,thick] (1.7,0) arc (0:180:0.2);
    \draw[dashed] (2.0,0) arc (0:180:0.5);
    \draw[dashed] (3.0,0) arc (0:180:1.5);
    \draw (1.8,0) arc (0:180:0.3);
    \draw (3.2,0) arc (0:180:1.7);
 }
\ \,
+
\tikz[scale=0.4,baseline=1.5ex]{
    \draw[dashed] (0,0) -- (1,0);
    \draw (-0.2,0.)-- (-0.2,-0.15);
    \draw (-0.2,-0.15)-- (1.,-0.7);
    \draw (1.2,-0.15)-- (1.2,0.0);
    \draw (1.2,-0.15)-- (0,-0.8);
    \draw[densely dotted,thick] (-0.3,0.)-- (-0.3,-0.4);
    \draw[densely dotted,thick] (1.3,-0.1)-- (1.3,0.);
    \draw[dashed] (2,0) -- (3,0);
    \draw (1.8,0.)-- (1.8,-0.15);
    \draw (1.8,-0.15)-- (3.,-0.7);
    \draw (3.2,-0.15)-- (3.2,0.);
    \draw (3.2,-0.15)-- (2,-0.7);
    \draw[densely dotted,thick] (1.7,0.)-- (1.7,-0.1);
    \draw[densely dotted,thick] (3.3,-0.1)-- (3.3,0.);
    \draw[dashed] (4,0) -- (5,0);
    \draw (3.8,0.)-- (3.8,-0.15);
    \draw (3.8,-0.15)-- (5.,-0.7);
    \draw (5.2,-0.15)-- (5.2,0.);
    \draw (5.2,-0.15)-- (4,-0.7);
    \draw[densely dotted,thick] (3.7,0.)-- (3.7,-0.1);
    \draw[densely dotted,thick] (5.3,-0.1)-- (5.3,0.);
    \draw[dashed] (6,0) -- (7,0);
    \draw (5.8,0.)-- (5.8,-0.15);
    \draw (5.8,-0.15)-- (7.,-0.8);
    \draw (7.2,-0.15)-- (7.2,0.);
    \draw (7.2,-0.15)-- (6,-0.7);
    \draw[densely dotted,thick] (5.7,0.)-- (5.7,-0.1);
    \draw[densely dotted,thick] (7.3,-0.4)-- (7.3,0.);
    \draw[densely dotted,thick] (1.3,-0.1)--(1.7,-0.1);
    \draw[densely dotted,thick] (3.3,-0.1)--(3.7,-0.1);
    \draw[densely dotted,thick] (5.3,-0.1)--(5.7,-0.1);
    \draw (1.0,-0.7)-- (2,-0.7);
    \draw (3.0,-0.7)-- (4,-0.7);
    \draw (5.0,-0.7)-- (6,-0.7);
    \draw[densely dotted,thick] (7.3,0) arc (0:180:3.8);
    \draw[densely dotted,thick] (1.7,0) arc (0:180:0.2);
    \draw[densely dotted,thick] (3.7,0) arc (0:180:0.2);
    \draw[densely dotted,thick] (5.7,0) arc (0:180:0.2);
    \draw[dashed] (6.0,0) arc (0:180:0.5);
    \draw[dashed] (4.0,0) arc (0:180:0.5);
    \draw[dashed] (2.0,0) arc (0:180:0.5);
    \draw[dashed] (7.0,0) arc (0:180:3.5);
    \draw (5.8,0) arc (0:180:0.3);
    \draw (3.8,0) arc (0:180:0.3);
    \draw (1.8,0) arc (0:180:0.3);
    \draw (7.2,0) arc (0:180:3.7);
 }
 \ \,
+
\cdots,
\end{align}
corresponding to effectively crossing and non-crossing diagrams of order $\frac{L_B}{L_{A_1}}$ and $\frac{L_B L_{A_2}}{L_{A_1}}$, respectively.

A few comments on the topology of the diagrams are in order. As we saw, PT involves a line crossing and the matrix ensemble $(XX^\dag)^{T_2}$ is not a Gaussian ensemble. Therefore, planar diagrams are not necessarily dominant terms to the resolvent function. As we explained in Appendix~\ref{app:genus}, one can assign two genus numbers $g_1$ and $g_2$ to the two subgraphs $A_1 B$ and $A_2 B$. In the $\Sigma_o$ and $\Sigma_e$ series shown above, the diagram are all planar with respect to $A_1B$, i.e., $g_1=0$, whereas $g_2$ varies between $0$  and $[\frac{k-1}{2}]$ depending on the diagrams. As discussed in Sec.~\ref{sec:diagrammatic} (below Eq.~(\ref{eq:LN_ME})), the dominant diagram corresponds to $g_2= [\frac{k-1}{2}]$. An alternative derivation of the SD equation using digarmatic approach without defining self-energies is discussed in Appendix~\ref{app:SD-alternative} where $g_1=0$ property looks more manifest. 

To derive the Schwinger-Dyson equation, we first define 
\begin{align}
\tikz[baseline=-0.5ex]{
    \draw[thick] (0,0) circle (0.25) node[anchor=center] {\footnotesize $\Sigma_o$};
    }
=
\ \,
\tikz[scale=.9,baseline=-1.5ex]{
    \draw[dashed] (-0.15,0) -- (1.15,0);
    \draw[thick,fill=white] (0.5,0) circle (0.25) node[anchor=center] {\footnotesize$F_o$};
    \draw (-0.3,0.0)-- (-0.3,-0.15);
    \draw (-0.3,-0.15)-- (1.1,-0.7);
    \draw (1.3,-0.15)-- (1.3,0.0);
    \draw (1.3,-0.15)-- (-0.1,-0.7);
    \draw (-0.6,-0.7)-- (-0.1,-0.7);
    \draw (1.1,-0.7)-- (1.6,-0.7);
    \draw[densely dotted,thick]   (-0.4,0.0)-- (-0.4,-0.4)--  (-0.6,-0.4);
    \draw[densely dotted,thick] (1.4,0.0) -- (1.4,-0.4) -- (1.6,-0.4) ;
    \draw[dashed] (1.15,0) arc (0:180:0.65);
    \draw (1.3,0) arc (0:180:0.8);
    \draw[densely dotted,thick] (1.4,0) arc (0:180:0.9);
    }\ ,
\end{align}
and
\begin{align}
\tikz[baseline=-0.5ex]{
    \draw[thick] (0,0) circle (0.25) node[anchor=center] {\footnotesize $\Sigma_e$};
    }
=
\ \,
\tikz[scale=.9,baseline=-1.5ex]{
    \draw[dashed] (-0.15,0) -- (1.15,0);
    \draw[thick,fill=white] (0.5,0) circle (0.25) node[anchor=center] {\footnotesize$F_e$};
    \draw (-0.3,0.0)-- (-0.3,-0.35)-- (1.3,-0.35) -- (1.3,0);
    \draw (-0.6,-0.55)-- (1.6,-0.55);
    \draw[densely dotted,thick]   (-0.4,0.0)-- (-0.4,-0.4)--  (-0.6,-0.4);
    \draw[densely dotted,thick] (1.4,0.0) -- (1.4,-0.4) -- (1.6,-0.4) ;
    \draw[dashed] (1.15,0) arc (0:180:0.65);
    \draw (1.3,0) arc (0:180:0.8);
    \draw[densely dotted,thick] (1.4,0) arc (0:180:0.9);
    }\ ,
\end{align}
which lead to the following algebraic relations,
\begin{align}
\Sigma_o(z)  &= \alpha F_o(z), \\
\Sigma_e(z)  &= \beta F_e(z), 
\end{align}
where the Hilbert space dimension ratios  are given by
\begin{align}
\alpha = \frac{L_B}{L_{A_1}}, \qquad
\beta= \frac{L_B L_{A_2}}{L_{A_1}}.
\end{align}

Next, we write self-consistent equations for $F$-functions as in
\begin{align}
\tikz[scale=.8,baseline=-1.5ex]{
    \draw[dashed] (-0.15,0) -- (1.15,0);
    \draw[thick,fill=white] (0.45,0) circle (0.25) node[anchor=center] {\footnotesize$F_o$};
    \draw (-0.3,0.0)-- (-0.3,-0.15);
    \draw (-0.3,-0.15)-- (1.1,-0.7);
    \draw (1.3,-0.15)-- (1.3,0.0);
    \draw (1.3,-0.15)-- (-0.1,-0.7);
    \draw (-0.6,-0.7)-- (-0.1,-0.7);
    \draw (1.1,-0.7)-- (1.6,-0.7);    \draw[densely dotted,thick]   (-0.4,0.0)-- (-0.4,-0.4)--  (-0.6,-0.4);
    \draw[densely dotted,thick] (1.4,0.0) -- (1.4,-0.4) -- (1.6,-0.4) ;
    }
=& \ 
\tikz[scale=0.7,baseline=-1.5ex]{
    \draw[dashed] (0,0) -- (1,0);
    \draw (-0.2,0.0)-- (-0.2,-0.15);
    \draw (-0.2,-0.15)-- (1.2,-0.5);
    \draw (1.2,-0.15)-- (1.2,0.0);
    \draw (1.2,-0.15)-- (-0.2,-0.5);
    \draw[densely dotted,thick] (-0.3,0.0)-- (-0.3,-0.4);
    \draw[densely dotted,thick] (1.3,-0.4)-- (1.3,0.0);
    }
  \,
+
 \,
\tikz[scale=0.8,baseline=-1.5ex]{
    \draw[dashed] (0,0) -- (1,0);
    \draw[thick,fill=white] (0.5,0) circle (0.25) node[anchor=center] {\footnotesize$F_e$};
    \draw (-0.2,0.)-- (-0.2,-0.35);
    \draw (1.1,0.)-- (1.1,-0.35);
    \draw[densely dotted,thick] (-0.3,0.)-- (-0.3,-0.4);
    \draw[densely dotted,thick] (1.2,-0.3)-- (1.2,0.);
    \draw[dashed] (2.5,0) -- (3.5,0);
    \draw (2.3,0.)-- (2.3,-0.15);
    \draw (2.3,-0.15)-- (3.8,-0.5);
    \draw (3.7,-0.15)-- (3.7,0.);
    \draw (3.7,-0.15)-- (2.3,-0.5);
    \draw[densely dotted,thick] (2.2,0.)-- (2.2,-0.3);
    \draw[densely dotted,thick] (3.8,-0.4)-- (3.8,0.);
    \draw[densely dotted,thick] (1.2,-0.3)--(2.2,-0.3);
    \draw (-0.2,-0.35)-- (1.1,-0.35);
    \draw (-0.2,-0.5)-- (2.3,-0.5);
    \draw[thick,fill=white] (1.7,-0.35) circle (0.25) node[anchor=center] {\footnotesize$G$};
    \draw[densely dotted,thick] (2.2,0) arc (0:180:0.5);
    \draw[dashed] (2.4,0) arc (0:180:0.7);
    \draw (2.3,0) arc (0:180:0.6);
 }\, , \\
\tikz[scale=.8,baseline=-1.5ex]{
    \draw[dashed] (-0.15,0) -- (1.15,0);
    \draw[thick,fill=white] (0.5,0) circle (0.25) node[anchor=center] {\footnotesize$F_e$};
    \draw (-0.3,0.0)-- (-0.3,-0.35)-- (1.3,-0.35) -- (1.3,0);
    \draw (-0.6,-0.55)-- (1.6,-0.55);
    \draw[densely dotted,thick]   (-0.4,0.0)-- (-0.4,-0.4)--  (-0.6,-0.4);
    \draw[densely dotted,thick] (1.4,0.0) -- (1.4,-0.4) -- (1.6,-0.4) ;
    }
=& 
\ \,
\tikz[scale=0.8,baseline=-1.5ex]{
    \draw[dashed] (0,0) -- (1,0);
    \draw[thick,fill=white] (0.5,0) circle (0.25) node[anchor=center] {\footnotesize$F_o$};
    \draw (-0.2,0)-- (-0.2,-0.2)-- (1.1,-0.5);
    \draw (1.1,0)-- (1.1,-0.2)-- (-0.2,-0.5);
    \draw[densely dotted,thick] (-0.3,0.)-- (-0.3,-0.4);
    \draw[densely dotted,thick] (1.2,-0.3)-- (1.2,0.);
    \draw[dashed] (2.5,0) -- (3.5,0);
    \draw (2.3,0.)-- (2.3,-0.15);
    \draw (2.3,-0.15)-- (3.8,-0.5);
    \draw (3.7,-0.15)-- (3.7,0.);
    \draw (3.7,-0.15)-- (2.3,-0.5);
    \draw[densely dotted,thick] (2.2,0.)-- (2.2,-0.3);
    \draw[densely dotted,thick] (3.8,-0.4)-- (3.8,0.);
    \draw[densely dotted,thick] (1.2,-0.3)--(2.2,-0.3);
    \draw (1.1,-0.5)-- (2.3,-0.5);
    \draw[thick,fill=white] (1.7,-0.35) circle (0.25) node[anchor=center] {\footnotesize$G$};
    \draw[densely dotted,thick] (2.2,0) arc (0:180:0.5);
    \draw[dashed] (2.4,0) arc (0:180:0.7);
    \draw (2.3,0) arc (0:180:0.6);
 }\, ,
\end{align}
which lead to the following algebraic relations
\begin{align}
F_o(z) &= 1 + F_e(z) G(z), \\
F_e(z) & = F_o(z) G(z).
\end{align}
They can be solved in terms of $G(z)$ as in
\begin{align}
\label{eq:SD-fsym}
F_e(z) = G(z)\cdot F_o(z) = \frac{G(z)}{1-G^2(z)}.
\end{align}
Putting everything together and solving for $G(z)$, we obtain a cubic equation
\begin{align}
\label{eq:SD-cubic}
z G^3 + (\beta-1) G^2 + (\alpha -z ) G +1 =0.
\end{align}
The proper solution to the above equation can be written as
\begin{align}
\label{eq:Gz_cubic_sol}
	G(z) =&  \frac{e^{-i\theta} Q_1(z)}{(Q_2(z)+\sqrt{D(z)})^{1/3}}
- e^{i\theta} (Q_2(z)+\sqrt{D(z)})^{1/3}
\nonumber \\
			& +\frac{1-\beta}{3z},
\end{align}
where $\theta=\pi/3$, and
\begin{align}
Q_1(z) &= \frac{3z(\alpha-z)-(\beta-1)^2}{9z^2}  , \\
Q_2(z) &= \frac{9z(\beta-1)(\alpha-z)-27z^2-2(\beta-1)^3}{54z^3} , \\
D(z) &=Q_1^3(z)+ Q_2^2(z).
\end{align}
Our cubic equation (\ref{eq:SD-cubic}) fully matches the spectral density derived from the Stieltjes transform in Ref.~\cite{Banica_resolvent}. 
We further note that the moments of $\rT$ is identical to those of  the difference of two independent random Wishart matrices 
(see Appendix~\ref{app:2mp} and Ref.~\cite{Banica_resolvent} for details). 

Despite the applicability of the above resolvent function to a wide range of parameters (since we only assume $N_{A_1}>N_{A_2}$), it has a drawback that it is given in terms of a solution to a cubic equation where taking the imaginary part to obtain the spectral density (as in Eq.\,(\ref{eq:dos})) leads to quite an involved expression. 
Nevertheless, we can numerically implement Eq.\,(\ref{eq:dos}) for the solution (\ref{eq:Gz_cubic_sol}) and compare with matrix simulations.
We confirm that LN matches our earlier result in Eq.~(\ref{eq:LN_regimes}) when computed via the spectral density by Eq.~(\ref{eq:neg_dist}).
Furthermore, Figure~\ref{fig:NS_vs_Lb} shows a good agreement between the analytical formula (\ref{eq:Gz_cubic_sol}) and numerical simulations as we go along a vertical line in the phase diagram (Fig.\,\ref{fig:phasediag}) from a maximally entangled state deep in the NPT regime towards a PPT state. Along this path we encounter two transitions: First, maximally entangled to saturated entangled transition (Fig.~\ref{fig:NS_vs_Lb}(c)) and second, NPT to PPT transition (Fig.~\ref{fig:NS_vs_Lb}(i)).
As we see in Figs.~\ref{fig:NS_vs_Lb}(a)-(c), the spectral density  in the maximally entangled regime  where $\alpha,\beta \ll 1$ decomposes into two disjoint distributions. Using the aforementioned property that moments of $\rT$ and difference of two independent Wishart matrices are identical, we find that each of the two distributions in this regime can be approximated by an MP distribution in the form of Eq.~(\ref{eq:MPdist}) (see Appendix~\ref{app:2mp} for more details). From this, we also find that $\braket{{\cal E}_{A_1:A_2}}=\log L_{A_2}$ which justifies our naive leading-order expansion in Eq.~(\ref{eq:RN_largeLA1}).

\begin{figure}
\includegraphics[scale=0.6]{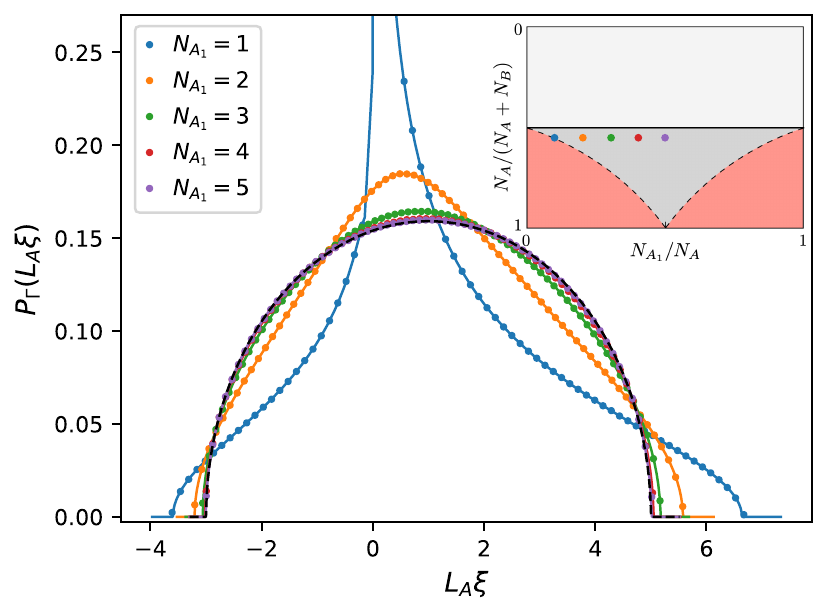}
\caption{\label{fig:Neg_vs_LA1} Negativity spectrum of a random mixed state $\rho_A$ (\ref{eq:red_den}) for different partitions of subsystem $A$. Inset shows the location of each curve (color coded) in the phase diagram. Here,
$N_B=8$, $N_A=10$, and $N_{A_2}=N_A-N_{A_1}$. Circles are numerical simulations (ensemble averaged over $10^4$ samples), solid lines are the analytical solution (\ref{eq:Gz_cubic_sol}), and dashed line shows the semi-circle law~(\ref{eq:semicirc}). Note that for $N_{A_1}=3,4,5$ where both subsystems are  smaller than half of the full system, i.e., $N_{A_1},N_{A_2}< \frac{N}{2}$, the spectrum approaches the semi-circle law regardless of the size of partitions. }
\end{figure}

Furthermore, we can reproduce the semi-circle law from the cubic equation in the right regime of parameters.
We recall that saturated entanglement corresponds to the limit $L_{A_1}\ll L_B L_{A_2}$, i.e., $\beta\gg 1$. Upon appropriate rescaling of variable $z\to y L_{A_2}$ which also implies $G(z)\to L_{A_2}^{-1}\tilde G(y)$ where $\tilde{G}(y):= G(y L_{A_2})$, we obtain
\begin{align}
 \frac{y}{L_{A_2}^2} \tilde G^3 +(q-\frac{1}{L_{A_2}^2})\tilde G^2 + (q -y ) \tilde G +1 =0,
\end{align}
in which $q=L_B/L_A$. The $1/L_{A_2}$ terms are negligible and hence we recover the SD equation for the semi-circle approximation~(\ref{eq:sd_semicirc}). We elucidate this behavior in Fig.~\ref{fig:Neg_vs_LA1} by fixing $L_A$ and $L_B$ while changing $L_{A_1}$ and $L_{A_2}$. We consider $N=10$ qubits and partition them as $N_A=10$ and $N_B=8$. As we see in this figure, for $3 \leq N_{A_i} \leq 7$ the spectral density can very well be approximated by the semi-circle law.  

\subsection{Phase diagram}

In addition, the fact that the spectral density at $z=0$ undergoes certain changes as we transition from one phase to another led us to introduce the spectral density at zero, 
\begin{align}
P(0)= -\frac{L_A}{\pi} \text{Im} \lim_{\epsilon \to 0^+} G(z=0+i\epsilon),
\end{align}
as an ``order parameter" to map out the phase diagram in Fig.~\ref{fig:phasediag}. Putting $z=0$ in Eq.\,(\ref{eq:SD-cubic}) and solving for $G$, we get
\begin{align}
\label{eq:G0}
G(0)= \frac{-\alpha + \sqrt{\alpha^2-4(\beta-1)}}{2 (\beta-1)}.
\end{align}
There are two types of phase boundaries in the phase diagram: First, the boundary shown as the dashed lines which separates the two entangled regimes with different scaling behaviors, second, the horizontal solid line which shows a transition from  saturated entangled (semi-circle law) states to PPT states. The former phase boundary is a continuation of the Page transition and characterized by a diverging spectral density. For instance, the right dashed curve corresponds to a vanishing denominator in Eq.\,(\ref{eq:G0}),
\begin{align}
\beta = 1 \quad \Rightarrow  \quad N_{A_1}^\ast= \frac{N}{2}.
\end{align}
where $N=N_A+N_B$ denotes the total number of qubits.
This means that for $N_{A_1}<N_{A_1}^\ast$ both $A_1$ and $A_2$ are smaller than half the system and we are in semi-circle regime, whereas for $N_{A_1}>N_{A_1}^\ast$ we are in the maximally entangled regime. Similar analysis can be done for the left dashed curve in Fig.~\ref{fig:phasediag}.
Moreover, we can find the diverging behavior of $G(z)$ near $z=0$ by putting $\beta=1$ in the cubic equation (\ref{eq:SD-cubic}),
\begin{align}
z G^3 + (\alpha -z ) G +1 =0.
\end{align}
The solution near $z=0$ is given by the following series expansion
\begin{align}
    G(z) =  -i \frac{{\alpha}^{1/2}}{z^{1/2}} + \frac{1}{2\alpha} + \frac{i (-3 + 4 \alpha^2)}{8 \alpha^{5/2}} z^{1/2} + O(z).
\end{align}
which implies that $P_\Gamma(\xi)\sim \xi^{-1/2}$ near $\xi=0$ similar to the Page transition for pure states.

The NPT-PPT boundary is characterized by the onset of vanishing spectral density at zero, that is given by
\begin{align}
\alpha^2 = 4 (\beta -1).
\end{align}
In terms of number of qubits, this condition can be recast as
\begin{align}
N_B = 2+ N_A + \log_2 \left( 1- 2^{2N_{A_1}-N}  \right).
\end{align}
We note that the last term is absent in the NPT-PPT transition point calculated from the semi-circle law, however, this difference is negligible in the thermodynamic limit $N\to \infty$.


In the next section, we discuss the two-point function which reveals further structure in the eigenvalues of $\rTc$ in the saturated entangled regime and indicates a clear deviation from the GUE ensemble already at the leading order.


\section{Two-Point Function}
\label{sec:2pt function}

To further characterize the negativity spectrum of random mixed states and uncover potential differences between $\rTc$ and the GUE ensemble, we compute the two-point function in this section. This calculation also provides an example where the trace normalization factor $\Tr(XX^\dagger)$ in the density matrix (\ref{eq:red_den}) can not be neglected. 
The two point function $G(z,w)$ generally provides more information than the one point function (propagator) $G(z)$ about correlations between the eigenvalues. It is defined by 
\begin{align}
	G(z,w)=\frac{1}{L_A^2}\left\langle\Tr\left(\frac{1}{z-H}\right)\Tr \left(\frac{1}{w-H}\right)\right\rangle_c
\end{align}
where $\braket{AB}_c\equiv \braket{AB}-\braket{A}\braket{B}$ denotes a connected ensemble average. We will restrict to the large-$L$ limit where $L_A,L_B\rightarrow\infty$ and $q=L_B/L_A$ is held fixed with $0<q<\infty$, such that $G(z,w)$ has a perturbative expansion in $1/L_A$. In what follows,  we first calculate the leading-order connected two-point function for the partially transposed Wishart matrix $H_0=L_B XX^\dagger$ and next, we consider $H=L_B\rho^{T_2}$ by taking into account the fluctuations of $\Tr(XX^\dagger)$.  
Our main goal here is to compute the two-point function and show that $\rT$ is different from GUE. Nevertheless, our formulas can be used to calculate joint probability distribution of two eigenvalues via the following relation
\begin{align}
    P (\xi,\zeta) &=
    \frac{1}{L_A^2}\left\langle\Tr\delta(\xi-H)\, \Tr\delta(\zeta-H)\right\rangle_c  \\
    &=
    -\frac{1}{4\pi^2}  \left(G(++)+G(--)-G(-+)-G(+-) \right)
    \nonumber
\end{align}
where $G(\pm,\pm)= G(\xi \pm i 0^+ ,\zeta \pm i 0^+)$.
We note that the diagrammatic approach is capable of addressing the eigenvalue correlations on scales larger than the spacing between the eigenvalues (where the physics is essentially controlled by level repulsion). This is because in the diagrammatic method we first take $L_A$ (or similar parameters) to infinity which gives us a systematic way to choose dominant diagrams and calculate $G(z,w)$. Eventually, we take $z$ and $w$ to approach the real axis and derive $P(\xi,\zeta)$. 
The fact that we take $L_A$ to infinity first, makes the discrete set of poles of $G(z,w)$ on the real axis merge into a branch cut, and we lose all the fine structures~\cite{Brezin:1993qg}. Therefore, the study of level statistics of $\rT$ is beyond the scope of this paper as it requires a different technology which could be subject of a separate study.

\subsection{Partially transposed Wishart matrix}

We begin by considering the two point function of the properly normalized partially transposed Wishart matrix~(\ref{eq:Wishart_normal}), $H_0= L_B (XX^\dag)^{T_2}$. Following Refs.~\onlinecite{Zee1995,Jurkiewicz}, we expand 
\begin{align}
	N_A^2 G_0(z,w)=\partial_z\partial_w\sum_{n=1}^\infty\sum_{k=1}^\infty\frac{1}{z^n w^k}\left\langle \frac{1}{n}\Tr H_0^n \frac{1}{k}\Tr H_0^k\right\rangle_c
\end{align}
Each term in the expansion can be computed diagramatically. The most general diagram looks like  a wheel
\begin{align}
\vcenter{\hbox{\includegraphics[scale=0.75]{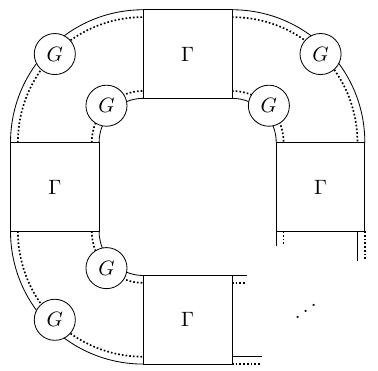}}},
\end{align}
where the inner and outer big circles represent the two traces, $G$ denotes the one-point function that we have computed to the leading order and $\Gamma$ is a four-point vertex function yet to be determined. We find that the following diagrams of $\Gamma$ contribute to the leading order of the two-point function. 
\begin{align}
\begin{tikzpicture}
	\begin{pgfonlayer}{nodelayer}
		\node [style=large box] (0) at (0, 0) {$\Gamma$};
		\node [style=none] (1) at (-0.5, 1.5) {};
		\node [style=none] (2) at (-0.5, 0.75) {};
		\node [style=none] (3) at (-0.5, -0.75) {};
		\node [style=none] (4) at (-0.5, -1.5) {};
		\node [style=none] (5) at (0.5, 1.5) {};
		\node [style=none] (6) at (0.5, 0.75) {};
		\node [style=none] (7) at (0.5, -0.75) {};
		\node [style=none] (8) at (0.5, -1.5) {};
		\node [style=none] (9) at (-3, 1.5) {};
		\node [style=none] (10) at (-3, 0.75) {};
		\node [style=none] (11) at (-3, -0.75) {};
		\node [style=none] (12) at (-3, -1.5) {};
		\node [style=none] (13) at (3, -1.5) {};
		\node [style=none] (14) at (3, -0.75) {};
		\node [style=none] (15) at (3, 0.75) {};
		\node [style=none] (16) at (3, 1.5) {};
	\end{pgfonlayer}
	\begin{pgfonlayer}{edgelayer}
		\draw (9.center) to (1.center);
		\draw (5.center) to (16.center);
		\draw (12.center) to (4.center);
		\draw (8.center) to (13.center);
		\draw [style=dotted] (10.center) to (2.center);
		\draw [style=dotted] (11.center) to (3.center);
		\draw [style=dotted] (7.center) to (14.center);
		\draw [style=dotted] (6.center) to (15.center);
	\end{pgfonlayer}
\end{tikzpicture}\quad=\quad\begin{tikzpicture}
	\begin{pgfonlayer}{nodelayer}
		\node [style=medium box] (0) at (0, 1) {$c/p$};
		\node [style=medium box] (1) at (0, -1) {$c/p$};
		\node [style=none] (2) at (0.5, 0.5) {};
		\node [style=none] (3) at (-0.5, 0.5) {};
		\node [style=none] (4) at (-0.5, 1.5) {};
		\node [style=none] (5) at (0.5, 1.5) {};
		\node [style=none] (6) at (-0.5, 0.75) {};
		\node [style=none] (7) at (0.5, 0.75) {};
		\node [style=none] (8) at (-0.5, -1.5) {};
		\node [style=none] (9) at (0.5, -1.5) {};
		\node [style=none] (10) at (-0.5, -0.5) {};
		\node [style=none] (11) at (0.5, -0.5) {};
		\node [style=none] (12) at (-0.5, -0.75) {};
		\node [style=none] (13) at (0.5, -0.75) {};
		\node [style=none] (14) at (1.5, 0.5) {};
		\node [style=none] (15) at (1.5, -0.5) {};
		\node [style=none] (16) at (-1.5, 0.5) {};
		\node [style=none] (17) at (-1.5, -0.5) {};
		\node [style=none] (18) at (1.75, 0.75) {};
		\node [style=none] (19) at (-1.75, 0.75) {};
		\node [style=none] (20) at (-1.75, -0.75) {};
		\node [style=none] (21) at (1.75, -0.75) {};
		\node [style=none] (23) at (-3, 1.5) {};
		\node [style=none] (24) at (-3, -1.5) {};
		\node [style=none] (25) at (3, -1.5) {};
		\node [style=none] (26) at (3, 1.5) {};
		\node [style=none] (27) at (2, 0.75) {};
		\node [style=none] (28) at (2, -0.75) {};
		\node [style=none] (29) at (3, 0.75) {};
		\node [style=none] (30) at (3, -0.75) {};
		\node [style=none] (31) at (-3, -0.75) {};
		\node [style=none] (32) at (-2, -0.75) {};
		\node [style=none] (33) at (-2, 0.75) {};
		\node [style=none] (34) at (-3, 0.75) {};
	\end{pgfonlayer}
	\begin{pgfonlayer}{edgelayer}
		\draw [style=dashed] (16.center) to (3.center);
		\draw [style=dashed] (16.center) to (17.center);
		\draw [style=dashed] (17.center) to (10.center);
		\draw [style=dashed] (2.center) to (14.center);
		\draw [style=dashed] (14.center) to (15.center);
		\draw [style=dashed] (15.center) to (11.center);
		\draw (7.center) to (18.center);
		\draw (18.center) to (21.center);
		\draw (21.center) to (13.center);
		\draw (12.center) to (20.center);
		\draw (20.center) to (19.center);
		\draw (19.center) to (6.center);
		\draw (23.center) to (4.center);
		\draw (5.center) to (26.center);
		\draw (24.center) to (8.center);
		\draw (9.center) to (25.center);
		\draw [style=dotted] (27.center) to (29.center);
		\draw [style=dotted] (27.center) to (28.center);
		\draw [style=dotted] (28.center) to (30.center);
		\draw [style=dotted] (34.center) to (33.center);
		\draw [style=dotted] (33.center) to (32.center);
		\draw [style=dotted] (32.center) to (31.center);
	\end{pgfonlayer}
\end{tikzpicture},
\end{align}
where the small squares on the right labeled by $c$ or $p$ are defined as 
\begin{align}
	\begin{tikzpicture}
	\begin{pgfonlayer}{nodelayer}
		\node [style=medium box] (1) at (0, 0) {$c$};
		\node [style=none] (2) at (-1.5, 0.5) {};
		\node [style=none] (4) at (-1.5, 0.25) {};
		\node [style=none] (5) at (-0.5, 0.25) {};
		\node [style=none] (6) at (-1.5, -0.5) {};
		\node [style=none] (7) at (-0.5, -0.5) {};
		\node [style=none] (8) at (0.5, 0.25) {};
		\node [style=none] (9) at (1.5, 0.25) {};
		\node [style=none] (10) at (0.5, -0.5) {};
		\node [style=none] (11) at (1.5, -0.5) {};
		\node [style=none] (12) at (0.5, 0.5) {};
		\node [style=none] (13) at (1.5, 0.5) {};
		\node [style=none] (14) at (-0.5, 0.5) {};
	\end{pgfonlayer}
	\begin{pgfonlayer}{edgelayer}
		\draw [style=dashed] (2.center) to (14.center);
		\draw [style=dashed] (12.center) to (13.center);
		\draw (4.center) to (5.center);
		\draw (6.center) to (7.center);
		\draw (8.center) to (9.center);
		\draw (10.center) to (11.center);
	\end{pgfonlayer}
\end{tikzpicture}\quad&=\quad\begin{tikzpicture}
	\begin{pgfonlayer}{nodelayer}
		\node [style=none] (0) at (-1, 0.25) {};
		\node [style=none] (1) at (-1, -0.5) {};
		\node [style=none] (2) at (1, -0.5) {};
		\node [style=none] (3) at (1, 0.25) {};
		\node [style=none] (4) at (-1, 0.5) {};
		\node [style=none] (5) at (1, 0.5) {};
	\end{pgfonlayer}
	\begin{pgfonlayer}{edgelayer}
		\draw (0.center) to (2.center);
		\draw (1.center) to (3.center);
		\draw [style=dashed] (4.center) to (5.center);
	\end{pgfonlayer}
\end{tikzpicture}\quad=\quad 1, \\
	\begin{tikzpicture}
	\begin{pgfonlayer}{nodelayer}
		\node [style=medium box] (0) at (0, 0) {};
		\node [style=medium box] (1) at (0, 0) {$p$};
		\node [style=none] (2) at (-1.5, 0.5) {};
		\node [style=none] (4) at (-1.5, 0.25) {};
		\node [style=none] (5) at (-0.5, 0.25) {};
		\node [style=none] (6) at (-1.5, -0.5) {};
		\node [style=none] (7) at (-0.5, -0.5) {};
		\node [style=none] (8) at (0.5, 0.25) {};
		\node [style=none] (9) at (1.5, 0.25) {};
		\node [style=none] (10) at (0.5, -0.5) {};
		\node [style=none] (11) at (1.5, -0.5) {};
		\node [style=none] (12) at (0.5, 0.5) {};
		\node [style=none] (13) at (1.5, 0.5) {};
		\node [style=none] (14) at (-0.5, 0.5) {};
	\end{pgfonlayer}
	\begin{pgfonlayer}{edgelayer}
		\draw [style=dashed] (2.center) to (14.center);
		\draw [style=dashed] (12.center) to (13.center);
		\draw (4.center) to (5.center);
		\draw (6.center) to (7.center);
		\draw (8.center) to (9.center);
		\draw (10.center) to (11.center);
	\end{pgfonlayer}
\end{tikzpicture}\quad&=\quad\begin{tikzpicture}
	\begin{pgfonlayer}{nodelayer}
		\node [style=basic] (0) at (0, 0) {$G$};
		\node [style=none] (1) at (-0.25, 0.25) {};
		\node [style=none] (2) at (-0.25, -0.5) {};
		\node [style=none] (3) at (-1.5, 0.25) {};
		\node [style=none] (4) at (-1.5, -0.5) {};
		\node [style=none] (5) at (0.25, 0.25) {};
		\node [style=none] (6) at (0.25, -0.5) {};
		\node [style=none] (7) at (1.5, 0.25) {};
		\node [style=none] (8) at (1.5, -0.5) {};
		\node [style=none] (9) at (-2.75, 0.25) {};
		\node [style=none] (10) at (-2.75, -0.5) {};
		\node [style=none] (12) at (2.75, -0.5) {};
		\node [style=none] (13) at (-1.75, 0.25) {};
		\node [style=none] (15) at (1.75, 0.25) {};
		\node [style=none] (16) at (-2.75, 0.5) {};
		\node [style=none] (18) at (2.75, 0.25) {};
		\node [style=none] (19) at (2.75, 0.5) {};
		\node [style=none] (20) at (2, 0.5) {};
		\node [style=none] (21) at (-2, 0.5) {};
		\node [style=none] (22) at (-1.75, 0.5) {};
		\node [style=none] (23) at (1.75, 0.5) {};
		\node [style=none] (24) at (-1.5, 0.5) {};
		\node [style=none] (26) at (1.5, 0.5) {};
	\end{pgfonlayer}
	\begin{pgfonlayer}{edgelayer}
		\draw [style=dotted] (3.center) to (1.center);
		\draw (4.center) to (2.center);
		\draw [style=dotted] (5.center) to (7.center);
		\draw (6.center) to (8.center);
		\draw (4.center) to (9.center);
		\draw (10.center) to (13.center);
		\draw (8.center) to (18.center);
		\draw (15.center) to (12.center);
		\draw [style=dashed] (16.center) to (21.center);
		\draw [style=dashed] (20.center) to (19.center);
		\draw [style=dashed, bend left=90, looseness=1.75] (21.center) to (20.center);
		\draw [bend left=90, looseness=1.75] (22.center) to (23.center);
		\draw (13.center) to (22.center);
		\draw (15.center) to (23.center);
		\draw [style=dotted, bend left=90, looseness=1.75] (24.center) to (26.center);
		\draw [style=dotted] (3.center) to (24.center);
		\draw [style=dotted] (7.center) to (26.center);
	\end{pgfonlayer}
\end{tikzpicture}\quad=\quad \frac{1}{L_{A_2}}G. 
\end{align}
These two channels are not of the same order, but one has to consider both since they differ in how the terminals are connected: $c$ for crossing means that the two solid lines switch positions after going through the $c$-type channel, while $p$ for parallel means that the solid lines preserve their positions. Let $m$ be the number of $\Gamma$ insertions in the wheel diagram above. We find that when $m=1$, each of the two small squares for $\Gamma$ can either be $c$-type or $p$-type. However, when $m>1$, only $c$-type channels contribute at the leading order. Summing up all these diagrams, we obtain
\begin{align}
L_A^2 G_0(z,w)=\partial_z\partial_w \big\{ -&\ln\left[ 1-qG(z)G(w) \right]+q G(z)G(w) \nonumber \\
& \times \left[ G(z)+G(w)+G(z)G(w) \right] \big\}. 
\end{align}
To compare this with the two-point function of GUE, it would be more insightful to recast the above expression in the following form,
\begin{align}
	L_A^2 G_0(z,w)=&\frac{G'(z)G'(w)}{[G(z)-G(w)]^2}-\left(\frac{1}{z-w}\right)^2
	+2qG'(z)
		\nonumber \\
&\times G'(w) [G(z)+G(w)+2G(z)G(w)]. 
\label{eq:Gzw_Wishart}
\end{align}
The first two terms are nothing but the GUE result whereas the last term signatures a deviation. 
This further substantiates that the partially transposed Wishart matrix $(XX^\dag)^{T_2}$ in the semi-circle regime is not a GUE matrix, besides what we found in the previous section that the $1/L$ corrections to the negativity spectrum is different from those of GUE. 
In deriving (\ref{eq:Gzw_Wishart}), we make use of the identity
\begin{align}
	1-q G(z)G(w)=\frac{(z-w)G(z)G(w)}{G(w)-G(z)},
\end{align}
which is in turn obtained from $1/G(z)=z-q(1+G)$ by combining  Eqs.~(\ref{eq:geometric}) and (\ref{eq:SD}).

\subsection{Partially transposed random mixed state}
Here, we would like to investigate the effect of the normalization factor in Eq.\,(\ref{eq:red_den}). We consider $H=L_B\rho^{T_2}$ which differs from $H_0 = L_B (XX^\dag)^{T_2}$ by the trace fluctuation. We write 
$H=H_0/(1+\delta)$ where
\begin{align}
	\delta =\Tr(XX^\dagger)-1=\frac{1}{L_B}\quad \begin{tikzpicture}
	\begin{pgfonlayer}{nodelayer}
		\node [style=none] (0) at (-0.5, 0.5) {};
		\node [style=none] (1) at (0.5, 0.5) {};
		\node [style=none] (2) at (-0.5, -0.5) {};
		\node [style=none] (3) at (0.5, -0.5) {};
		\node [style=none] (4) at (-0.75, 0.5) {};
		\node [style=none] (5) at (-0.75, -0.75) {};
		\node [style=none] (6) at (0.75, 0.5) {};
		\node [style=none] (7) at (0.75, -0.75) {};
		\node [style=none] (8) at (-1, 0.5) {};
		\node [style=none] (9) at (-1, -1) {};
		\node [style=none] (10) at (1, -1) {};
		\node [style=none] (11) at (1, 0.5) {};
	\end{pgfonlayer}
	\begin{pgfonlayer}{edgelayer}
		\draw [style=dashed] (0.center) to (2.center);
		\draw [style=dashed] (2.center) to (3.center);
		\draw [style=dashed] (3.center) to (1.center);
		\draw (4.center) to (5.center);
		\draw (5.center) to (7.center);
		\draw (7.center) to (6.center);
		\draw [style=dotted] (8.center) to (9.center);
		\draw [style=dotted] (9.center) to (10.center);
		\draw [style=dotted] (10.center) to (11.center);
	\end{pgfonlayer}
\end{tikzpicture}\quad. 
\end{align}
The $-1$ subtraction indicates that, when applying the right-most diagrammatic representation, one shall avoid self contraction loops of $\delta$. 
Let $f(z)=\Tr[1/(z-H_0)]$, we have the Taylor expansion: 
\begin{align}
	\Tr \left(\frac{1}{z-H}\right)=\sum_{n=0}^\infty [\mathcal{L}_n(z)f(z)]\delta^n, 
\end{align}
where $\mathcal{L}_n$ is a differential operator defined as 
\begin{align}
	n>0:&\quad\mathcal{L}_n(z)=\frac{1}{n!}z^n\frac{\partial^n}{\partial z^n}+\frac{1}{(n-1)!}z^{n-1}\frac{\partial^{n-1}}{\partial z^{n-1}}, \\
	n=0:&\quad\mathcal{L}_0=1. 
\end{align}
Therefore, the two-point function has the expansion: 
\begin{align}
	L_A^2 G(z,w)=\sum_{n,m}
	&\mathcal{L}_n(z)\mathcal{L}_m(w)\big[ \left\langle  f(z)f(w) \delta ^{n+m}\right\rangle 
	\nonumber \\
	&- \left\langle f(z)\delta^n \right\rangle \left\langle f(w)\delta^m \right\rangle\big]. 
\end{align}
Let us now explain how to compute expectation values with $\delta$ insertions. Take $\braket{f(z)\delta^n}$ for example. Before adding the $\delta^n$ term, we already have a set of diagrams contributing to $\braket{f(z)}$, which we call the main diagrams. Inserting the $\delta$'s represented by the staple-shaped figure may modify the original diagrams in two ways: the staples may either attach to contraction lines of the main diagram, each contributing a factor $1/(L_A L_B)$, or they can form loops among themselves (but not single loops which have been subtracted away). Obviously the latter type is more significant whenever possible because more loops are formed which give more dominant large-$L$ factors. We computed the following expectation values which contribute to the leading order of the two-point function: 
\begin{align}
	\braket{f(z)\delta }&=\frac{1}{L_A L_B}[-\mathcal{L}_1(z)]\braket{f(z)},
\end{align}
\begin{align}	
	\braket{f(z)\delta^2}&=\frac{1}{L_A L_B}\braket{f(z)}+{\rm higher}, \\
	\braket{f(z)f(w)\delta }&=\frac{1}{L_A L_B}[-\mathcal{L}_1(z)-\mathcal{L}_1(w)]\braket{f(z)f(w)},\\
	\braket{f(z)f(w)\delta^2 }&=\frac{1}{L_A L_B}\braket{f(z)f(w)}+{\rm higher}. 
\end{align}
The first line is obtained using the fact that the operator $-\mathcal{L}_1(z)$ extracts the number of contraction lines in a main diagram for $\braket{f(z)}$, similar for the third line. Combining all these results, we find 
\begin{align}
	L_A^2G(z,w)= L_A^2 &G_0(z,w) 
	-\frac{1}{q}[G(z)+zG'(z)]\nonumber \\
	&\times[G(w)+wG'(w)].
\end{align}
We see that the two-point function of $\rTc$ differs from that of GUE by the last terms in the above equation and Eq.~(\ref{eq:Gzw_Wishart}). We note that
the former contribution is proportional to $q^{-1}$ while the latter is proportional to $q$. 

\section{Discussion}
\label{sec:discussion}

In conclusion, we investigated the entanglement properties of random mixed states (of two parties $A_1$ and $A_2$ which are coupled to the bath $B$) through the window of the partial transpose and logarithmic negativity. To this end, we developed a graphical representation of a random mixed state and its partial transpose which provides a systematic large-$L$ perturbation theory to compute the R\'enyi entanglement negativities, resolvent function, and other useful quantities to characterize the density matrix.

Studying random mixed states can be thought of as adding a new axis to the 1d phase diagram of random pure states to account for how mixed the random state is due to the coupling to an external bath. In other words, we were dealing with two knobs: the relative ratio of number of qubits within our system $A$, $\frac{N_{A_1}}{N_{A_2}}$, and the relative size of the system to the bath $\frac{N_B}{N_A}$. This results in a 2d phase diagram (Fig.~\ref{fig:phasediag}) which is mirror symmetric with respect to the middle line $N_{A_1}=N_{A_2}$ similar to the 1d phase diagram of Page states which is symmetric with respect to the middle point. There are three distinct phases in the 2d phase diagram: maximally entangled, saturated entangled, and PPT (unentangled) states. In the case of NPT (entangled) states, the bath has to be smaller than the system by two qubits, i.e., $N_{B}<N_A+2$, otherwise the bath destroys the entanglement in $A$ and fully decoheres $\rho_A$ into a PPT state. The maximally entangled states are realized when either $A_1$ or $A_2$ are larger than their complement (e.g., $N_{A_s}>N_{A_{\bar s}}+ N_B$), and the logarithmic negativity is bounded by the size of smaller subsystem. 
As we increase the number of qubits in $B$, the entanglement negativity does not change until the limit $N_B = N_{A_s}-N_{A_{\bar s}}$ is reached where the system undergoes a transition and the spectral density of $\rT$ diverges at zero. The rigidity of the maximally entangled phase against increasing the bath size may be related to the fact that entanglement within $A$ is encoded in a complex way which cannot be reduced by weak perturbations. This phenomenon is reminiscent of what has recently been observed in late-time states of random quantum circuits~\cite{Gullans2020,Fan2021} where the state is said to be a quantum error-protected code in the sense that the entanglement is robust against weak projective measurements.
In the saturated entangled states both subsystems are smaller than their complements and the logarithmic negativity is related to the difference of the system size and the bath size, i.e., $N_A-N_B$. Interestingly, the entanglement in the latter case does not depend on how $\rho_A$ is partitioned. We also calculated the spectral density of $\rTa$ or the entanglement negativity spectrum for these phases in the thermodynamic limit where $N_{A_1}, N_{A_2}, N_B\to \infty$ while the ratios are held fixed. We found that to the leading order in the Hilbert size dimensions, the maximally entangled states give two disconnected MP distributions, while both saturated entangled and PPT states give a semi-circle law spectral density. The semi-circle law regime has no analog in random pure states, especially in the saturated entangled regime which describes a multipartite entangled state among $A_1$, $A_2$, and $B$.

A complementary way of describing these three entanglement phases is in terms of replica symmetry. A replica symmetry transformation corresponds to the cyclic permutation of replicas (i.e., density matrices) in the moments $\Tr[(\rT)^n]$. Hence, $\Tr[(\rT)^n]$ is manifestly invariant under this transformation. When viewed diagrammatically, the $n$-th moment looks like an $n$-site chain and the $\mathbb{Z}_n$ cyclic permutation corresponds to a shift by one lattice constant (or a discrete translation symmetry). The dominant diagrams in the maximally entangled and PPT phases given in Eqs.(\ref{eq:RT4-diag-LA1gg}) and (\ref{eq:LAll}) are invariant under the cyclic permutation; however, each term in the saturated entangled phase is not invariant (c.f.~diagrams in Appendix~\ref{app:catalan}) and the saturated entangled phase can be regarded as a replica symmetry breaking phase \footnote{This was recently emphasized in Ref.~\cite{2021arXiv210111029D}.}. This phenomenon may be helpful in understanding critical properties at the transition points. We should note that of course the overall sum remains invariant in all three phases.

 Having a systematic technique at hand to do calculations, we rigorously address potential confusions regarding the following numerical observation: The appearance of the semi-circle law spectral density at the leading order may suggest that $\rTa$ belongs to a (shifted) GUE ensemble. However, we showed that this similarity stops beyond the leading order, and the higher order corrections to the resolvent function and the connected part of two-point function of eigenvalues are manifestly different between $\rTa$ and GUE.
 
\begin{figure}
\includegraphics[scale=0.65]{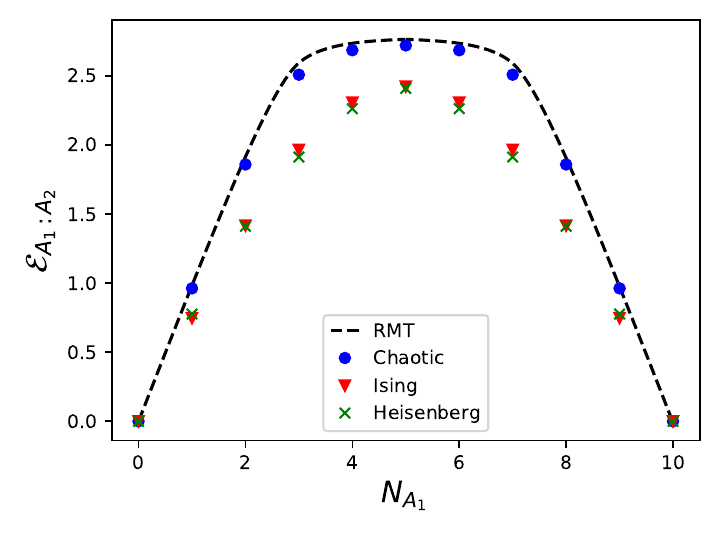}
\caption{\label{fig:spinchain} Logarithmic negativity of two adjacent intervals of length $N_{A_1}$ and $N_{A_2}$ in a highly excited state of a chaotic quantum spin-$1/2$ chain (blue circles). Here, $N_{A_1}+N_{A_2}=10$ and $N_B=4$.
For reference, we also show the corresponding plot for two integrable spin chains: Ising chain (red triangles) and Heisenberg chain (green crosses). See Appendix~\ref{app:spin-chain} for details of the numerics.
The agreement between the random matrix theory (dashed line) and chaotic spin chain is evident.
To calculate LN from random matrix theory (RMT), we numerically take the imaginary part of the solution to the cubic equation  (\ref{eq:Gz_cubic_sol}) and then calculate the negativity via Eq.~(\ref{eq:neg_dist}).
}
\end{figure}

There are several open questions and new avenues for future research.
The diagrammatic implementation of partial transpose proposed in this paper may be adapted to implement other manipulations of the random density matrix such as realignment~\cite{realign1,realign2,Aubrun_realign} and 
reflected entropy~\cite{Dutta2019} which are also considered as candidates to detect entanglement in mixed states. It would be interesting to map out the phase diagram of random mixed states by means of these measures and find out possible similarities and differences.

It is well known that random pure states share a lot of similarities with the late-time state of a system which is evolved under a chaotic Hamiltonian or an arbitrary high energy eigenstate of such Hamiltonian (which satisfies the eigenstate thermalization hypothesis~\cite{ETH_rev}).
It would be interesting to check to what extents our findings for random matrices apply to the actual states (See Ref.~\cite{Grover2020} for a recent study in this direction). To motivate this idea, we numerically checked our prediction for the logarithmic negativity curve (colored blue in  Fig.~\ref{fig:Neg_vs_L}(b)) in a highly excited state of a spin chain Hamiltonian. See Appendix~\ref{app:spin-chain} for details of the numerics. The result is plotted in Fig.~\ref{fig:spinchain} which shows good agreement. As a reference, we also provided the results for two integrable spin chains: Ising and Heisenberg, which show a stark difference, especially in the plateau regime (See also Appendix~\ref{app:spin-chain} for more examples in XXZ spin chain). More numerical investigations of this type deserves a separate study and worth pursuing. Beyond such theoretical analyses, the PPT-NPT and plateau transitions could in principle be experimentally studied in trapped-ion quantum simulators~\cite{Brydges2019} thanks to recent advances in calculating R\'enyi negativity using the correlations of randomized measurements~\cite{Elben2020}.
Another closely related setup which realizes a volume-law entangled state is time-evolution under random unitary circuits~\cite{Nahum1,*Nahum2}. Given the correspondence between the late-time states under such time evolution and random pure states~\cite{HunterJones2019}, studying the time-dependent entanglement negativity spectrum could be an interesting direction (see \cite{KudlerFlam2020} for preliminary results).
More generally, it would be interesting to apply our results to chaotic Hamiltonians using the equilibrated pure state formalism of \cite{Vardhan2020}. This may provide a direct connection between our diagrammatic approach and the replica wormholes of \cite{Shenker2019}, leading to a more detailed understanding of the entanglement structure present in the black hole evaporation process. We leave this to future work.

In this paper, we studied tripartite random Page states. It would be interesting to apply the formalism developed here to other types of random states with different structures and probability distributions such as states on random graphs~\cite{Collins_2010,*Collins_2013}, random matrix product states~\cite{Collins_mps_2013}, and
states in constrained Hilbert spaces~\cite{PhysRevLett.124.050602}.

Finally, it is worth comparing the leading order terms in the logarithmic negativity (\ref{eq:LN_regimes}) with a more commonly used quantity, namely, the mutual information (albeit it is not an entanglement measure because it is sensitive to classical correlations). First of all, the mutual information remains non-zero throughout the phase diagram with a smooth behavior across the NPT to PPT transition. Second, we can use the Page formula (\ref{eq:EE_page}) to calculate the leading order term of the mutual information as follows
\begin{align}
    \braket{I_{A_1:A_2}}  \approx  0, & \qquad  N_{A} < N_{B},
    \label{eq:MI_PPT}
\end{align}
while if $N_A > N_{B}$,
\begin{align}
    \label{eq:MI_regimes}
     \braket{I_{A_1:A_2}}  \approx \ 
    \left\{
    \begin{matrix}
    N_A-N_B, &   N_{A_s} < \frac{N}{2},
    \\ \\
    2 \min(N_{A_1},N_{A_2}), &   \text{otherwise}.
    \end{matrix}
    \right.
\end{align}
Using Eq.~(\ref{eq:LAgg_Renyi}) and (\ref{eq:LAll_Renyi}), we note that the R\'enyi mutual information also yields the same leading order result. Hence, we may write ${\cal E}=\frac{1}{2} I^{(\alpha)}$ for random mixed states where $\alpha$ is the R\'enyi index.
Although such relation for $\alpha=1/2$ typically holds for tripartite pure states of $(1+1)$d conformal field theories and integrable models in and out of equilibrium (where quasi-particle description is applicable~\cite{ac-18b,wen-2015}), it does not necessarily hold in chaotic systems (potentially due to the breakdown of quasi-particle picture \cite{2020arXiv200811266K}). From this point of view, it is a bit surprising to see that random mixed states obey this relation despite being representative chaotic states. Further investigations along this line would be illuminating.

\emph{Note Added.$-$}
On completion of this work, Ref.~\cite{Grover2020} appeared, which proposed interesting probes of quantum dynamics based on entanglement negativity, and which has some modest overlap with our results in  section~\ref{sec:diagrammatic} on the R\'enyi negativity, albeit in the limits $L_{A_1}=L_{A_2}$ and  $L_A/L_B \to 0$ or $\infty$.

\acknowledgements
The authors would like to acknowledge insightful discussions with Xiao  Chen, Tarun  Grover, Tsung-Cheng Lu, Max Metlitski, Ramis  Movassagh, Shinsei Ryu, Xue-Yang Song, Sagar Vijay, and Xueda Wen.
HS, SL, and AV were supported by a Simons Investigator award (AV) and by the Simons Collaboration on Ultra-Quantum Matter, which is a grant from the Simons Foundation (651440, AV). JKF is supported through a Simons Investigator Award to Shinsei Ryu from the Simons Foundation.

\appendix

\renewcommand\theequation{A\arabic{equation}}

\section{Derivation of Catalan number in R\'enyi negativity}
\label{app:catalan}

Here, we show how the Catalan numbers appear in the R\'enyi negativity (Eq.\,(\ref{eq:RN_catalan})) in the saturated entangled (semi-circle) regime. This can be easily seen by noting how loops of subsystem $B$ are formed. For instance, we can write
\begin{align}
  \Tr(\rT)^2=\vcenter{\hbox{\includegraphics[scale=0.24]{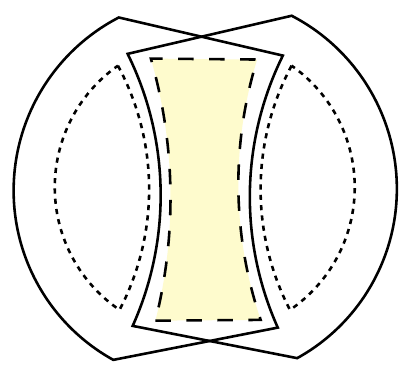}}},
\end{align}
\begin{align}
  \Tr(\rT)^4=\vcenter{\hbox{\includegraphics[scale=0.3]{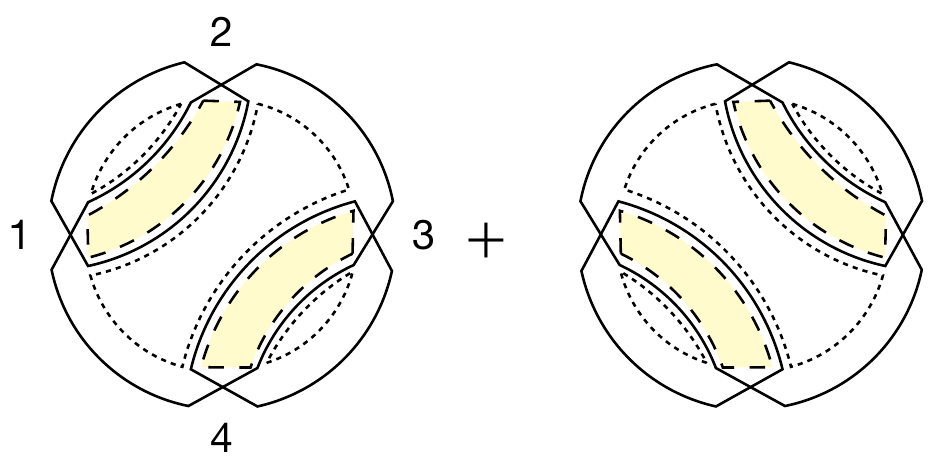}}},
\end{align}
\begin{align}
   \vcenter{\hbox{\includegraphics[scale=0.3]{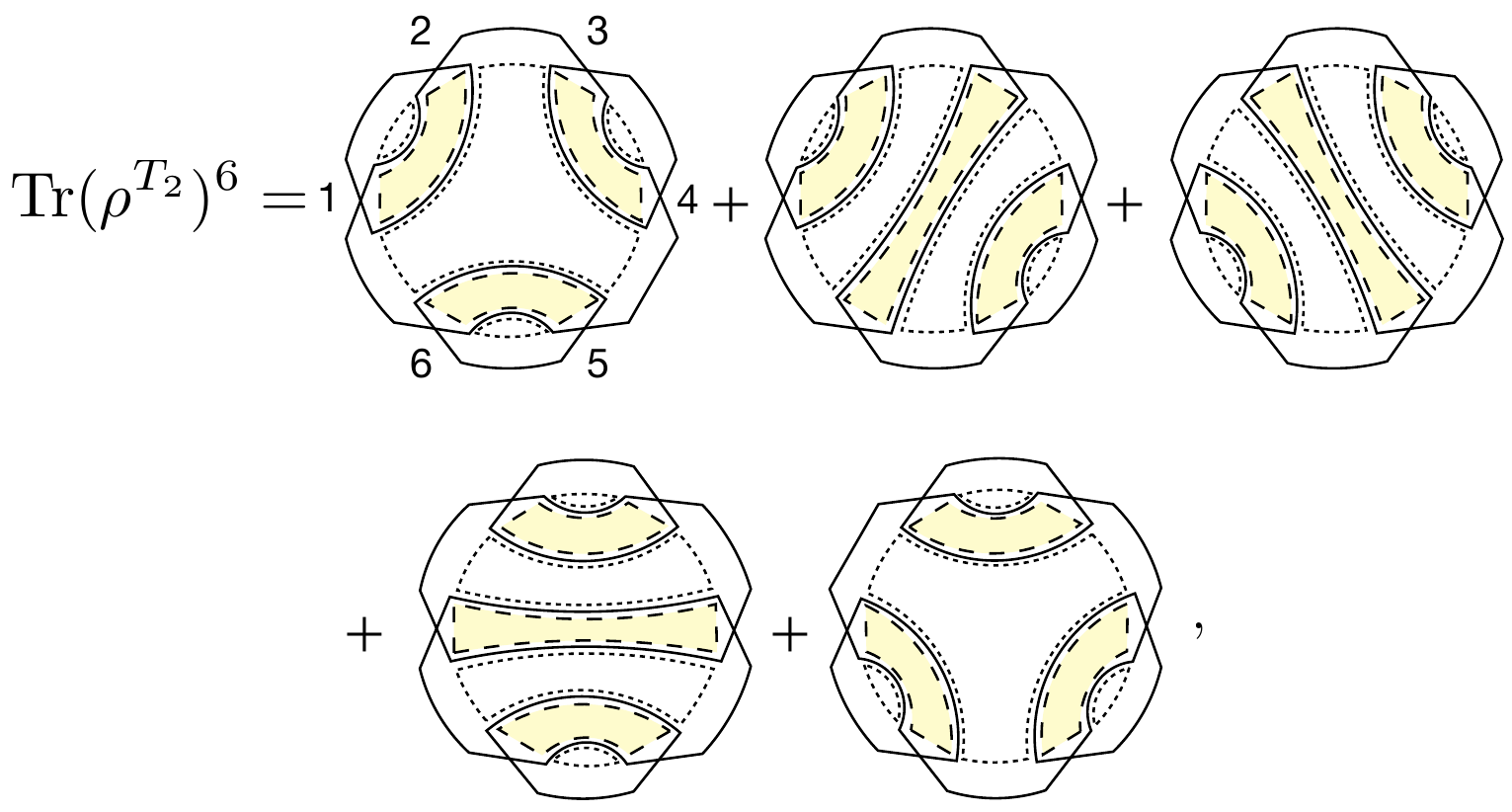}}}
\end{align}
\begin{align}
   \vcenter{\hbox{\includegraphics[scale=0.3]{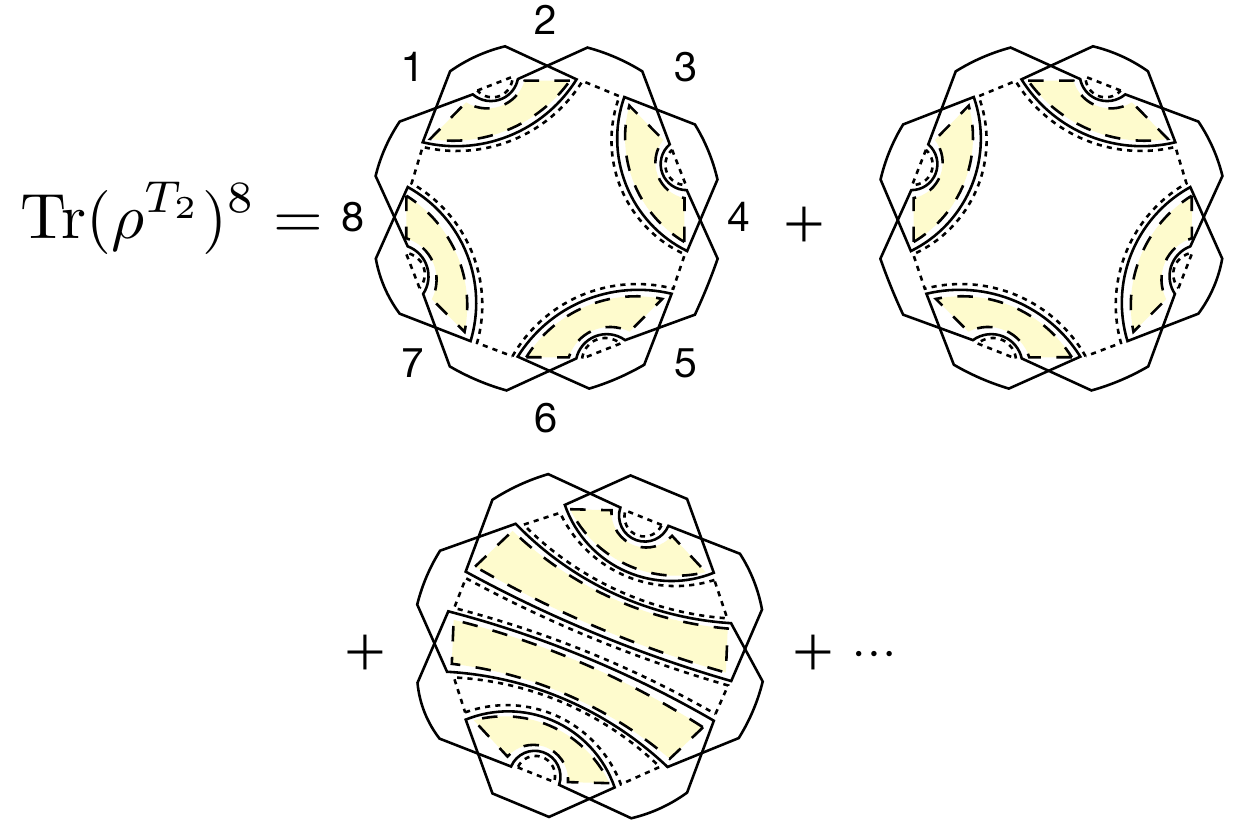}}}
\end{align}
where we show all diagrams for the first three even moments and a few for the $8$th moment all of which are of order $L_A/(L_A L_B)^k$ for $n=2k$. To emphasize the $B$ loops, we mark them by a yellow shading.
 Number of the diagrams are then equal to the number of non-crossing partitions of $k$ points, where there are two density matrices per point. For example, group the density matrices as $\{(1,2), (3,4), \cdots (2k-1,2k)\}$ where $(i,i+1)$ represents a point. Such a non-crossing partition of points is by definition the Catalan number $C_k$. A similar derivation can be carried out for $n=2k+1$ where we single out one density matrix and treat the rest similar to what is done above.

\renewcommand\theequation{B\arabic{equation}}

\section{Diagrammatic approach and the genus expansion}
\label{app:genus}

In this appendix, we discuss the genus expansion of the moments of the partially transposed density matrix. We should note that we already have the exact expression for the moments of $(XX^\dag)^{T_2}$ (see Eq.~(\ref{ExactExpression})). Here, we wish to assign a genus number to each term such as those shown in Table~\ref{tab:genus}.

In this context, a Feynman diagram is viewed as a graph. The genus of a graph is the minimal integer $g$ such that the graph can be placed on a sphere with $g$ handles (i.e., an oriented surface of genus $g$) without crossing itself. The mapping from a Feynman diagram to a graph works as follows: the lower part of the diagram is contracted to a point and the arcs associated with the ensemble averaging are regarded as edges. For instance, $k$-th moment of the density matrix has $k$ (double-line) edges and one vertex. Using Euler's relation $2-2g = V-E + F$, we have
\begin{align}
 2 - 2g = 1 - k + F, 
\end{align}
where $F$ is the number of loops in a given diagram (or faces of the corresponding graph)~\cite{Forrester}. Therefore, knowing either genus or number of loops determines the other.

As a warm-up example, let us consider the moments of the reduced density matrix (i.e., the Wishart matrix). It can be expanded as follows,
\begin{align}
    \label{eq:wishart-genus}
    \Tr ({\rho^k}) = \frac{1}{(L_A L_B)^k} \sum_{g=0}^{[(k-1)/2]} \sum_{m,n>0} a_{g,k}(m,n)\, L_B^m L_A^n ,
\end{align}
where number of loops are $F = m+ n$ or alternatively,
\begin{align}
    \label{eq:wishart-genus-exponent}
    n+m = k+1 - 2g,
\end{align}
Here, $a_{g,k}(m,n)$ denotes the number of different pairings (or degeneracy of a diagram).
To find the dominant diagram we need to maximize the exponent in  $2^{mN_B+nN_A}$ where $m$ and $n$ are subject to the constraint that $m+n\leq k+1$ because the genus $g$ in Eq.~(\ref{eq:wishart-genus-exponent}) is a non-negative integer.
For $N_A>N_B$, the maximum value of the exponent corresponds to $m=1$ and $n=k$ which implies $g=0$ (e.g., diagram of Eq.~(\ref{eq:LAgg_Renyi})). Similarly, for $N_B>N_A$ the maximum is reached by $m=k$, $n=1$, and $g=0$ (e.g., diagram of Eq.~(\ref{eq:LAll_Renyi})). When $N_A = N_B$, the exponent is proportional to $k+1-2g$ so all $g=0$ terms contribute at leading order. Therefore, the dominant term is always a zero-genus (planar) diagram. 


Now, let us consider the diagrams associated with the moments of $\rT$. They look more complicated because of the crossings at the base of the diagram. Nevertheless, we can characterize these diagrams in terms of the topology of two subgraphs $A_1B$ and $A_2 B$ and write
\begin{align}
    \label{eq:pt-genus}
     \Tr [(\rT)^k] =& \frac{1}{(L_A L_B)^k} \sum_{g_1,g_2=0}^{[(k-1)/2]} \nonumber \\
     &\sum_{m,n_1,n_2>0} b_{g_1,g_2,k}(m,n_1,n_2)\, L_B^m L_{A_1}^{n_1}  L_{A_2}^{n_2},
\end{align}
where
\begin{align}
    \label{eq:pt-genus-relation}
    n_s+m &= k+1 - 2g_s \qquad s = 1, 2,
\end{align}
and $b_{g_1,g_2,k}(m,n_1,n_2)$ is the number of different pairings with the same topology. Here, $g_s,$ denotes the genus of the subgraph composed of $A_s$ and $B$ lines after removing the $A_{\bar s}$ lines. 
Eq.~(\ref{eq:pt-genus-relation}) and the fact that exponents are positive implies that $g_s \leq (k-1)/2$. The number of loops and genus numbers for a few small moments are summarized in Table~\ref{tab:genus}. It is worth noting that there is an additional constraint on $(m,n_1,n_2,g_1,g_2)$ that the diagrams must be drawn as triple-line with $k$ crossings (due to the partial transpose). By inspection, we find that the necessary and sufficient condition is
\begin{align}
\label{eq:pt-constraint3}
    n_1 + n_2 \leq k+2 - (k\ \text{mod}\ 2).
\end{align}
We conjecture that three constraints given by the genus numbers in Eq.~(\ref{eq:pt-genus-relation}) and the above inequality produce all terms in the exact expression (\ref{ExactExpression}). The genus number equations can be explicitly derived from diagrams; however, Eq.~(\ref{eq:pt-constraint3}) was derived by inspection. To check the validity of the latter, we verified that the possible diagrams according to the genus expansion are in one-to-one correspondence with the exact expression up to $12$-th power in $\Tr(\rho^{T_2})^n$. A rigorous proof would be subject of a future work.

Now, let us discuss which term dominates in each phase. In terms of number of qubits, each term in the expansion scales as 
$2^{m N_B + n_1 N_{A_1}+ n_2 N_{A_2}}$.
First, we consider the limit $N_B, N_{A_2}< N_{A_1}$ as a representative of the maximally entangled regime. In this case, we need to maximize $n_1$ where we get $n_1=k$ and $m=1$. Using Eq.~(\ref{eq:pt-constraint3}), we find that the maximum value for $n_2$ is $1$ ($2$) for odd (even) values of $k$. Hence, the dominant diagram here has $g_1=0$ and $g_2 = [\frac{k-1}{2}]$. As for the subleading terms, we have summed over all planar diagrams with respect to $A_1 B$ subgraph as is done in Eqs.~(\ref{eq:So-expansion}) and (\ref{eq:Se-expansion}).
Similar results can be derived for the regime $N_{A_1}< N_{A_2}$ by exchanging $A_1$ and $A_2$.

The semi-circle law (entanglement saturation) regime is described as when all three subsystems are smaller than their complement. Therefore, we need to maximize all three exponents. To maximize the left-hand side of Eq.~(\ref{eq:pt-genus-relation}), we need to put $g_1=g_2=0$, which implies planar diagrams. For odd moments, the solution to Eq.~(\ref{eq:pt-genus-relation}) is given by $n_1=n_2=m = \frac{k+1}{2}$. 
For even moments, the two candidate solutions are $n_1=n_2=m+1=\frac{k}{2}+1$ and $n_1=n_2=m-1=\frac{k}{2}$, but the former term is the dominant term since it is larger by a factor of $2^{N_A-N_B}$. The corresponding diagrams are shown in Eqs.~(\ref{eq:LAgg}) and (\ref{eq:LAgg-odd}) for even and odd moments, respectively.
Lastly, we should note that $g_1=g_2=0$ diagrams are already included as subleading terms in the summation for the resolvent function in the ME phase, which means that the SD equation in this case contains the semi-circle law solution in the right limit.

Finally, in the PPT regime characterized by $N_{A}<N_B$, we need to maximize $m$. Hence, we obtain $m=k$, $n_1=n_2=1$, which implies $g_1=g_2=0$.

\begin{table}[]
    \centering
    \begin{tabular}{ccc|ccc}
         $\ k\ $ & $\ g_1\ $ &$\ g_2\ $ & $\ m\ $ & $\ n_1\ $ & 
         $\ n_2\ $  \\
         \hline
         \hline
         1 & 0 & 0 & 1 & 1 & 1  \\
         \hline
         2 &  0 & 0 & 2 & 1 & 1 \\
           & 0 & 0 & 1 & 2 & 2  \\
        \hline
         3 & 0 & 0 & 3 & 1 & 1 \\
           & 0 & 0 & 2 & 2 & 2  \\
           & 0 & 1 & 1 & 3 & 1  \\
           & 1 & 0 & 1 & 1 & 3  \\
        \hline
         4 & 0 & 0 & 4 & 1 & 1 \\
          & 0 & 0 & 3 & 2 & 2 \\
          & 0 & 0 & 2 & 3 & 3 \\
          & 1 & 0 & 2 & 1 & 3 \\
          & 0 & 1 & 2 & 3 & 1 \\
          & 1 & 1 & 2 & 1 & 1 \\
           & 1 & 0 & 1 & 2 & 4  \\
           & 0 & 1 & 1 & 4 & 2  \\
           & 1 & 1 & 1 & 2 & 2  \\
         \hline
    \end{tabular}
    \caption{Subgraph genus and number of loops for each subsystem in $\Tr[(\rT)^k]$ diagrams in Eq.~(\ref{eq:pt-genus}). For instance, the corresponding diagrams for $k=3$ are shown in Eq.~(\ref{eq:tr_rT3}).}
    \label{tab:genus}
\end{table}


\renewcommand\theequation{C\arabic{equation}}

\section{Alternative derivation of the Schwinger-Dyson equation}

\label{app:SD-alternative}

In this appendix, we present another diagrammatic way of deriving the cubic SD equation (\ref{eq:SD-cubic}) in the limit $L_{A_1}\gg L_{A_2}$. Therefore, the result is applicable to the right half of the phase diagram when $N_{A_1}>N_{A_2}$. Similar results can be obtained for the other half by exchanging $A_1$ and $A_2$.
We start with the matrix elements of the resolvent function,
\begin{align}
    \label{eq:radiation-exp}
    & \tikz[baseline=-0.5ex]{
    \draw[densely dotted,thick] (0.35,0.05)--(0.65,0.05);
    \draw[densely dotted,thick] (-0.35,0.05)--(-0.65,0.05);
    \draw (0.35,-0.05)--(0.65,-0.05);
    \draw (-0.35,-0.05)--(-0.65,-0.05);
    \draw[thick] (0,0) circle (10pt) node[anchor=center] {$\hat G$};
    } \nonumber \\
    = & \ \,
   \tikz[baseline=-0.5ex]{
    \draw[densely dotted,thick] (-0.4,0.05)--(0.35,0.05);
    \draw (-0.4,-0.05)--(0.35,-0.05);
    }
\ \,    
+\vcenter{\hbox{\includegraphics[scale=0.45]{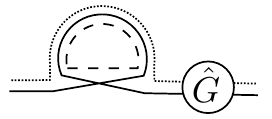}}}
\nonumber \\
&+\vcenter{\hbox{\includegraphics[scale=0.45]{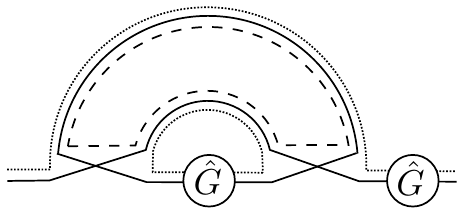}}}
\nonumber
\\
&+\vcenter{\hbox{\includegraphics[scale=0.45]{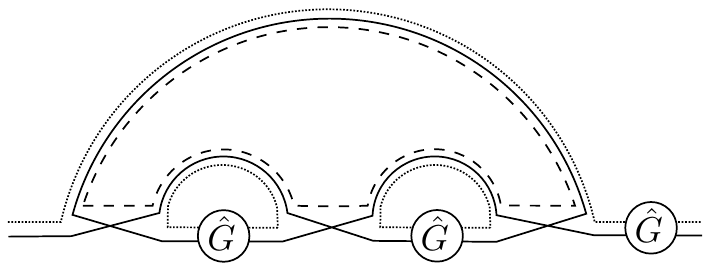}}}
\nonumber
\\
&+\vcenter{\hbox{\includegraphics[scale=0.45]{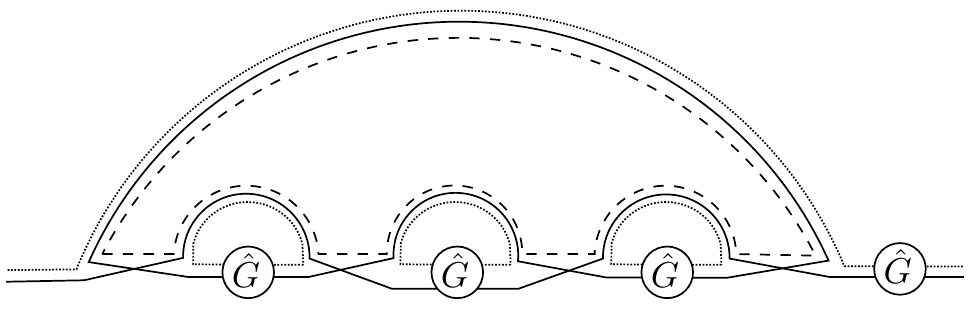}}}\nonumber \\
& + \cdots \end{align}
Hence, we get the following algebraic relation
\begin{align}
    \hat G_{ij} =& \frac{\delta_{ij}}{z} +
    \frac{1}{z L_A} \hat G_{ij} +  \frac{G}{z L_A^2 L_B} \hat G_{ij}\nonumber \\ 
    &+ \frac{1}{z L_A^3 L_B^2} [\hat G_2^2]_{i_2 k_2} \hat G_{i_1 k_2,j_1 j_2} \nonumber \\ 
    &+ \frac{1}{z L_A^4 L_B^3} [\hat G_2]_{i_2 k_2} \hat G_{i_1 k_2,j_1 j_2} \Tr(\hat G_2^2)
    + \cdots
\end{align}
where we define partial traces over the resolvent matrix
\begin{align}
    [\hat G_1]_{i_1,j_1} &= \sum_{i_2} \hat G_{i_1 i_2,j_1 i_2} = \Tr_{A_2} \hat G , \\
    [\hat G_2]_{i_2,j_2} &= \sum_{i_1} \hat G_{i_1 i_2,i_1 j_2} =\Tr_{A_1} \hat G,
\end{align}
which implies 
\begin{align}
    G(z) =  \sum_{i} \hat G_{ii}(z) = \Tr_{A_1} \hat G_1  = \Tr_{A_2} \hat G_2.
\end{align}
We may drop the subscript for the trace unless it is unclear which subsystem we are tracing out.
Performing the trace over subsystem $A_1$ and $A_2$, we arrive at
\begin{align}
    \label{eq:sd_rhoT2}
    z G = L_A &+ \sum_{k=1}  \frac{L_B}{ (L_A L_B)^{2k-1}} \Tr (G_2^{2k-1}) 
    \nonumber \\ &+ \sum_{k=1} \frac{L_B}{ (L_A L_B)^{2k}} (\Tr (G_2^{k}))^2 .
\end{align}
As mentioned in the main text, because of the Haar symmetry $\hat G$ is proportional to the identity matrix; therefore, we may write
\begin{align}
    [\hat G_2]_{i_2,j_2}= \frac{G}{L_{A_2}} \delta_{i_2,j_2},
\end{align}
which gives
\begin{align}
    \Tr(\hat G_2^n) =  \frac{G^n}{L_{A_2}^{n-1}}.
\end{align}
Hence, the SD equation above is further simplified into
\begin{align}
    \label{eq:sd_rhoT2_simplified}
    z G = L_A &+ \sum_{k=1}  \frac{L_B}{ (L_A L_B)^{2k-1}} \frac{G^{2k-1}}{L_{A_2}^{2(k-1)}} \nonumber \\
    &+ \sum_{k=1} \frac{L_B}{ (L_A L_B)^{2k}} \frac{G^{2k}}{L_{A_2}^{2(k-1)}} .
\end{align}
Carrying out the geometric series and rescaling the variables as
$z\to \frac{z}{L_{A_2}L_B}$ and $G\to L_A L_{A_2} L_B G$, we recover Eq.~(\ref{eq:SD-cubic}).
\section{Decomposing resolvent function in maximally entangled regime}
\label{app:2mp}

\renewcommand\theequation{D\arabic{equation}}

According to the free probability theory, the free cumulants $R_n(H)$ of the random matrix $H$ in (\ref{eq:Gzdef}) can be read off from the inverse of $G(z)$ (denoted by $G^{-1}(z)$) as follows
\begin{align}
    G^{-1}(z) = \frac{1}{z} + \sum_{n=1}^\infty R_n(H) z^{n-1},
\end{align}
where the sum in the right hand side is usually denoted by $R(z)$ and called $R$-transform~\cite{Speicher}. 
Free convolution theorem states that the free cumulants of a sum of two independent random Hermitian matrices is simplified into a sum of free cumulants, i.e.,
\begin{align}
    R_n(H_1+ H_2) = R_n (H_1) + R_n(H_2),
\end{align}
where $H_i$'s denote the two random matrices. Hence, the role of $R(z)$ in free probability theory is very similar to that of the log of the Fourier transform of probability distribution in regular probability theory, where the distribution of two independent random variables $x+y$ is given by the convolution of the respective probability distributions, i.e., $f(x+y)= f_1(x) * f_2(y)$ and its Fourier transform is given by a product as in $F(\omega)=F_1(\omega)F_2(\omega)$.

From the definition of self energy in Eq.~(\ref{eq:geometric}), it is easy to see that $R(G)=\Sigma(G)$.  In other words, $R$-transform is identical to the self-energy function $\Sigma(G(z))$ when expressed in terms of $G(z)$. In our case, from Eq.\,(\ref{eq:SD-fsym}) we obtain
\begin{align}
    \label{eq:Sigma}
    \Sigma(G) = \frac{\beta G+\alpha}{1-G^2}.
\end{align}
Thus, $\Sigma$ can be written as $\Sigma = \Sigma_1 + \Sigma_2$ where
\begin{align}
    \Sigma_1 &= \frac{\beta+\alpha}{2(1- G)}, \nonumber \\
    \Sigma_2 &= -\frac{\beta-\alpha}{2(1+ G)}.
\end{align}
We should note that the self energies of the form $\Sigma= t/(1- G)$ corresponds to the resolvent function of a Wishart matrix and the respective MP distribution is given by
\begin{align}
    \label{eq:MP_appendix}
    P(\lambda) &=   \frac{\sqrt{(\lambda_+-\lambda)(\lambda-\lambda_-)}}{2\pi \lambda},
\end{align}
where $\lambda_\pm = (1\pm \sqrt{t})^2$.
Similarly, $\Sigma=-t/(1+G)$ belongs to minus times a Wishart matrix with the same distribution as above up to flipping the sign of $\lambda$.
Using the free convolution theorem, we deduce that the spectral density given by the solution of the cubic equation (\ref{eq:SD-cubic}) is the same as the spectral density of the difference of two Wishart matrices, $W_1- W_2$, with parameters $t_1=\frac{1}{2}(\alpha+\beta)$ and $t_2=\frac{1}{2}(\beta-\alpha)$, respectively.
A similar property was also discussed in Ref.~\cite{Banica_resolvent}.

We can use the above simplification in terms of difference of two Wishart matrices to calculate the leading order term for the negativity in the maximally entangled regime.
As we take the thermodynamic limit, i.e., $\alpha \ll \beta \ll 1$, there is a simplification in the form of spectral density. The rank of each matrix can be computed by the area under the curve $P(\lambda)$, which is found to be $\beta L_A/2$. This means that there are $L_{A}(1-\beta/2) \gg 1$ zero eigenvalues in both $W_1$ and $W_2$. Therefore, the fact that subspaces of non-zero eigenvalues are random implies that projection operators into such subspaces for the two independent matrices $W_1$ and $W_2$ are orthogonal with high probability. This means that these two matrices are simultaneously diagonalizable or in other words they commute. A direct consequence of this argument is that the spectral density of $\rTa$ can be approximated by a sum of two MP distributions over positive and negative domains associated with $W_1$ and $W_2$, respectively. We numerically check that this property holds in Figs.~(\ref{fig:NS_vs_Lb})(a)-(c). Using the fact that $\int \lambda P(\lambda) d\lambda = t$ for the MP distribution (\ref{eq:MP_appendix}), we can further find the negativity in this limit 
\begin{align}
    \braket{{\cal N}_{A_1:A_2}} = \frac{L_A}{L_{B} L_{A_2}} t_2 = \frac{1}{2} (L_{A_2}-1),
\end{align}
where the extra factor in the first identity comes from including the correct normalization factor for the density matrix. 

\section{Chaotic spin chain}
\label{app:spin-chain}
\renewcommand\theequation{E\arabic{equation}}

In this appendix, we provide details of the numerical simulation of a highly excited state of a chaotic spin chain. We consider the following Hamiltonian for a spin-$\frac{1}{2}$ system
\begin{align}
    H_1 &= -\sum_{i=1}^N\left(Z_iZ_{i+1}+h_x X_i+h_z Z_i \right), 
\end{align}
as a representative model which realizes both integrable and non-integrable regimes~\cite{Banuls2011}.
Here, $X_i$ and $Z_i$ are Pauli matrices, we set $h_x = 1.05$, and
periodic boundary condition, $Z_{i+N}\equiv Z_i$, is imposed. We study the above Hamiltonian for two choices of parameter $h_z$. At $h_z = 0$, the Hamiltonian is nothing but the transverse-field Ising model which is integrable. Consequently, its energy level spacings have Poisson statistics. We use this limit as a reference. At $h_z=0.5$, the Hamiltonian is chaotic in the sense that its energy level spacing has Wigner-Dyson statistics. We study highly excited states of this Hamiltonian near zero energy and compare to the RMT results. We find a very close correspondence in the chaotic regime but not the integrable regime (see Fig.~\ref{fig:spinchain}). Given that our exact diagonalization scheme is limited to $14$ spins, we smoothen the results by averaging over a small band of energy eigenstates near zero.

\begin{figure}
\includegraphics[scale=0.6]{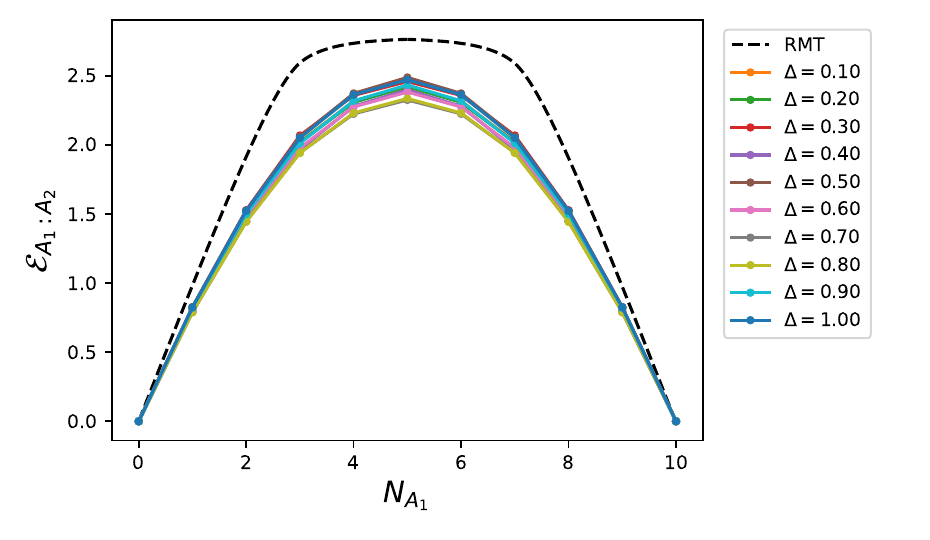}
\caption{\label{fig:xxzchain} Logarithmic negativity of two adjacent intervals of length $N_{A_1}$ and $N_{A_2}$ in a highly excited state of a XXZ spin chain (\ref{eq:xxz}) for various values of $ZZ$ coupling $\Delta$. Here, $N_{A_1}+N_{A_2}=10$ and $N_B=4$.
The deviation from the random matrix theory (dashed line) is consistent with the fact that this model is integrable.
}
\end{figure}

We also studied the anti-ferromagnetic $XXZ$ chain as another canonical example of an integrable system,
\begin{align}
    H_2 &=  \sum_{i=1}^N \left(X_iX_{i+1}+Y_i Y_{i+1}+ \Delta Z_i Z_{i+1} \right).
    \label{eq:xxz}
\end{align}
Figure \ref{fig:xxzchain} shows that the entanglement negativity curves deviate from the random matrix theory predictions as expected.

\bibliography{refs.bib}

\end{document}